\documentclass[aps,pre,onecolumn,longbibliography,nofootinbib,showkeys,showpacs,amssymb]{revtex4-1}
\usepackage{latexsym}
\usepackage{amsfonts}
\usepackage{color}
\usepackage{graphicx}
\usepackage{mathptmx}      % use Times fonts if available on your TeX system
\usepackage{bm}% bold math symbols \bm\alpha
\usepackage{enumerate}
\usepackage{subfigure}
\usepackage{listings}
\usepackage{grffile}
\begin{document}
\title{Dynamics of open quantum spin systems:
An assessment of the quantum master equation approach}

\author{P. Zhao}
\affiliation{Zernike Institute for Advanced Materials,\\
University of Groningen, Nijenborgh 4, NL-9747AG Groningen, The Netherlands}
\author{H. De Raedt}
\email{h.a.de.raedt@rug.nl}
\thanks{Corresponding author}
\affiliation{Zernike Institute for Advanced Materials,\\
University of Groningen, Nijenborgh 4, NL-9747AG Groningen, The Netherlands}
\author{S. Miyashita}
\affiliation{
Department of Physics, Graduate School of Science,\\
University of Tokyo, Bunkyo-ku, Tokyo 113-0033, Japan
}
\author{F. Jin}
%\email{f.jin@fz-juelich.de}
\affiliation{Institute for Advanced Simulation, J\"ulich Supercomputing Centre,\\
Forschungszentrum J\"ulich, D-52425 J\"ulich, Germany}
\author{K. Michielsen}
\affiliation{Institute for Advanced Simulation, J\"ulich Supercomputing Centre,\\
Forschungszentrum J\"ulich, D-52425 J\"ulich, Germany}
\affiliation{RWTH Aachen University, D-52056 Aachen, Germany}

\date{\today}

\begin{abstract}
Data of the numerical solution of the time-dependent Schr\"odinger equation of a system containing one spin-1/2 particle
interacting with a bath of up to
32 spin-1/2 particles is used to construct a Markovian quantum master equation describing the dynamics of the system spin.
The procedure of obtaining this quantum master equation, which takes the form of a Bloch equation with time-independent coefficients,
accounts for all non-Markovian effects in as much the general structure of the quantum master equation allows.
Our simulation results show that, with a few rather exotic exceptions, the Bloch-type equation with time-independent coefficients
provides a simple and accurate description of the dynamics of a spin-1/2 particle in contact with a thermal bath.
A calculation of the coefficients that appear in the
Redfield master equation in the Markovian limit shows
that this perturbatively derived equation quantitatively differs from
the numerically estimated Markovian master equation, the results of which agree
very well with the solution of the time-dependent Schr\"odinger equation.
\end{abstract}

%% insert suggested PACS numbers in braces on next line
%% insert suggested keywords - APS authors don't need to do this

\maketitle

\section{Introduction}\label{section1}

In general, a physical system can seldom be considered as completely isolated from its environment.
Such closed systems can and should, of course, be studied in great detail.
However, as they lack the ability to interact with the environment in which they are embedded
or with the apparatus that is used to perform measurements on it,
such studies do not include the effects of the, usually uncontrollable,
environment which may affect the dynamics of the system in a non-trivial manner.
The alternative is to consider the system of interest as an open system
that is a system interacting with its environment.

The central idea of theoretical treatments of open quantum systems
is to derive approximate equations of motion of the system by elimination
of the environmental degrees of freedom~\cite{REDF57,NAKA58,ZWAN60,BREU02}.
In 1928, Pauli derived a master equation
for the occupation probabilities of a quantum subsystem
interacting with the environment~\cite{PAUL28}.
Since then, various methods have been developed to derive
quantum master equations starting from the Liouville-von Neumann equation for the density matrix
of the whole system~\cite{REDF57,NAKA58,ZWAN60,PLEN98,BREU02}.
In order to obtain an equation of motion for the system which
is tractable and readily amenable to detailed analysis,
it is customary to make the so-called Markov approximation,
which in essence assumes that the correlations of
the bath degrees of freedom vanish on a short time span.

Without reference to any particular model system, in 1970,
Lindblad derived a quantum master equation
which is Markovian and which preserves positivity (a non-negative definite density matrix)
during the time evolution~\cite{LIND76,BREU02}.
The applicability of the Lindblad master equation is restricted to
baths for which the time correlation functions of the operators
that couple the system to the bath are essentially $\delta$-functions~\cite{GASP99},
an assumption that may be well justified in quantum optics~\cite{PLEN98}.

Using second-order perturbation theory, Redfield derived a master equation which does
not require the bath correlations to be approximately $\delta$-functions in time~\cite{REDF57}.
The Redfield master equation has found many applications to problems where the
dynamics of the bath is faster than that of the system, for instance to the case of
nuclear magnetic resonance in which the system consists of one spin coupled to other spins and/or to phonons.
This approach and variations of it have been successfully applied to study
the natural linewidth of a two-level system~\cite{LOUI73,ABRA61,COHE92},
systems of interacting spins~\cite{HAMA84} and nonlinear spin relaxation~\cite{SHIB80}.

The Redfield master equation can be systematically derived from the principles
of quantum theory but only holds for weak coupling.
However, the Redfield master equation
may lead to density matrices that are not always positive,
in particular when the initial conditions are such that
they correspond to density matrices that close to the boundary of physically admissible
density matrices~\cite{SUAR92,PECH94}.

Obviously, the effect of the finite correlation time of the
thermal bath becomes important when the time scale of the system is comparable
to that of the thermal bath.
Then the Markovian approximation may no longer be adequate and
in deriving the quantum master equation, it becomes necessary
to consider the non-Markovian aspects and to treat the
initial condition correctly~\cite{SASS90,WEIS99,BREU02,TANI06,BREU06,MORI08,SAEK08,UCHI09,MORI14,CHEN15}.

By introducing the concept of slippage in the initial conditions,
it was shown that the Markovian equations of motion obtained in the weak coupling regime
are a consistent approximation to the actual reduced dynamics and that slippage captures the effects of
the non-Markovian evolution that takes place in a short transient time, of the order of
the relaxation time of the isolated bath~\cite{SUAR92}.
Provided that nonlocal memory effects that take place on a very short time scale
are included, the Markovian approximation that preserves the symmetry of the Hamiltonian
yields an accurate description of the system dynamics~~\cite{SUAR92}.
Following up on this idea, a general form of a slippage operator
to be applied to the initial conditions of the Redfield master equation was derived~\cite{GASP99}.
The slippage was expressed in terms of an operator describing
the non-Markovian dynamics of the system during the time in which the bath relaxes on its own, relatively short, time scale.
It was shown that the application of the slippage superoperator to the
initial density matrix of the system yields a Redfield master equation that preserves positivity~\cite{GASP99}.
Apparently, the difference between the non-Markovian dynamics and its Markovian approximation
can be reduced significantly by first applying the slippage operator and then
letting the system evolve according to the Redfield master equation~\cite{GASP99}.

The work discussed and cited earlier almost exclusively focuses on models
of the environment that are described by a collection of harmonic oscillators.
In contrast, the focus of this paper is on the description
of the time evolution of a quantum system with one spin-1/2 degree of freedom
coupled to a larger system of similar degrees of freedom, acting as a thermal bath.
Our reasons for focusing on spin-1/2 models are twofold.

First, such system-bath models are relevant for
the description of relaxation processes in nuclear magnetic and electron spin resonance~\cite{KUBO57,REDF57,ABRA61}
but have also applications to, e.g. the field of quantum information processing,
as most of the models used in this field are formulated in terms of qubits (spin-1/2 objects)~\cite{NIEL10,JOHN11}.

Secondly, the aim of the present work is to present a quantitative assessment of
the quantum master equation approach by comparing the results with those
obtained by an approximation-free, numerical solution
of the time-dependent Schr\"odinger equation of the system+bath.
The work presented in this paper differs from earlier numerical work on
dissipative quantum dynamics~\cite{WANG00,NEST03,GELM03,GELM04,KATZ08}
by accounting for the non-trivial many-body dynamics of the bath without resorting to approximations,
at the expense of using much more computational resources.
Indeed, with state-of-the-art computer hardware, e.g the IBM BlueGene/Q, and corresponding simulation software~\cite{RAED07X},
it has become routine to solve the time-dependent
Schr\"odinger equation for systems containing up to 36 spin-1/2 objects.
As we demonstrate in this paper, this allows us to mimic a large thermal bath
at a specific temperature and solve for the full dynamic evolution of a
spin-1/2 object coupled to the thermal bath of spin-1/2 objects.

From the numerically exact solution of the Schr\"odinger dynamics
we compute the time-evolution of the density matrix of the system
and by least-square fitting, obtain the ``optimal'' quantum master equation
that approximately describes the same time-evolution.
For a system of one spin-1/2 object, this quantum master equation
takes the form of a Bloch equation with time-independent coefficients.
Clearly, this procedure of obtaining the quantum master equation is free of any approximation
and accounts for all non-Markovian effects in as much the general structure of the quantum master equation allows.
Our simulation results show that, with a few rather exotic exceptions,
the Bloch-type equation with time-independent coefficients
provides a very simple and accurate description of the dynamics of a spin-1/2 object in contact with a thermal bath.

The paper is organized as follows.
In section~\ref{section2}, we give the Hamiltonians that specify the system, bath and system-bath interaction.
Section~\ref{section3} briefly reviews the numerical techniques that we use to solve the
time-dependent Schr\"odinger equation, to compute the density matrix, and to prepare the
bath in the thermal state at a given temperature.
We also present simulation results that demonstrate that the method of preparation
yields the correct thermal averages.
For completeness, Sec.~\ref{section4} recapitulates the standard derivation of
the quantum master equation, writes the formal solution of the latter in a form
that is suited for our numerical work and shows that the Redfield % and Lindblad quantum master
equations have this form.
We then use the simulation tool to compute the correlations of the bath-operators that
determine the system-bath interaction and discuss their relaxation behavior.
Section~\ref{section5} explains the least-square procedure of extracting,
from the solution of the time-dependent Schr\"odinger equation,
the time-evolution matrix and the time-independent contribution that
determine the ``optimal'' quantum master equation.
This least-square procedure is validated by its application to data
that originate from the Bloch equation, as explained in Appendix~\ref{section6}.
In Sec.~\ref{section7}, we specify the procedure by which
we fit the quantum master equation to the data
obtained by solving the time-dependent Schr\"odinger equation
and present results of several tests.
The results of applying the fitting procedure to baths
containing up to 32 spins are presented in Sec.~\ref{section8}.
Finally, in Sec.~\ref{section9}, we discuss some exceptional cases
for which the quantum master equation is not expected to provide
a good description.
The paper concludes with the summary, given in Sec.~\ref{section10}.

%%%%%%%%%%%

%%%%%%%%%%%%%%%%%%%%%%%%%%%%%%%%%%%%%%%%%%%%%%%%%%%%%%%%%%%%%%%%%%%%%%%%%%%%%%%%%%%%%%%%%%%%%%%%%%%%%%%%%%%%%%%%%%%%%%%%%%

\section{System coupled to a bath: Model}\label{section2}

The Hamiltonian of the system (S) + bath (B) takes the generic form
\begin{eqnarray}
H&=& H_\mathrm{S} + H_\mathrm{B} + \lambda H_\mathrm{SB}
.
\label{s40}
\end{eqnarray}
The overall strength of the system-bath interaction is controlled by the parameter $\lambda$.
In this work, we limit ourselves to a system which consists of one spin-1/2 object described by the Hamiltonian
\begin{eqnarray}
%H_\mathrm{S}&=& -h^x \sigma^x_{0} -h^z \sigma^z_{0}
H_\mathrm{S}&=& -h^x \sigma^x_{0} %\frac{1}{2}
,
\label{s41}
\end{eqnarray}
where $\bm{\sigma}_{n}=(\sigma^x_{n},\sigma^y_{n},\sigma^z_{n})=(\sigma^1_{n},\sigma^2_{n},\sigma^3_{n})$
denote the Pauli-spin matrices for spin-1/2 object $n$,
and $h^x$ is a time-independent external field.
Throughout this paper, we adopt units such that $\hbar=1$ and
$h^x=1/2$ and express time in units of $\pi/h^x$.
We will use the double notation with the $(x,y,z)$ and $(1,2,3)$ superscripts
because depending on the situation, it simplifies the writing considerably.

The Hamiltonian for the system-bath interaction is chosen to be
\begin{eqnarray}
H_\mathrm{SB}&=& -\sum_{n=1}^{N_{\mathrm{B}}}\left( % \frac{1}{4}
J^x_{n}\sigma^x_{n}\sigma^x_{0}+
J^y_{n}\sigma^y_{n}\sigma^y_{0}+
J^z_{n}\sigma^z_{n}\sigma^z_{0}\right)
= \sum_{\alpha=x,y,z} \sigma^\alpha_{0} B_\alpha % \frac{1}{2}
= \sum_{i=1}^3 \sigma^i_{0} B_i % \frac{1}{2}
,
\label{s43}
\end{eqnarray}
where $N_{\mathrm{B}}$ is the number of spins in the bath,
the $J^\alpha_{n}$ are real-valued random numbers in the range $[-J,+J]$ and
\begin{eqnarray}
B_x&=&B_1=-\sum_{n=1}^{N_{\mathrm{B}}} J^x_{n}\sigma^x_{n} % \frac{1}{2}
%\nonumber \\
\quad,\quad
B_y=B_2=-\sum_{n=1}^{N_{\mathrm{B}}} J^y_{n}\sigma^y_{n} % \frac{1}{2}
%\nonumber \\
\quad,\quad
B_z=B_3=-\sum_{n=1}^{N_{\mathrm{B}}} J^z_{n}\sigma^z_{n} % \frac{1}{2}
\label{s43a}
\end{eqnarray}
are the bath operators which, together with
the parameter $\lambda$, define the system-bath interaction.
As the system-bath interaction strength is controlled by $\lambda$, we may set $J=1/4$
without loss of generality.

As a first choice for the bath Hamiltonian $H_\mathrm{B}$ we take
\begin{eqnarray}
H_\mathrm{B}&=&
-K \sum_{n=1}^{N_{\mathrm{B}}}\left( \sigma^x_{n}\sigma^x_{n+1}+\sigma^y_{n}\sigma^y_{n+1}+\Delta\sigma^z_{n}\sigma^z_{n+1}\right) % \frac{1}{4}
-\sum_{n=1}^{N_{\mathrm{B}}}\left( h^x_{n} \sigma^x_{n}+h^z_{n} \sigma^z_{n}\right) % \frac{1}{2}
.
\label{s42}
\end{eqnarray}
The fields $h^x_{n}$ and $h^z_{n}$ are real-valued random numbers in the range
$[-h^x_\mathrm{B},+h^x_\mathrm{B}]$ and $[-h^z_\mathrm{B},+h^z_\mathrm{B}]$, respectively.
In our simulation work, we use periodic boundary conditions $\sigma^\alpha_{n}=\sigma^\alpha_{n+N_{\mathrm{B}}}$
for $\alpha=x,y,z$. Note that we could have opted equally well to use open-end boundary conditions but
for the sake of simplicity of presentation, we choose the periodic boundary conditions.
For $\Delta=1$, the first term in Eq.~(\ref{s42}) is the Hamiltonian of the one-dimensional (1D) Heisenberg model on a ring.

As a second choice, we consider the 1D ring with Hamiltonian
\begin{eqnarray}
H_\mathrm{B}&=& -\sum_{n=1}^{N_{\mathrm{B}}}\left( % \frac{1}{4}
K^x_n\sigma^x_{n}\sigma^x_{n+1}
+K^y_n\sigma^y_{n}\sigma^y_{n+1}
+K^z_n\sigma^z_{n}\sigma^z_{n+1}
\right)
-\sum_{n=1}^{N_{\mathrm{B}}}\left( h^x_{n} \sigma^x_{n}+h^z_{n} \sigma^z_{n}\right) % \frac{1}{2}
,
\label{s42a}
\end{eqnarray}
where the $K^x_n$'s, $K^y_n$'s, and $K^z_n$'s are uniform random numbers in the range $[-K,K]$.
Because of the random couplings,
it is unlikely that it is integrable (in the Bethe-ansatz sense)
or has any other special features such as conserved magnetization etc.

The bath Hamiltonians (\ref{s42}) and (\ref{s42a}) both share the property that
the distribution of nearest-neighbor energy levels is of Wigner-Dyson-type,
suggesting that the correspondig classical baths exhibit chaos.
Earlier work along the lines presented in this paper has shown that
spin baths with a Wigner-Dyson-type distribution are more effective as
sources for fast decoherence than spin baths with Poisson-type distribution~\cite{LAGE05}.
Fast decoherence is a prerequisite for a system to exhibit fast relaxation to the
thermal equilibrium state~\cite{YUAN09,JIN10x}.
Extensive simulation work on spin-baths with very different degrees of connectivity~\cite{YUAN07,YUAN08,JIN13a,NOVO16}
suggest that as long as there is randomness in the system-bath coupling and
randomness in the intra-bath couplings, the simple models (\ref{s42}) and (\ref{s42a})
may be considered as generic spin baths.

Finally, as a third choice, we consider
\begin{eqnarray}
H_\mathrm{B}&=& -\sum_{\langle n,n'\rangle}\left( % \frac{1}{4}
K^x_{n,n'}\sigma^x_{n}\sigma^x_{n'}
+K^y_{n,n'}\sigma^y_{n}\sigma^y_{n'}
+K^z_{n,n'}\sigma^z_{n}\sigma^z_{n'}
\right)
-\sum_{n=1}^{N_{\mathrm{B}}}\left( h^x_{n} \sigma^x_{n}+h^z_{n} \sigma^z_{n}\right) % \frac{1}{2}
,
\label{s42b}
\end{eqnarray}
where the $K^x_{n,n'}$'s, $K^y_{n,n'}$'s, and $K^z_{n,n'}$'s are uniform random numbers in the range $[-K,K]$,
and $\sum_{\langle n,n'\rangle}$ denotes the sum over all pairs of nearest neighbors on a
three-dimensional (3D) cubic lattice.
Again, because the random couplings and the 3D connectivity,
it is unlikely that it is integrable or has any other special features such as conserved magnetization etc.
As the solution of the time-dependent Schr\"odinger equation (TDSE)
for the 3D model Eqs.~(\ref{s42b}) takes about a factor of 2
more CPU time than in the case of a 1D model with the same number of bath spins,
in most of our simulations we will use the 1D models and
only use the 3D model to illustrate that the connectivity of the bath is not a relevant factor.

%%%%%%%%%%%%%%%%%%%%%%%%%%%%%%%%%%%%%%%%%%%%%%%%%%%%%%%%%%%%%%%%%%%%%%%%%%%%%%%%%%%%%%%%%%%%%%%%%%%%%%%%
\section{Quantum dynamics of the whole system}\label{section3}

The time evolution of a closed quantum system defined
by Hamiltonian (\ref{s40}) is governed by the TDSE
\begin{eqnarray}
i\frac{\partial}{\partial t} |\Psi(t)\rangle &=& H|\Psi(t)\rangle
.
\label{s44}
\end{eqnarray}%
The pure state $|\Psi(t)\rangle$ of the whole system $\mathrm{S}+\mathrm{B}$ evolves in time according to
\begin{eqnarray}
|\Psi(t)\rangle
&=&
e^{-itH}|\Psi(0)\rangle=\sum_{i=1}^{D_{\mathbf{S}}} \sum_{p=1}^{D_{\mathrm{B}}} c(i,p,t)|i,p\rangle
,
\label{s45}
\end{eqnarray}%
where $D_{\mathrm{S}}=2$ and $D_{\mathrm{B}}=2^{{N_\mathrm{B}}}$ are the dimensions of the Hilbert space
of the system and bath, respectively.
The coefficients $\{c(i,p,t)\}$ are the complex-valued amplitudes of
the corresponding elements of the set $\{ |i,p\rangle \}$ which denotes
the complete set of the orthonormal states in up--down basis of the system and bath spins.

The size of the quantum systems that can be simulated, that is the
size for which Eq.~(\ref{s45}) can actually be computed, is primarily
limited by the memory required to store the pure state.
Solving the TDSE requires storage of all the numbers $\{ c(i,p,t)|i=1,2\;,p=1,\ldots,2^{N_\mathrm{B}}\}$.
Hence the amount of memory that is required is proportional to $2^{N_\mathrm{B}+1}$,
that is it increases exponentially with the number of spins of the bath.
As the number of arithmetic operations also increases exponentially,
it is advisable to use 13 - 15 digit floating-point arithmetic
(corresponding to $16=2^4$ bytes for each pair of real numbers).
Therefore, representing a pure state of $N_\mathrm{B}+1$ spin-$1/2$ objects
on a digital computer requires at least $2^{N_{\mathrm{B}}+5}$ bytes.
For example, for $N_\mathrm{B}=23$ ($N_\mathrm{B}=35$) we need at least 256~MB (1~TB) of memory to store a single state $|\Psi(t)\rangle$.
In practice we need storage for three vectors, and memory for communication buffers, local variables and the code itself.

The CPU time required to advance the pure state by one time step $\tau$
is primarily determined by the number of operations to be performed on the state vector,
that is it also increases exponentially with the number of spins.
The elementary operations performed by the computational kernel
can symbolically be written as $|\Psi \rangle \leftarrow U |\Psi\rangle$
where the $U$'s are sparse unitary matrices with a relatively complicated structure.
A characteristic feature of the problem at hand is that for most of the $U$'s,
all elements of the set $\{c(i,p,t)|i=1,2\;,p=1,2^{N_\mathrm{B}}\}$ are involved in the operation.
This translates into a complicated scheme for efficiently accessing memory,
which in turn requires a sophisticated communication scheme \cite{RAED07X}.

We can exclude that the conclusions that we draw from the numerical results are
affected by the algorithm used to solve the TDSE by performing
the real-time propagation by $e^{-itH}$ by means
of the Chebyshev polynomial algorithm~\cite{TALE84,LEFO91,IITA97,DOBR03}.
This algorithm is known to yield results that are very accurate
(close to machine precision), independent of the time step used~\cite{RAED06}.
A disadvantage of this algorithm is that, especially when
the number of spins exceeds 28, it consumes significantly
more CPU and memory resources than a Suzuki-Trotter product-formula based algorithm~\cite{RAED06}.
Hence, once it has been verified that the numerical results
of the latter are, for practical purposes, as good
as the numerically exact results, we use the latter for the
simulations of the large systems.

\subsection{Density matrix}\label{section3a}

According to quantum theory, observables are represented by Hermitian matrices
and the correspondence with measurable quantities is through their averages defined as~\cite{BALL03}
\begin{equation}
\langle {\cal A} \rangle = \mathbf{Tr\;} \rho(t) {\cal A}
,
\label{s46a}
\end{equation}
where ${\cal A}$ denotes a Hermitian matrix representing the observable,
$\rho(t)$ is the density matrix of the whole system $S+B$ at time $t$
and $\mathbf{Tr\;}$ denotes the trace over all states of the whole system $\mathrm{S}+\mathrm{B}$.
If the numerical solution of the TDSE for a pure state of
$N_\mathrm{B}+1$ spins already requires  resources that
increase exponentially with the number of spins of the bath,
computing Eq.~(\ref{s46a}) seems an even more daunting task.
Fortunately, we can make use of the ``random-state technology''
to reduce the computational cost to that of solving the TDSE for one pure state~\cite{HAMS00}.
The key is to note that if $|\Phi\rangle$ is a pure state, picked randomly
from the $2^{{N_\mathrm{B}+1}}$-dimensional unit hypersphere,
one can show in general that for Hermitian matrices ${\cal A}$~\cite{HAMS00}
\begin{equation}
\mathbf{Tr\;} {\cal A} = D\langle\Phi| {\cal A}|\Phi\rangle \pm {\cal O}(D^{-1/2})
,
\label{s46b}
\end{equation}
where $D$ is the number of diagonal elements of the matrix ${\cal A}$ (= the dimension of the Hilbert space)
and $\pm {\cal O}(x)$ should be read as saying that the standard deviation
is of order $x$.
For the case at hand $D=2^{N_\mathrm{B}+1}$, hence Eq.~(\ref{s46b}) indicates that
for a large bath, the statistical errors resulting from approximating
$\mathbf{Tr\;} {\cal A}$ by $\langle\Phi| {\cal A}|\Phi\rangle$ vanishes
exponentially with the number of bath spins.
For large baths, this property makes the problem amenable to numerical simulation.
Therefore, from now on, we replace the ``$\mathbf{Tr\;}$'' by
a matrix element of a random pure state whenever the trace operation
involves a number of states that increases exponentially with the
number of spins (in the present case, bath spins only).

The state of the system $S$ is completely described by the reduced density matrix
\begin{equation}
\rho_{\mathrm{S}}(t)\equiv\mathbf{Tr}_{\mathrm{B}}\rho(t)
,
\label{s46}
\end{equation}
where $\rho(t)$ is the density matrix of the whole system $S+B$ at time $t$,
$\mathbf{Tr}_{\mathrm{B}}$ denotes the trace over the degrees of freedom of the bath,
and $\mathbf{Tr}_{\mathrm{S}}\rho_{\mathrm{S}}(t)=\mathbf{Tr\;}\rho(t)=1$.
In practice, as the dimension of the Hilbert space of the bath may be assumed to be large,
we can, using the ``random-state technology'', compute the trace over the bath degrees of freedom as
\begin{equation}
\left(\mathbf{Tr}_{\mathrm{B}} {\cal A}\right)_{i,j} \approx
\sum_{p=1}^{D_{\mathrm{B}}} c^{\ast}(i,p,t)c(j,p,t)\;\langle i,p|{\cal A} |j,p\rangle
.
\label{s46c}
\end{equation}

In the case that the system contains only one spin, which is the case that we consider in the present work,
the reduced density matrix can, without loss of generality, be written as
\begin{eqnarray}
\rho_{\mathrm{S}}(t)&=&\frac{1}{2}\sum_{\alpha=x,y,z} \left[\openone+\rho_\alpha(t)\sigma^\alpha_{0}\right]
=\frac{1}{2}\sum_{k=1}^3 \left[\openone+\rho_k(t)\sigma_0^k\right]
,
\label{s48}
\end{eqnarray}
where $\rho_x(t)=\rho_1(t)$, $\rho_y(t)=\rho_2(t)$ and $\rho_z(t)=\rho_3(t)$ are real numbers.
Making use of the ``random-state technology'',  it follows immediately from  Eq.~(\ref{s48}) that
\begin{eqnarray}
\rho_1(t)&=&\rho_x(t)=\mathbf{Tr}_{\mathrm{S}}\rho_{\mathrm{S}}(t)\sigma^x_{0}=\mathbf{Tr\;} \rho(t)\sigma^x_{0}\approx  \langle\Psi(t)|\sigma^x_{0}|\Psi(t)\rangle
\nonumber \\
\rho_2(t)&=&\rho_y(t)=\mathbf{Tr}_{\mathrm{S}}\rho_{\mathrm{S}}(t)\sigma^y_{0}=\mathbf{Tr\;} \rho(t)\sigma^y_{0}\approx\langle\Psi(t)|\sigma^y_{0}|\Psi(t)\rangle
\nonumber \\
\rho_3(t)&=&\rho_z(t)=\mathbf{Tr}_{\mathrm{S}}\rho_{\mathrm{S}}(t)\sigma^z_{0}=\mathbf{Tr\;} \rho(t)\sigma^z_{0}\approx\langle\Psi(t)|\sigma^z_{0}|\Psi(t)\rangle
.
\label{s49}
\end{eqnarray}
Therefore, to obtain (accurate approximations to) the expectation values of the system operators
we compute the expressions that appear in the left-hand side of Eq.~(\ref{s49}) using
the numerical solution of the TDSE in the form given by Eq.~(\ref{s45}).

\subsection{Thermal equilibrium state}\label{section3b}

As a first check on the numerical method, it is of interest to simulate the case in which the system+bath are
initially in thermal equilibrium and study the effects of the bath size $N_\mathrm{B}$
and system-bath interaction strength $\lambda$ on the expectation values of the system spin.

The procedure is as follows.
First we generate a random state of the whole system, meaning that
\begin{eqnarray}
|\Phi(\beta)\rangle&=& \frac{e^{-\beta H/2}|\Phi\rangle}{ \langle\Phi|e^{-\beta H}|\Phi\rangle^{1/2} }
,
\label{s4k0}
\end{eqnarray}
where $\beta$ denotes the inverse temperature.
As one can show that for any observable ${\cal A}(t)$~\cite{HAMS00}
\begin{equation}
\langle {\cal A}(t)\rangle=\frac{\mathbf{Tr\;} e^{-\beta H}{\cal A}(t)}{\mathbf{Tr\;} e^{-\beta H}}
= \langle\Phi(\beta)| {\cal A}(t)|\Phi(\beta)\rangle \pm {\cal O}(D^{-1/2})
,
\label{s4h1}
\end{equation}
we can use $\langle\Phi(\beta)| {\cal A}|\Phi(\beta)\rangle$ to estimate $\langle {\cal A}(t)\rangle$.
As $e^{-\beta H}$ commutes with $e^{-it H}$,
$\langle {\cal A}(t)\rangle=\langle {\cal A}(t=0)\rangle$ is time independent.
Excluding the trivial case that $[H,{\cal A}(t)]=0$,
$\langle\Phi(\beta)| {\cal A}(t)|\Phi(\beta)\rangle=
\langle\Phi(\beta)|e^{+it H} {\cal A}e^{-it H}|\Phi(\beta)\rangle$ depends on time:
indeed, in general the random state $|\Phi(\beta)\rangle$ is unlikely to be an eigenstate of $H$.
Therefore, the simulation data obtained by solving the TDSE with
$|\Phi(\beta)\rangle$ as the initial state should display some time dependence.
However, from Eq.~(\ref{s4h1}), it follows directly that the time dependent contributions
will vanish very fast, namely as $D^{-1/2}$.
Hence this time dependence, an artifact of using ``random state technology'',
reveals itself as statistical fluctuations and can be ignored.

\begin{figure}[t]
\begin{center}
\includegraphics[width=0.48\hsize]{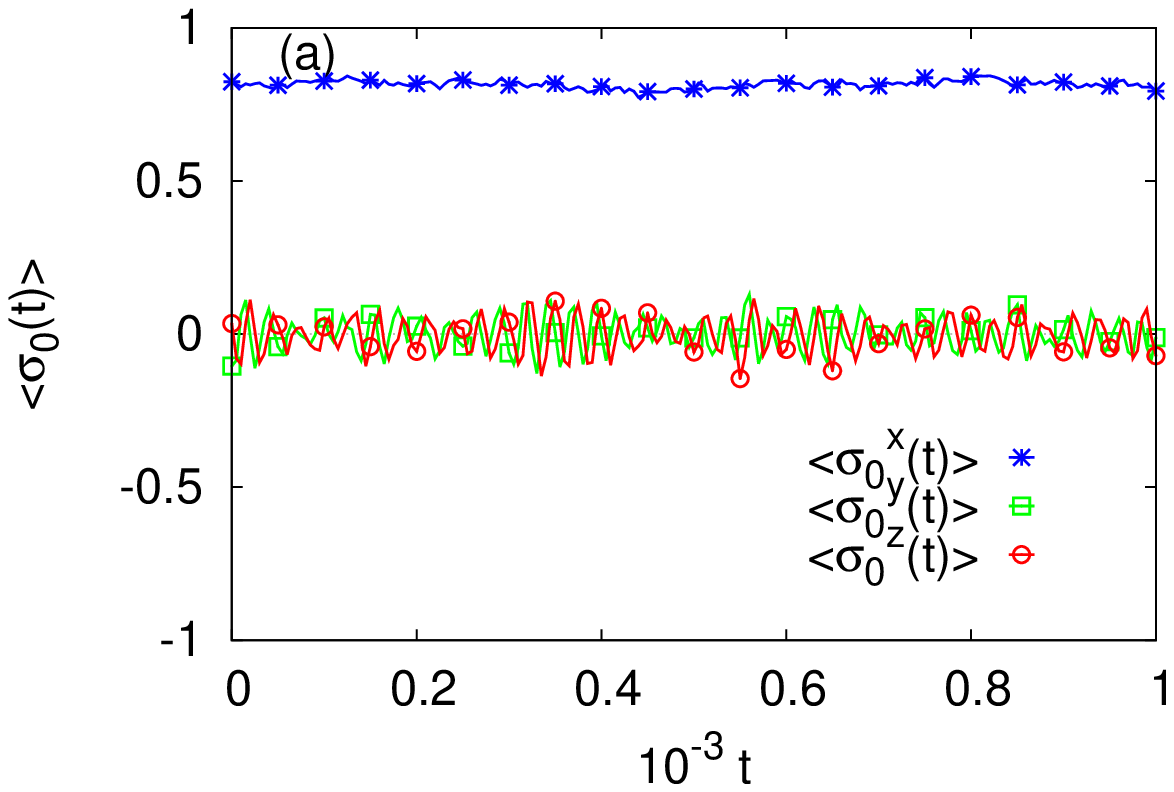}
\includegraphics[width=0.48\hsize]{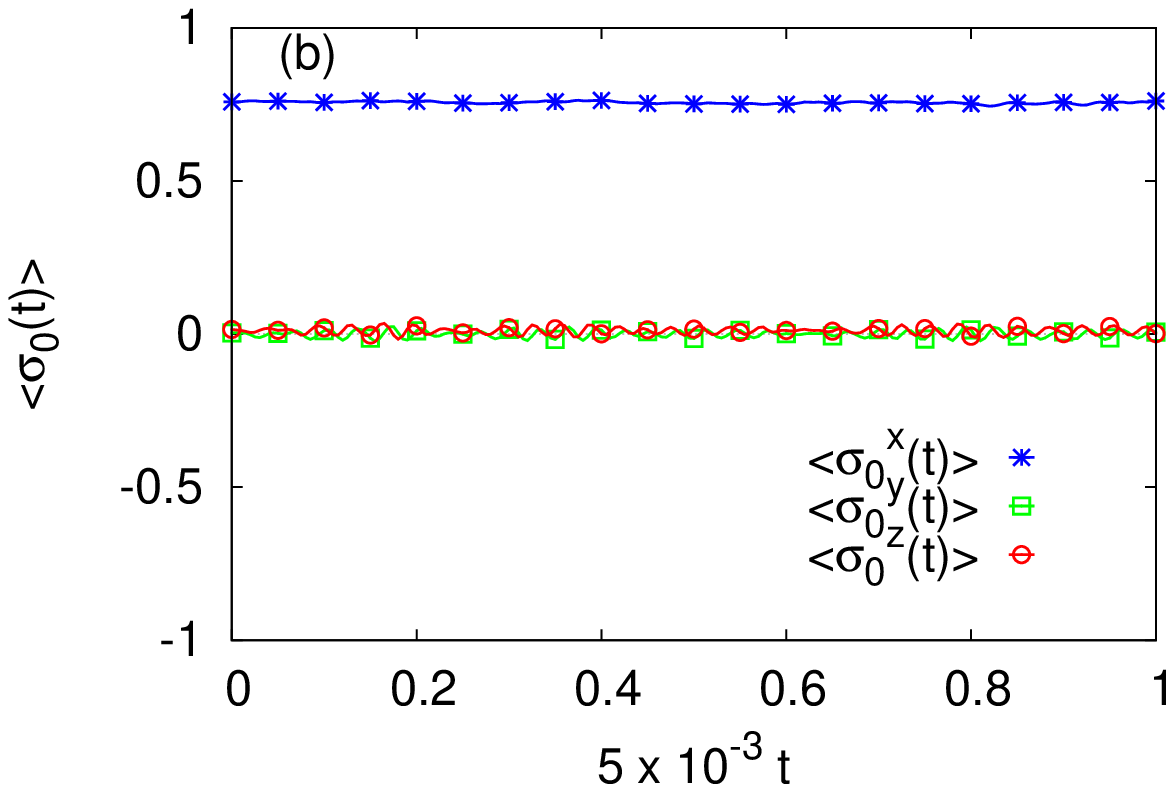}
\caption{(color online) %
Time evolution of the average of the system spin as obtained by solving the TDSE
with a random thermal state at $\beta=2$ as the initial state.
The Hamiltonian of the bath is given by Eq.~(\ref{s42}) with $K=-1/4$ and $\Delta=1$ (antiferromagnetic Heisenberg model),
The parameters of the system-bath Hamiltonian Eq.~(\ref{s43}) are $J=1/4$ and $h^x_\mathrm{B}=h^z_\mathrm{B}=1/8$.
The system-bath interaction $\lambda=0.1$.
(a) $N_\mathrm{B}=13$; (b) $N_\mathrm{B}=28$.
Lines connecting the data points are guide to the eye.
}
\label{fig3k}
\end{center}
\end{figure}

For the system in thermal equilibrium at the inverse temperature $\beta$ we have
\begin{equation}
\langle \sigma^x_0 \rangle=\tanh(\beta h^x) % +{\cal O}(\lambda^2)
\quad,\quad
\langle \sigma^y_0 \rangle=0\quad,\quad
\langle \sigma^z_0 \rangle=0
.
\label{s4h2}
\end{equation}
In Fig.~\ref{fig3k} we show simulation results for a bath at $\beta=2$
for $N_{\mathrm{B}}=13$ (left) and $N_{\mathrm{B}}=28$ spins (right).
If the system-bath interaction is sufficiently weak then, from Eq.~(\ref{s4h2}),
we expect that $\langle \sigma^x_0 \rangle\approx\tanh \beta h^x$
which for $\beta h^x=1$ yields $\langle \sigma^x_0 \rangle\approx0.762$.
From the TDSE solution with $N_{\mathrm{B}}=13$, it is clear that the spin averages
fluctuate (due to the use of the random thermal state which is not an eigenstate of $H$).
%For instance, we have
%$\langle\Phi(\beta)| \sigma^x_0(t=1000)|\Phi(\beta)\rangle = 0.79$,
%$\langle\Phi(\beta)| \sigma^y_0(t=1000)|\Phi(\beta)\rangle = -0.01$,
%and $\langle\Phi(\beta)| \sigma^z_0(t=1000)|\Phi(\beta)\rangle = -0.07$.
As expected, for $N_{\mathrm{B}}=28$ the fluctuations are much smaller, in concert with Eq.~(\ref{s4h1}).

Computing the time averages for a bath with $N_{\mathrm{B}}=13$ and for the time interval $[0,T]$ with $T=1000$ yields
\begin{eqnarray}
\frac{1}{T}\int_0^T dt\;\langle\Phi(\beta)| \sigma^x_0(t)|\Phi(\beta)\rangle &=& 0.81(0.14)
\nonumber \\
\frac{1}{T}\int_0^T dt\;\langle\Phi(\beta)| \sigma^y_0(t)|\Phi(\beta)\rangle &=& 0.00(0.05)
\nonumber \\
\frac{1}{T}\int_0^T dt\;\langle\Phi(\beta)| \sigma^z_0(t)|\Phi(\beta)\rangle &=& -0.01(0.05)
,
\label{s4h3}
\end{eqnarray}
where the numbers in parenthesis give the standard deviation.
For $N_{\mathrm{B}}=28$ and for the time interval $[0,T]$ with $T=200$ we find
\begin{eqnarray}
\frac{1}{T}\int_0^T dt\;\langle\Phi(\beta)| \sigma^x_0(t)|\Phi(\beta)\rangle &=& 0.76(0.01)
\nonumber \\
\frac{1}{T}\int_0^T dt\;\langle\Phi(\beta)| \sigma^y_0(t)|\Phi(\beta)\rangle &=& 0.00(0.01)
\nonumber \\
\frac{1}{T}\int_0^T dt\;\langle\Phi(\beta)| \sigma^z_0(t)|\Phi(\beta)\rangle &=& 0.00(0.01)
,
\label{s4h4}
\end{eqnarray}
indicating that for most practical purposes, a bath of  $N_{\mathrm{B}}=28$
spin may be sufficiently large to mimic an infinitely large bath.
The numbers in Eq.~(\ref{s4h4}) also give an indication of the statistical fluctuations
that we may expect for a bath containing $N_{\mathrm{B}}=28$ spins.
For the model parameters and the value of $\lambda$ chosen,
the second-order corrections in $\lambda$ are of the order of $0.01$
and are hidden in the statistical fluctuations,
suggesting that values of $\lambda\le 0.1$ are within the perturbative regime.

The latter statement is not as obvious as it may seem.
To first order in $\lambda$, we have
\begin{eqnarray}
\langle\sigma_0^x\rangle &=& \langle\sigma_0^x\rangle_{\mathrm{S}}
-\beta\lambda\left(\langle\sigma_0^x\rangle_{\mathrm{S}}-1\right) \langle B^x\rangle_{\mathrm{B}}
,
\label{notobv0}
\end{eqnarray}
where $\langle . \rangle_{\mathrm{S}}$ and $\langle . \rangle_{\mathrm{B}}$ denote the
thermal equilibrium averages with respect to the system and bath, respectively.
For the sake of argument, consider the case that $K=0$, $h_n^z=0$ and $h_n^x=h_{\mathrm{B}}^x$ for all $n=1,\ldots,N_{\mathrm{B}}$
(the same reasoning applies to the contributions of second order in $\lambda$).
Then, Eq.~(\ref{notobv0}) becomes
\begin{eqnarray}
\langle\sigma_0^x\rangle &=& \tanh(\beta h^x) +\beta\lambda N_{\mathrm{B}} (1-\tanh(\beta h^x))\tanh(\beta h_{\mathrm{B}}^x)
,
\label{notobv1}
\end{eqnarray}
showing that the contribution of the ``perturbation term'' increases with the number of spins in the bath.
In other words, it is not sufficient to consider small values of $\lambda$.
For the perturbation by the bath to be weak, it is necessary that $\lambda N_{\mathrm{B}}$ is small.
In this respect the spin bath considered in this paper is not different from e.g. the
standard spin-boson model~\cite{BREU02}.
In our simulation work, we adopt a pragmatic approach: we simply compute the
averages and compare them with the theoretical results of the isolated system (as we did above).
The coupling $\lambda$ is considered to be small enough if the corrections are hidden in the statistical fluctuations.

%%%%%%%%%%%%%%%%%%%%%%%%%%%%%%%%%%%%%%%%%%%%%%%%%%%%%%%%%%%%%%%%%%%%%%%%%%%%%%%%%%%%%%%%%%%%%%%%%%%%
\section{Quantum master equation: generalities}\label{section4}

We are interested in the dynamics of a system, the degrees of freedom of which interact with
other degrees of freedom of a ``bath'', ``environment'', etc.
The combination of system + bath forms a closed quantum system.
When we consider the system only, we say that we are dealing with an open quantum system.
The quantum state of the system + bath is represented by the density matrix $\rho=\rho(t)$ which evolves in time
according to %the Liouville equation (von Neumann equation, Heisenberg equation of motion)

\begin{eqnarray}
\frac{\partial\rho(t)}{\partial t}&=& i[\rho(t),H]
,
\label{s0e0}
\end{eqnarray}
where $H$ is the Hamiltonian of the system + bath (recall that we adopt units such that $\hbar=1$).

The ``relevant'' part of the dynamics may formally be separated from the ``uninteresting'' part
by using the Nakajima-Zwanzig projection operator formalism~\cite{NAKA58,ZWAN60}.
Let ${\cal P}$ be the projector onto the ``relevant'' part
and introduce the Liouville operator ${\cal L}A=i[A,H]$.
Denoting by ${\cal Q}=\openone-{\cal P}$ the projector on the ``uninteresting'' part,
it follows that
\begin{eqnarray}
\frac{\partial{\cal P}\rho(t)}{\partial t}&=& {\cal P}{\cal L} {\cal P}\rho(t)+{\cal P}{\cal L} {\cal Q}\rho(t)
,
\label{s0e1a}
\\
\frac{\partial{\cal Q}\rho(t)}{\partial t}&=& {\cal Q}{\cal L} {\cal P}\rho(t)+{\cal Q}{\cal L} {\cal Q}\rho(t)
.
\label{s0e1b}
\end{eqnarray}
Note that because $H$ is Hermitian, $i{\cal L}$,
$i{\cal P}{\cal L}{\cal P}$ and $i{\cal Q}{\cal L}{\cal Q}$ are Hermitian too.
The formal solution of the matrix-valued, inhomogeneous, linear, first-order differential equation Eq.~(\ref{s0e1b}) reads as
\begin{eqnarray}
{\cal Q}\rho(t)&=& e^{t{\cal Q}{\cal L}{\cal Q}}{\cal Q}\rho(t=0) + \int_0^t du\; e^{u{\cal Q}{\cal L}{\cal Q}}{\cal Q}{\cal L} {\cal P}\rho(t-u)
,
\label{s0e1c}
\end{eqnarray}
as can be verified most easily by calculating its derivative with respect to time and
using ${\cal P}{\cal P}={\cal P}$, ${\cal P}{\cal Q}={\cal Q}{\cal P}=0$ and ${\cal Q}{\cal Q}={\cal Q}$.
Substituting Eq.~(\ref{s0e1c}) into Eq.~(\ref{s0e1a}) yields
\begin{eqnarray}
\frac{\partial{\cal P}\rho(t)}{\partial t}&=&
{\cal P}{\cal L} {\cal P}\rho(t) +
{\cal P}{\cal L}{\cal Q}e^{t{\cal Q}{\cal L}{\cal Q}}{\cal Q}\rho(t=0) + \int_0^t du\;
{\cal P}{\cal L}{\cal Q} e^{u{\cal Q}{\cal L}{\cal Q}}{\cal Q}{\cal L} {\cal P}\rho(t-u)
.
\label{s0e2}
\end{eqnarray}

We are primarily interested in the time evolution of the system.
Therefore, we choose ${\cal P}$ such that it projects onto the system variables
and we perform the trace over the bath degrees-of-freedom.
A common choice for the projector ${\cal P}$ is~\cite{SUAR92,GASP99,BREU02,MORI08,UCHI09,MORI14}
\begin{equation}
{\cal P}A= \rho_\mathrm{B}\mathbf{Tr}_\mathrm{B}\; A
,
\label{s0e6}
\end{equation}
where
\begin{equation}
\rho_\mathrm{B}=\frac{e^{-\beta H_\mathrm{B}}}{\mathbf{Tr}_\mathrm{B}\;e^{-\beta H_\mathrm{B}}}
,
\label{s0e7}
\end{equation}
is the density matrix of the bath in thermal equilibrium. % ($\mathbf{Tr}_\mathrm{B}\; \rho_\mathrm{B}=1$).
Accordingly, the density matrix of the system is given by
\begin{equation}
\rho_\mathrm{S}(t)=\mathbf{Tr}_\mathrm{B} {\cal P}\rho(t) = \mathbf{Tr}_\mathrm{B} \rho(t)
,
\label{s0e2a}
\end{equation}
consistent with Eq.~(\ref{s46}).

In the present work, we will mostly consider initial states that are represented by the direct-product ansatz
\begin{equation}
\rho(t=0)=\rho_\mathrm{S}\rho_\mathrm{B},
.
\label{s0e8}
\end{equation}
but occasionally, we also consider as an initial state, the thermal equilibrium state
of the system + bath, that is $\rho(t=0)=e^{-\beta H}/\mathbf{Tr}\;e^{-\beta H}$, see Sec.~\ref{section3b}.
The direct-product ansatz Eq.~(\ref{s0e8}) not only implies ${\cal Q}\rho(t=0)=0$
but also defines the initial condition for Eq.~(\ref{s0e2}).
In general, this initial condition may be incompatible with the initial
condition for the TDSE of the whole system,
which may affect the dynamics on a time-scale comparable to the
relaxation time of the bath~\cite{UCHI09}.

Adopting Eq.~(\ref{s0e8}), Eq.~(\ref{s0e2}) simplifies to
\begin{eqnarray}
\frac{\partial \rho_\mathrm{S}(t)}{\partial t}&=&
\mathbf{Tr}_\mathrm{B}{\cal P}{\cal L}{\cal P}\rho(t)
+ \int_0^t du\;
\mathbf{Tr}_\mathrm{B} {\cal P}{\cal L}{\cal Q} e^{u{\cal Q}{\cal L}{\cal Q}}{\cal Q}{\cal L} {\cal P}\rho(t-u)
,
\label{s0e3}
\end{eqnarray}
which is not a closed equation for $\rho_\mathrm{S}(t)$ yet~\cite{MORI08}.

Using the explicit form of the Hamiltonian Eq.~(\ref{s40}),
the first term in Eq.~(\ref{s0e3}) may be written as
$\mathbf{Tr}_\mathrm{B}{\cal P}{\cal L}{\cal P}\rho(t)={\cal L}_0 \rho_\mathrm{S}(t) $ where
for any system operator $X_\mathrm{S}$,
\begin{eqnarray}
{\cal L}_0 X_\mathrm{S}
\equiv -i\left\{
\left[ H_S ,X_\mathrm{S}(t) \right]
+\sum_{i=1}^3 \langle B_i\rangle_\mathrm{B} \left[\sigma_0^i ,X_\mathrm{S}(t)\right] \right\}
,
\label{s0e3a}
\end{eqnarray}
and $\langle B_i\rangle_\mathrm{B}\equiv \mathbf{Tr}_\mathrm{B} \rho_\mathrm{B} B_i$.
Therefore, Eq.~(\ref{s0e3}) may be written as
\begin{eqnarray}
\frac{\partial \rho_\mathrm{S}(t)}{\partial t}&=&
{\cal L}_0\rho_\mathrm{S}(t)
 + \int_0^t du\;
\mathbf{Tr}_\mathrm{B} {\cal P}{\cal L}{\cal Q} e^{(t-u){\cal Q}{\cal L}{\cal Q}}{\cal Q}{\cal L} \rho_\mathrm{B}\rho_\mathrm{S}(u)
%={\cal L}_0\rho_\mathrm{S}(t) + \int_0^t du\;{\cal K}(t-u)\rho_\mathrm{S}(u)
.
\label{s0e4}
\end{eqnarray}

Using representation Eq.~(\ref{s48}), multiplying both sides of Eq.~(\ref{s0e4}) by $\sigma_0^j$,
performing the trace over the system degree of freedom, and
denoting $\bm{\rho}(t)=(\rho_1(t),\rho_2(t),\rho_3(t))$, Eq.~(\ref{s0e4}) can be written as
\begin{equation}
\frac{\partial \bm{\rho}(t)}{\partial t}= \mathbf{L}\bm{\rho}(t)
 + \int_0^t du\; \mathbf{M}(t-u)\bm{\rho}(u)
+ \int_0^t du\; \mathbf{K}(u)
,
\label{QMEQ5}
\end{equation}
where
\begin{eqnarray}
\mathbf{L}_{jk}&=&\frac{1}{2}\mathbf{Tr}_\mathrm{S} \sigma_0^j  {\cal L}_0  \sigma_0^k
\nonumber \\
\mathbf{M}_{jk}(u)
&=&
\frac{1}{2}
\mathbf{Tr}\; \sigma_0^j {\cal P}{\cal L}{\cal Q} e^{u{\cal Q}{\cal L}{\cal Q}}{\cal Q}{\cal L} \rho_\mathrm{B}\sigma_0^k
\nonumber \\
\mathbf{K}_{j}(u) &=& \frac{1}{2}
\mathbf{Tr}\; \sigma_0^j {\cal P}{\cal L}{\cal Q} e^{u{\cal Q}{\cal L}{\cal Q}}{\cal Q}{\cal L} \rho_\mathrm{B}
.
\label{QMEQ5a}
\end{eqnarray}
As we have only made formal manipulations, solving Eq.~(\ref{QMEQ5}) of the system is just as difficult
as solving Eq.~(\ref{s0e0}) of the whole system.
In other words, in order to make progress, it is necessary to make approximations.
A common route to derive an equation which can actually be solved is to assume that
$\lambda$ is sufficiently small such that perturbation theory may be used to approximate the second term in Eq.~(\ref{s0e4})
and that it is allowed to replace $\rho_\mathrm{S}(u)$ in Eq.~(\ref{s0e4}) by $\rho_\mathrm{S}(t)$~\cite{BREU02}.

As the purpose of the present work is to scrutinize the approximations just mentioned by comparing
the solution obtained from the Markovian quantum master equation with the one obtained by solving the TDSE,
we will not dwell on the justification of these approximations
and derivation of this equation itself, but merely state
that the result of making these approximations is an equation that may be cast in the form
\begin{center}
\framebox{
\parbox[t]{0.5\hsize}{%
%\begin{scriptsize}
%\lstinputlisting[language=Fortran]{simpleDS.tex}
%\end{scriptsize}
\begin{eqnarray}
\frac{\partial \bm{\rho}(t)}{\partial t}= \mathbf{A}\bm{\rho}(t) + \mathbf{b}
.
\label{GBE}
\end{eqnarray}
}}
\end{center}
In the following we will refer to Eq.~(\ref{GBE}) as ``the'' quantum master equation (QMEQ).
In Sec.~\ref{section4a} we give a well-known example of a quantum master equation that is
of the form Eq.~(\ref{GBE}).

The formal solution of Eq.~(\ref{GBE}) reads as
\begin{eqnarray}
\bm{\rho}(t) &=&e^{t\mathbf{A}}\bm{\rho}(0) + \int_{0}^t e^{(t-u)\mathbf{A}}\mathbf{b}\;du
%=e^{t\mathbf{A}}\bm{\rho}(0) + \left(1-e^{t\mathbf{A}}\right) \mathbf{A}^{-1}\mathbf{b}
,
\label{QMEQ6}
\end{eqnarray}
or, equivalently
\begin{eqnarray}
\bm{\rho}(t+\tau) &=&e^{\tau\mathbf{A}}\bm{\rho}(t) + \int_{0}^{\tau} e^{(\tau-u)\mathbf{A}}\mathbf{b}\;du
=e^{\tau\mathbf{A}}\bm{\rho}(t) + \mathbf{B}
,
\label{QMEQ7}
\end{eqnarray}
where
\begin{eqnarray}
\mathbf{B}=\int_{0}^{\tau} e^{(\tau-u)\mathbf{A}}\mathbf{b}\;du
,
\label{QMEQ7a}
\end{eqnarray}
does not depend on time.
Equation~(\ref{QMEQ7}) directly connects to the numerical work because in practice,
we solve the TDSE with a finite time step $\tau$.

Generally speaking, as a result of the coupling to the bath, the system is expected to exhibit relaxation
towards a stationary state, meaning that $\bm{\rho}(t)\approx\bm{\rho}(\infty)$ for $t$ sufficiently large.
If such a stationary state exists, it follows from Eq.~(\ref{QMEQ7}) that
$\bm{\rho}(\infty) \approx e^{\tau\mathbf{A}}\bm{\rho}(\infty) + \mathbf{B}$
or that $\mathbf{B} \approx (1-e^{\tau\mathbf{A}})\bm{\rho}(\infty)$,
yielding
\begin{eqnarray}
\bm{\rho}(t+\tau) - \bm{\rho}(\infty) &\approx&e^{\tau\mathbf{A}}(\bm{\rho}(t) -\bm{\rho}(\infty))
.
\label{QMEQ6a}
\end{eqnarray}
Equation~(\ref{QMEQ6a}) suggests that the existence of a stationary state
implies that there is no need to determine $\mathbf{B}$.
However, numerical experiments with the Bloch equation model (see Appendix~\ref{section6}) show that using
Eq.~(\ref{QMEQ6a}), a least-square fit
to solution of the Bloch equation often fails to yield the correct $e^{\tau\mathbf{A}}$.
Therefore, as explained in Sec.~\ref{section5}, we will use Eq.~(\ref{QMEQ7}) and determine both
$e^{\tau\mathbf{A}}$ and $\mathbf{B}$ by least-square fitting to TDSE or Bloch equation data.

%\begin{center}
%\framebox{
%\parbox[t]{0.9\hsize}{%
%%\begin{scriptsize}
%%\lstinputlisting[language=Fortran]{simpleDS.tex}
%%\end{scriptsize}
We can now formulate more precisely, the procedure to test whether or not a quantum master equation
of the form Eq.~(\ref{GBE}) provides a good approximation to the data $\rho_k(t)=\langle \sigma^k(t)\rangle$
obtained by solving the TDSE of the system interacting with the bath using a time step $\tau$.
To this end, we use the latter data to determine
the matrix $e^{\tau\mathbf{A}}$ and vector $\mathbf{B}$ such that, in a least square sense, the
difference between the data obtained by solving Eq.~(\ref{QMEQ7}) for a substantial interval of time
and the corresponding TDSE data is as small as possible.
If the values of $\bm\rho(t)$ computed according to Eq.~(\ref{QMEQ7})
are in good agreement with the data $\rho_k(t)$,
one might say that at least for the particular time interval studied, there exists a mapping
of the Schr\"odinger dynamics of the system onto the QMEQ Eq.~(\ref{GBE}).
%}}
%\end{center}
%

\subsection{Markovian quantum master equation: Example}\label{section4a}

We consider the Redfield master equation~\cite{REDF57}
under the Markovian assumption~\cite{GASP99,BREU02}
\begin{eqnarray}
\frac{d\rho_{\mathrm{S}}(t)}{dt}
&=&
-i[H_S,\rho_{\mathrm{S}}(t)]+\lambda^2 \sum_{j=1}^3 \left(
R_j^{\phantom{^\dagger}}\rho_{\mathrm{S}}(t)\sigma_j
+\sigma_j\rho_{\mathrm{S}}(t) R_j^\dagger
-\sigma_j R_j^{\phantom{^\dagger}}\rho_{\mathrm{S}}(t)
-\rho_{\mathrm{S}}(t) R_j^\dagger\sigma_j\right)
,
\label{QMEQ0}
\end{eqnarray}
where $\rho_{\mathrm{S}}(t)$ is the density matrix of the system.
The operators $R_j$ are given by~\cite{GASP99}
\begin{eqnarray}
R_j&=&\sum_{k=1}^3 \int_{0}^\infty\;dt\; C_{jk}(t)e^{-itH_s}\sigma_k e^{+itH_s}\quad,\quad j=1,2,3
,
\label{QMEQ1}
\end{eqnarray}
where $C_{jk}(t)=\mathbf{Tr}_\mathrm{B}\rho_\mathrm{B} B_j(t) B_k(0)$
are the correlations of the bath operators~\cite{GASP99}.
The specific form of $C_{jk}(t)$ is not of interest to us at this time (but also see Sec.~\ref{section8}).
For what follows, it is important that the specific form Eq.~(\ref{QMEQ1}) of the
operators $R_j$ allows us to write
\begin{eqnarray}
R_j&=&\sum_{k=1}^3 r_{jk}\sigma_k
,
\label{QMEQ2}
\end{eqnarray}
where
\begin{eqnarray}
r_{j1}&=&\int_{0}^\infty\;dt\; C_{j1}(t)
\nonumber \\
r_{j2}&=&\int_{0}^\infty\;dt\; \left( C_{j2}(t)\cos 2h^xt + C_{j3}(t)\sin 2h^xt \right)
\nonumber \\
r_{j3}&=&\int_{0}^\infty\;dt\; \left( C_{j3}(t)\cos 2h^xt - C_{j2}(t)\sin 2h^xt \right)
,
\label{QMEQ2b}
\end{eqnarray}
do not depend on time (due to the Markov approximation).

As a first step, we want to derive from Eq.~(\ref{QMEQ0}), the corresponding
equations in terms of the $\rho_k(t)$'s.
This can be done by
using representation Eq.~(\ref{s48}),
multiplying both sides of Eq.~(\ref{QMEQ0})
with $\sigma_k$ for $k=1,2,3$ and taking the trace, a calculation for which we resort to
Mathematica$^{\circledR}$.
We obtain
\begin{eqnarray}
\frac{d\rho_1}{dt}
&=&
+4\lambda^2 \left[ \left(r_{23}^\mathrm{I}-r_{32}^\mathrm{I}\right)
- \left(r_{22}^\mathrm{R}+r_{33}^\mathrm{R}\right) \rho_1
+ r_{21}^\mathrm{R} \rho_2 + r_{31}^\mathrm{R} \rho_3
\right]
\nonumber \\
\frac{d\rho_2}{dt}
&=& +h^x\rho_3
+4\lambda^2 \left[ \left(r_{31}^\mathrm{I}-r_{13}^\mathrm{I}\right) + r_{12}^\mathrm{R} \rho_1
- \left(r_{11}^\mathrm{R}+r_{33}^\mathrm{R}\right) \rho_2 + r_{32}^\mathrm{R} \rho_3
\right]
\nonumber \\
\frac{d\rho_3}{dt}
&=& -h^x\rho_2
+4\lambda^2 \left[ r_{12}^\mathrm{I}-r_{21}^\mathrm{I} + r_{13}^\mathrm{R} \rho_1 + r_{23}^\mathrm{R} \rho_2
- \left(r_{11}^\mathrm{R}+r_{22}^\mathrm{R}\right) \rho_3
\right]
,
\label{QMEQ4}
\end{eqnarray}
where we used the notation $z=z^\mathrm{R}+iz^\mathrm{I}$.
It directly follows that Eq.~(\ref{QMEQ4}) can be written in the form Eq.~(\ref{GBE}).
It is straightforward to show that this holds for quantum master equations of the Lindblad form as well.

%%%%%%%%%%%%%%%%%%%%%%%%%%%%%%%%%%%%%%%%%%%%%%%%%%%%%%%%%%%%%%%%%%%%%%%%%%%%%%%%%%%%%%%%%%%%%%%%%%%%%%%%%%%%%%%%%%%%%%%%

\begin{figure}[t]
\begin{center}
\includegraphics[width=0.48\hsize]{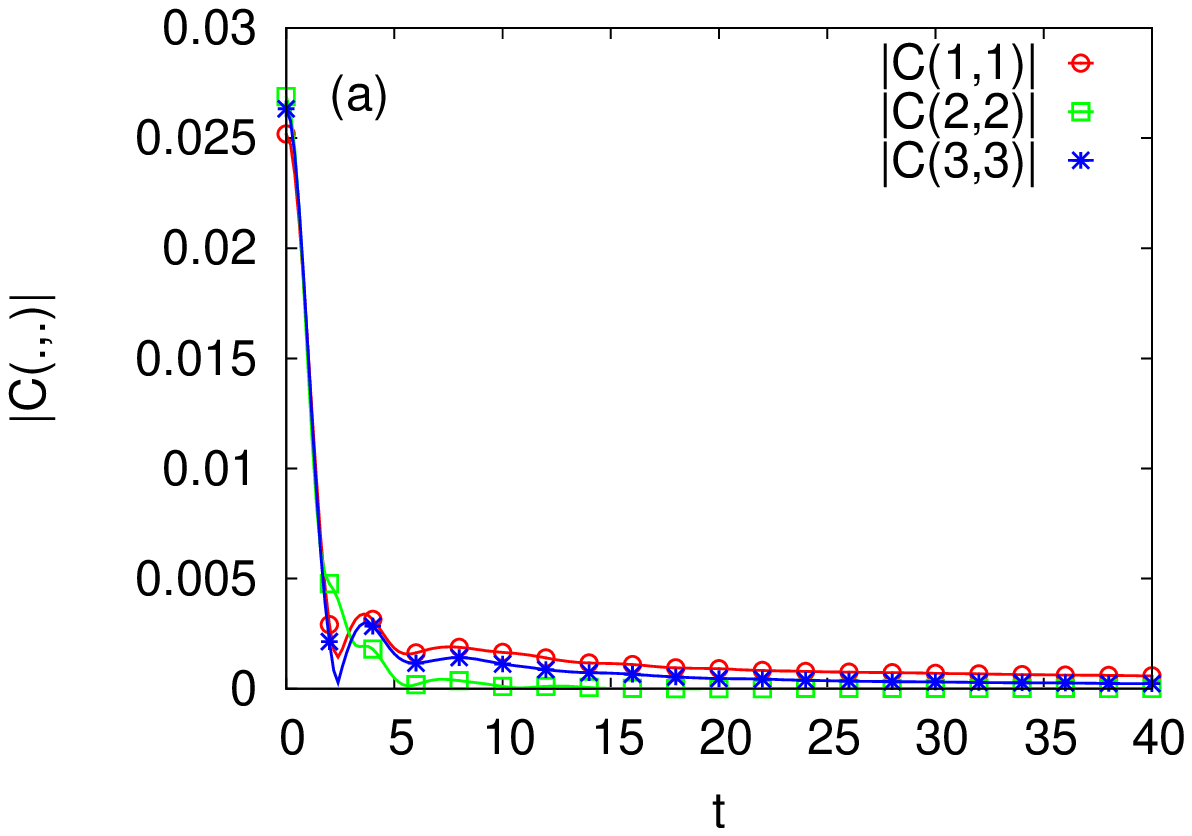}
\includegraphics[width=0.48\hsize]{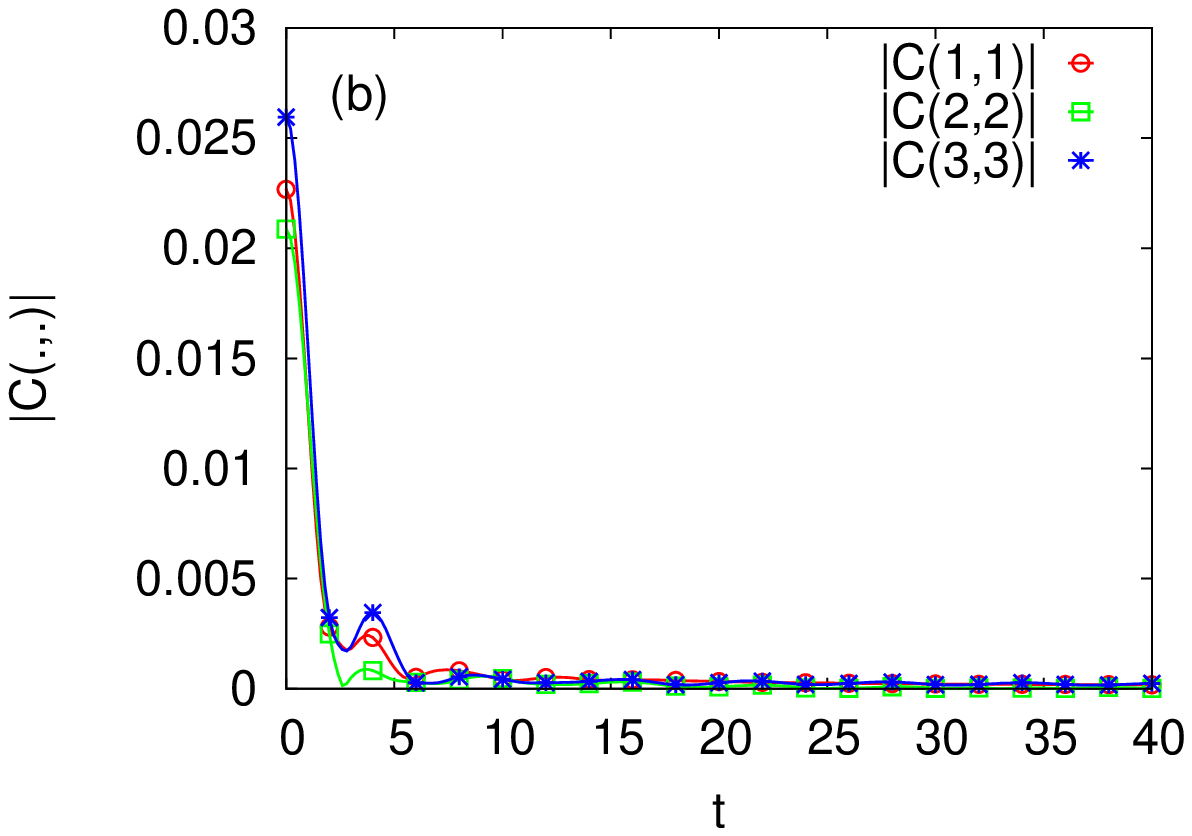}
\includegraphics[width=0.48\hsize]{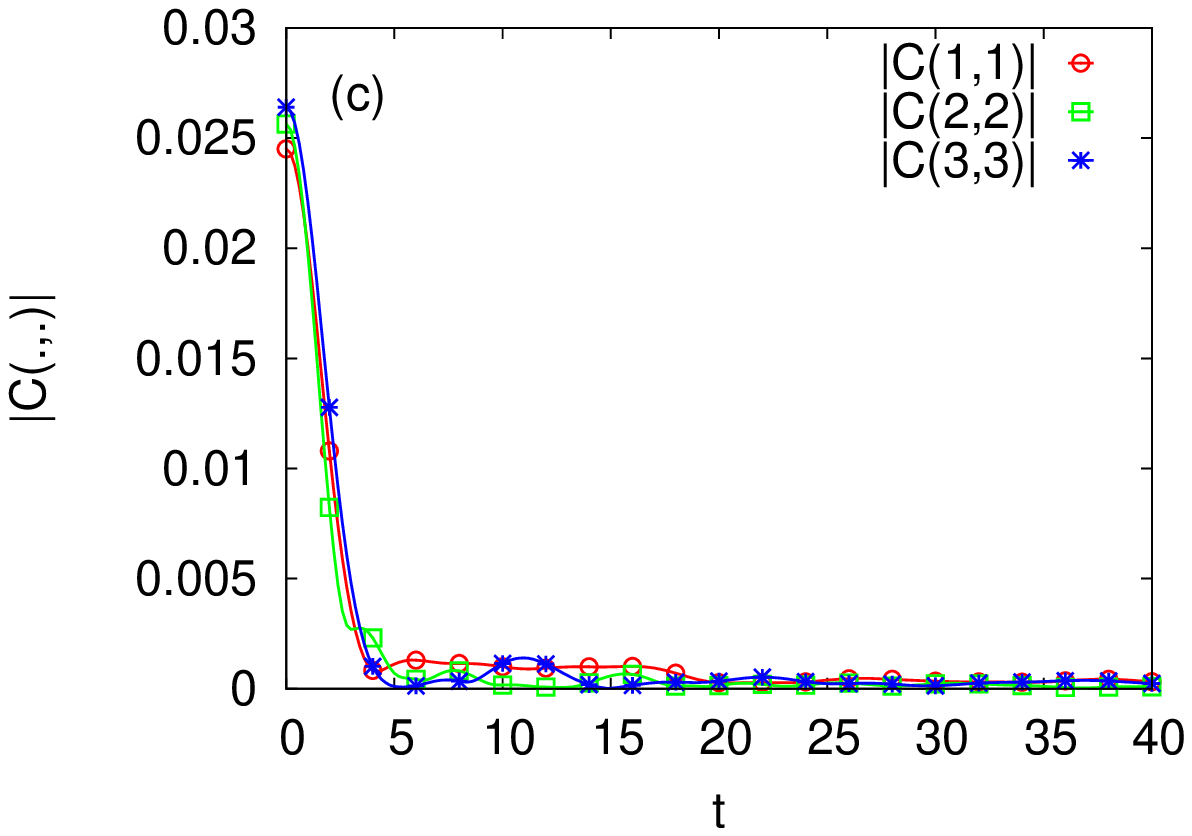}
\includegraphics[width=0.48\hsize]{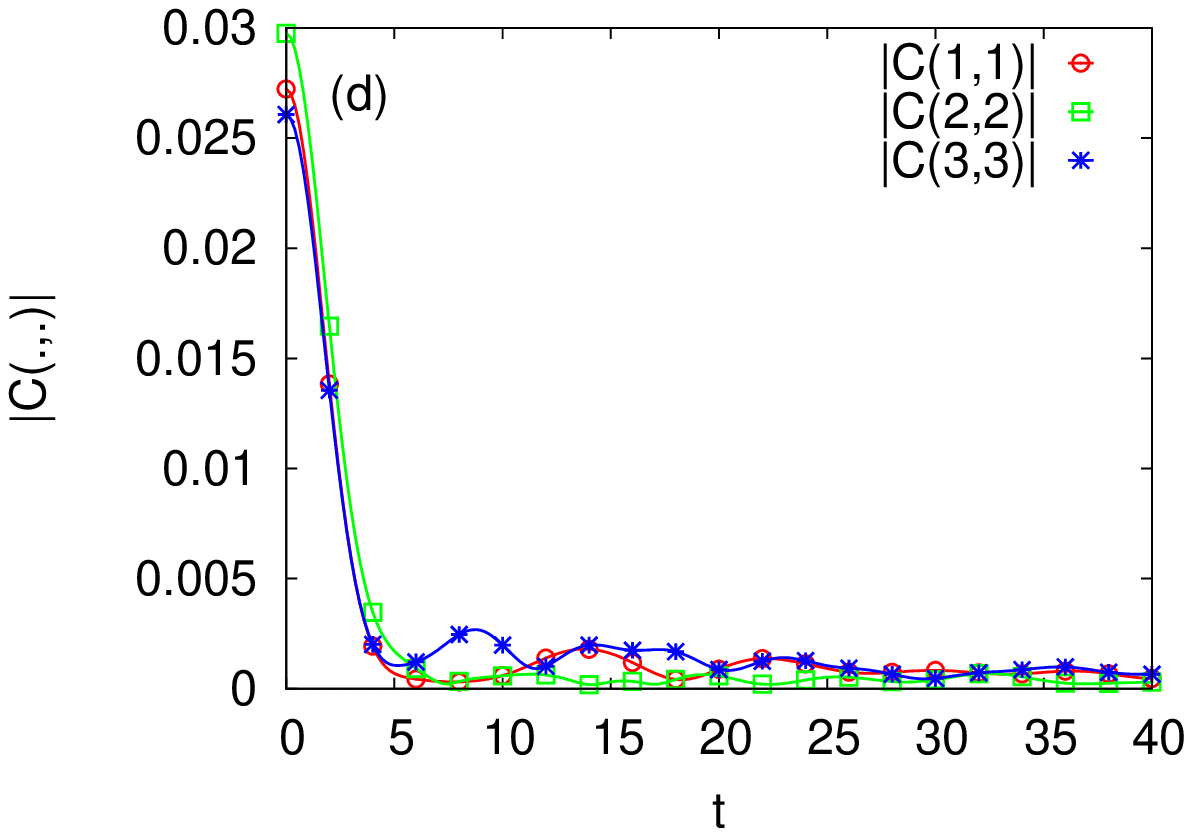}
\caption{(color online) %
The absolute values of three of the nine bath-operator correlations Eq.~(\ref{bc0}) as obtained by solving the TDSE
for a bath of $N_\mathrm{B}=32$ spins with a random thermal state at $\beta=1$ as the initial state.
The bath-operator correlations that have absolute values that are too small to be seen on the scale
of the plot have been omitted.
The parameters of the system-bath Hamiltonian $H_\mathrm{SB}$ are $J=1/4$ and $h^x_\mathrm{B}=h^z_\mathrm{B}=1/8$.
(a) the bath Hamiltonian $H_\mathrm{B}$ is given by Eq.~(\ref{s42}) with $K=-1/4$ and $\Delta=1$
(antiferromagnetic Heisenberg model) and $\lambda=0$;
(b) same as (a) except that $\lambda=0.1$;
(c) the bath Hamiltonian $H_\mathrm{B}$ is given by Eq.~(\ref{s42a}) with $K=1/4$ and $\lambda=0$;
(d) same as (a) except that $\lambda=0.1$.
}
\label{figbc0}
\end{center}
\end{figure}

\subsection{Bath correlations}\label{section4b}

A crucial assumption in deriving the QMEQ Eq.~(\ref{GBE})
from the exact equation Eq.~(\ref{s0e4}) is that the correlations of the bath
decay on a short time scale, short relative to the time scale of the motion
of the system spin~\cite{BREU02}.
Moreover, in the perturbative derivation of quantum master equations, such as the Redfield master equation,
it is assumed that the time evolution of the bath operators is governed by the bath Hamiltonian only~\cite{BREU02}.

Having the time evolution of the whole system at our disposal, we can compute, without additional assumptions
or approximations, the correlations
\begin{equation}
C(i,j,t)=\mathbf{Tr} \rho(t=0) B_i(t) B_j(0)\quad,\quad i,j=1,2,3
,
\label{bc0}
\end{equation}
of the bath operators Eq.~(\ref{s43a}).
Note that in general, Eq.~(\ref{bc0}) is complex-valued and that, because of the choice Eq.~(\ref{s0e8}), $C(i,j,t)=C_{ij}(t)$ if $\lambda=0$.
Of particular interest is the question whether, for the chosen value of the system-bath interaction $\lambda$,
the dynamics of the system spin significantly affects the bath dynamics.

In Fig.~\ref{figbc0} we present simulation results of the correlations $|C(i,i,t)|$
for a bath of $N_\mathrm{B}=32$ spins,
for different choices of the bath Hamiltonian, and with and without system-bath interaction.
The calculation of the nine correlations Eq.~(\ref{bc0})
requires solving four TDSEs simultaneously, using as
the initial states the random thermal state $|\Psi(\beta)\rangle$,
$B_1|\Psi(\beta)\rangle$, $B_2|\Psi(\beta)\rangle$, and $B_3|\Psi(\beta)\rangle$.
As the whole system contains $33$ spins,
these calculations are fairly expensive in terms of CPU and memory cost.
One such calculation needs somewhat less than 1TB memory to run
and takes about 5 hours using 65536 BlueGene/Q processors which, in practice, limits
the time interval that can be studied.

In all four cases, the absolute values of correlations for $i\not=j$ are much smaller
than those for $i=j$ and have therefore been omitted in Fig.~\ref{figbc0}.
The remaining three correlations decay rapidly but, on the time scale shown, are definitely non-zero at $t=20$.

Comparison of the top and bottom figures of Fig.~\ref{figbc0}
may suggest that the bath correlations decay faster
if the bath is described by the antiferromagnetic Heisenberg model (\ref{s42})
than if the bath Hamiltonian has random couplings [see Eq.~(\ref{s42a})].
However, this is a little misleading.
For the bath Hamiltonian with random couplings $K_n^\alpha$ in the range
$[-1/4,1/4]$, we have $\langle |K_n^\alpha|\rangle\approx 1/8$.
On the other hand, for the antiferromagnetic Heisenberg bath we have $K=-1/4$
roughly indicating that the bath dynamics may be about two times
faster than in the case of the bath Hamiltonian with random couplings.
The presence of random couplings renders the quantitative comparison of the relaxation times non-trivial.
However, from Fig.~\ref{figbc0} it is clear that as a bath,
the antiferromagnetic Heisenberg model performs better than the model with random interactions
in the sense that for $t>10$ the correlations of the former seem to have reached a stationary state
whereas in the case of the latter, they do not.
Moreover, using the full Hamiltonian ($\lambda=0.1$) instead of only the bath Hamiltonian to solve the TDSE,
for $t>10$ the changes to the correlations are less pronounced if
the bath is an antiferromagnetic Heisenberg model than if the bath has random interactions.
Based on these results, it seems advantageous to
adopt the antiferromagnetic Heisenberg model ($K=-1/4$) as the Hamiltonian of the bath.

Qualitatively, in all cases, the correlations are either small for all $t$ or decrease
by about order of magnitude on a short time scale ($t<10$), indicating that
the approximations that changed Eq.~(\ref{QMEQ5}) into Eq.~(\ref{GBE}) may apply
to the spin model we are considering.

%%%%%%%%%%%%%%%%%%%%%%%%%%%%%%%%%%%%%%%%%%%%%%%%%%%%%%%%%%%%%%%%%%%%%%%%%%%%%%%%%%%%%%%%
\section{Algorithm to extract $e^{\tau\mathbf{A}}$ and $\mathbf{B}$ from TDSE data}\label{section5}

Recall that our primary objective is to determine the Markovian master equation Eq.~(\ref{GBE})
which gives the best (in the least-square sense) fit to the solution of the TDSE.
Obviously, this requires taking into account the full motion of the system spin,
not only the decay envelope, over an extended period of time.

The numerical solution of the TDSE of the full problem yields the data $\rho_k(t)=\langle \sigma^k(t)\rangle$.
In this section, we consider these data as given and discuss the algorithm that takes as input
the values of $\rho_k(t)$ and returns the optimal choice of the
matrix $e^{\tau\mathbf{A}}$ and vector $\mathbf{B}$,
meaning that we minimize the least-square error between the data
$\{\rho_k(t)\}$ and the corresponding data, obtained by solving Eq.~(\ref{QMEQ7}).

Denoting $\rho_k(n)\equiv\rho_k(n\tau)$, it follows that if Eq.~(\ref{QMEQ7}) is assumed to hold, we must have
\begin{eqnarray}
\left(\begin{array}{cccc}
\rho_1(1)&\rho_1(2)&\ldots&\rho_1(N)\\
\rho_2(1)&\rho_2(2)&\ldots&\rho_2(N)\\
\rho_3(1)&\rho_3(2)&\ldots&\rho_3(N)\\
\end{array}\right)
&=&
\left(\begin{array}{cccc}
(e^{\tau\mathbf{A}})_{11} & (e^{\tau\mathbf{A}})_{12} & (e^{\tau\mathbf{A}})_{13} & (\mathbf{B})_{1}\\
(e^{\tau\mathbf{A}})_{21} & (e^{\tau\mathbf{A}})_{22} & (e^{\tau\mathbf{A}})_{23} & (\mathbf{B})_{2}\\
(e^{\tau\mathbf{A}})_{31} & (e^{\tau\mathbf{A}})_{32} & (e^{\tau\mathbf{A}})_{33} & (\mathbf{B})_{3}\\
\end{array}\right)
\left(\begin{array}{cccc}
\rho_1(0)&\rho_1(1)&\ldots&\rho_1(N-1)\\
\rho_2(0)&\rho_2(1)&\ldots&\rho_2(N-1)\\
\rho_3(0)&\rho_3(1)&\ldots&\rho_3(N-1)\\
1 &1 & 1 &1\\
\end{array}\right)
,
\nonumber \\
\label{QMEQ11}
\end{eqnarray}
where $N$ is the number of time steps for which the solution of the TDSE is known.
We may write Eq.~(\ref{QMEQ11}) in the more compact form
\begin{eqnarray}
\mathbf{Z}=\mathbf{Y}\mathbf{X}
,
\label{QMEQ8}
\end{eqnarray}
where $\mathbf{Z}$ is a $3\times N$ matrix of data,
$\mathbf{Y}$ is a $3\times 4$ matrix that we want to determine,
and $\mathbf{X}$ is a $4\times N$ matrix of data.

We determine $\mathbf{Y}$ by solving the linear least square problem,
that is we search for the solution of the problem $\min_\mathbf{Y} ||\mathbf{Z}-\mathbf{Y}\mathbf{X}||^2$.
A numerically convenient way to solve this minimization problem is to compute
the singular value decomposition~\cite{GOLU96,PRES03} of $\mathbf{X}=\mathbf{U}\bm{\Sigma}\mathbf{V}^T$
where $\mathbf{U}$ is an orthogonal $3\times3$ matrix,
$\bm{\Sigma}$ is the $3\times N$ matrix with the singular values of $\mathbf{X}$ on its
diagonal, and $\mathbf{V}^T$ is an orthogonal $N\times N$ matrix.
In terms of these matrices we have
\begin{eqnarray}
\mathbf{Y}=\mathbf{Z}\mathbf{V}\bm{\Sigma}^+\mathbf{U}^T
,
\label{QMEQ9}
\end{eqnarray}
where $\bm{\Sigma}^+$ is the pseudo-inverse of $\bm\Sigma$,
which is formed by replacing every non-zero diagonal entry of $\bm\Sigma$
by its reciprocal and transposing the resulting matrix.

Numerical experiments show that the procedure outlined above is not robust: it sometimes fails
to reproduce the known $e^{\tau\mathbf{A}}$ and $\mathbf{B}=0$, in particular in the case
that $e^{\tau\mathbf{A}}$ is (close to) an orthogonal matrix.
Fortunately, a straightforward extension renders the procedure very robust.
The key is to use data from three runs with different initial conditions.
This also reduces the chance that the estimates
of $e^{\tau\mathbf{A}}$ and $\mathbf{B}$ are good by accident.
In practice, we take the initial states to be orthogonal (see Sec.~\ref{section7} for the precise specification).

Labeling the data for different initial states by superscripts we have
\begin{eqnarray}
\left(\begin{array}{ccc}
\mathbf{Z}^{(1)}&\mathbf{Z}^{(2)}&\mathbf{Z}^{(3)}
\end{array}\right)
&=&
\mathbf{Y}
\left(\begin{array}{ccc}
\mathbf{X}^{(1)}&\mathbf{X}^{(2)}&\mathbf{X}^{(3)}
\end{array}\right)
,
\label{QMEQ10}
\end{eqnarray}
but now $\mathbf{Z}=(\mathbf{Z}^{(1)}\,\mathbf{Z}^{(2)}\,\mathbf{Z}^{(3)})$
and $\mathbf{X}=(\mathbf{X}^{(1)}\,\mathbf{X}^{(2)}\,\mathbf{X}^{(3)})$ are $3\times 3N$
and $4\times 3N$ matrices of data, respectively.
Using Eq.~(\ref{QMEQ9}) we compute
\begin{eqnarray}
\mathbf{Y}&=&
\left(\begin{array}{cccc}
(e^{\tau\mathbf{A}})_{11} & (e^{\tau\mathbf{A}})_{12} & (e^{\tau\mathbf{A}})_{13} & (\mathbf{B})_{1}\\
(e^{\tau\mathbf{A}})_{21} & (e^{\tau\mathbf{A}})_{22} & (e^{\tau\mathbf{A}})_{23} & (\mathbf{B})_{2}\\
(e^{\tau\mathbf{A}})_{31} & (e^{\tau\mathbf{A}})_{32} & (e^{\tau\mathbf{A}})_{33} & (\mathbf{B})_{3}\\
\end{array}\right)
,
\nonumber \\
\label{QMEQ12}
\end{eqnarray}
from which the matrix $e^{\tau\mathbf{A}}$ and vector $\mathbf{B}$  immediately follow.
In Appendix~\ref{section6}, we discuss the method that we used to validate the extraction method.

%%%%%%%%%%%%%%%%%%%%%%%%%%%%%%%%%%%%%%%%%%%%%%%%%%%%%%%%%%%%%%%%%%%%%%%%%%%%%%%%%%%%%%%%%%%%%%%%%%%%%%%%%%%%%%%%%%%%%%%%
%%%%%%%%%%%%%%%%%%%%%%%%%%%%%%%%%%%%%%%%%%%%%%%%%%%%%%%%%%%%%%%%%%%%%%%%%%%%%%%%%%%%%%%%%%%%%%%%%%%%%%%%%%%%%%%%%%%
\section{Fitting a quantum master equation to the solution of the TDSE}\label{section7}

The procedure to test the hypothesis as to whether the QMEQ Eq.~(\ref{GBE})
provides a good approximation to the exact TDSE
of a (small) system which is weakly coupled to a (large) environment
can be summarized as follows:

\begin{enumerate}
\item
Make a choice for the model parameters $h^x_\mathrm{B}$, $h^z_\mathrm{B}$, $K$, $\Delta$,
and the system-bath interaction $\lambda$, for the number of bath spins $N_\mathrm{B}$, the inverse temperature
$\beta$ of the bath, and the time step $\tau$ ($\tau=1$ unless mentioned explicitly).
\item
Prepare three initial states
$|\Psi(0)\rangle_x = |x\rangle |\phi\rangle$,
$|\Psi(0)\rangle_y = |y\rangle |\phi\rangle$, and
$|\Psi(0)\rangle_z = |\uparrow\rangle |\phi\rangle$
where  $|x\rangle = (|\uparrow\rangle + |\downarrow\rangle)/\sqrt{2}$,
$|y\rangle = (|\uparrow\rangle + i|\downarrow\rangle)/\sqrt{2}$,
and $|\phi\rangle$ denotes a pure state picked randomly from the $2^{{N_\mathrm{B}}}$-dimensional unit hypersphere.
For each of the three initial states we may or may not use different realizations of
$|\phi\rangle$.
If $\beta>0$, prepare typical thermal states by projection~\cite{HAMS00}, that is
set $|\Psi(0)\rangle_x=|x\rangle |\phi(\beta/2)\rangle/\langle\phi(\beta/2)|\phi(\beta/2)\rangle^{1/2}$ (and similarly
for the two other initial states)
where $|\phi(\beta/2)\rangle = e^{-\beta H_{\mathrm{B}}/2}|\phi\rangle$.
\item
For each of the three initial states, solve the TDSE for $0\le t=n\tau \le T=N\tau$.
The case of interest is when $T$ is large enough for the system-bath to reach a steady state.
For each of the three different initial states compute
$\rho_{i,j}(k)\equiv \langle\Psi(k\tau)|\sigma^i_{0}|\Psi(k\tau)\rangle_j$,
for $i,j=x,y,z$ and store this data.
\item
Use the data $\rho_{i,j}(k)$ to construct the $3\times3N$ matrix
$\mathbf{Z}=(\mathbf{Z}^{(1)}\,\mathbf{Z}^{(2)}\,\mathbf{Z}^{(3)})$
and  $4\times3N$ matrix $\mathbf{X}=(\mathbf{X}^{(1)}\,\mathbf{X}^{(2)}\,\mathbf{X}^{(3)})$
[see Eq.~(\ref{QMEQ10})]
and compute the $3\times4$ matrix $\mathbf{Y}$, yielding
the best (in the least-square sense) estimates of $e^{\tau\mathbf{A}}$ and $\mathbf{B}$.
\item
Use the estimates of $e^{\tau\mathbf{A}}$ and $\mathbf{B}$ to compute
the averages [denoted by $\widetilde\rho_{i,j}(k)$] of the three components of the system spin operators $\bm\sigma_0(t)$,
according to Eq.~(\ref{QMEQ7}) for each of the three different initial states.
Quantify the difference of the reconstructed data, i.e. the solution of the ``best'' approximation
in terms of the QMEQ, and the original data obtained by solving the TDSE by
the number
\begin{equation}
e_{\mathrm{max}}(t=k\tau)=\max_{i,j} \vert \rho_{i,j}(k) -\widetilde\rho_{i,j}(k)\vert
.
\label{emax}
\end{equation}
\item
Check if the approximate density matrix of the system, defined by $\widetilde\rho_{i,j}(k)$,
is non-negative definite.
In none of our simulation runs the approximate density matrix of the system failed this test.
\end{enumerate}

%

%%%%%%%%%%%%%%%%%%%%%%%%%%%%%%%%%%%%%%%%%%%%%%%%%%%%%%%%
\begin{figure}[t]
\begin{center}
\includegraphics[width=0.45\hsize]{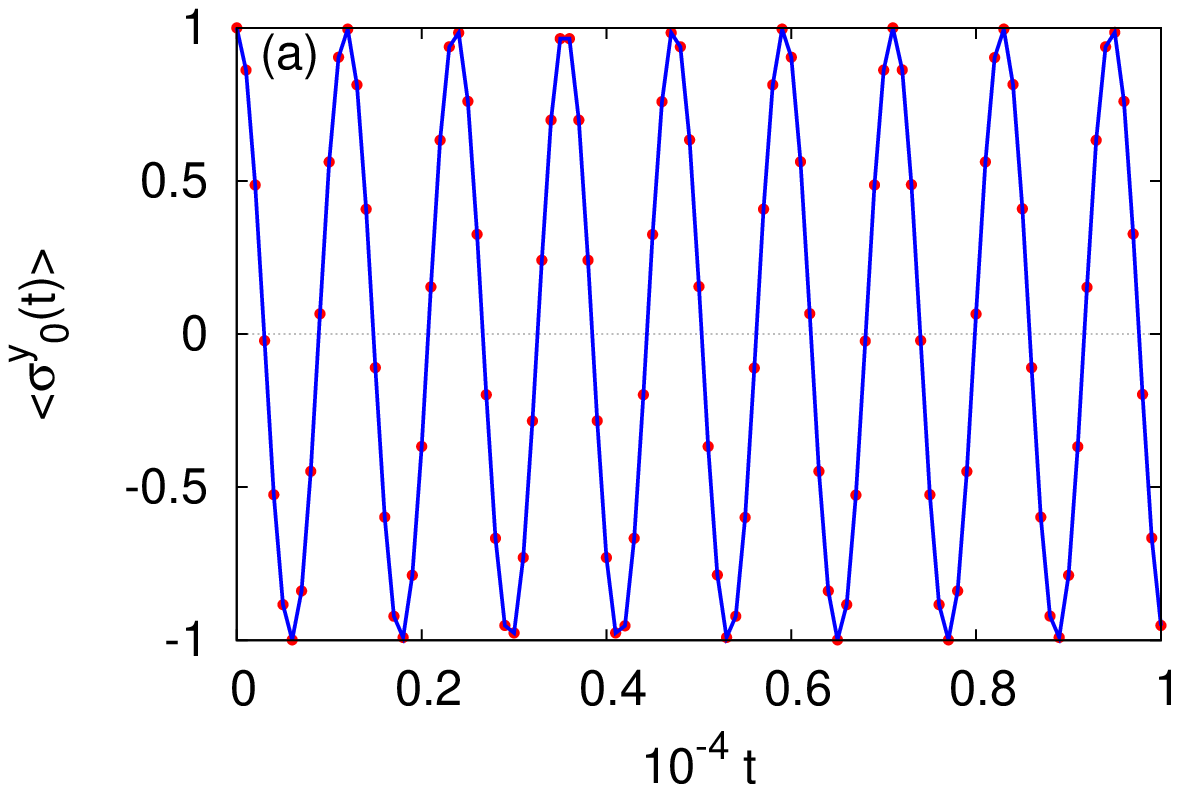}
\includegraphics[width=0.45\hsize]{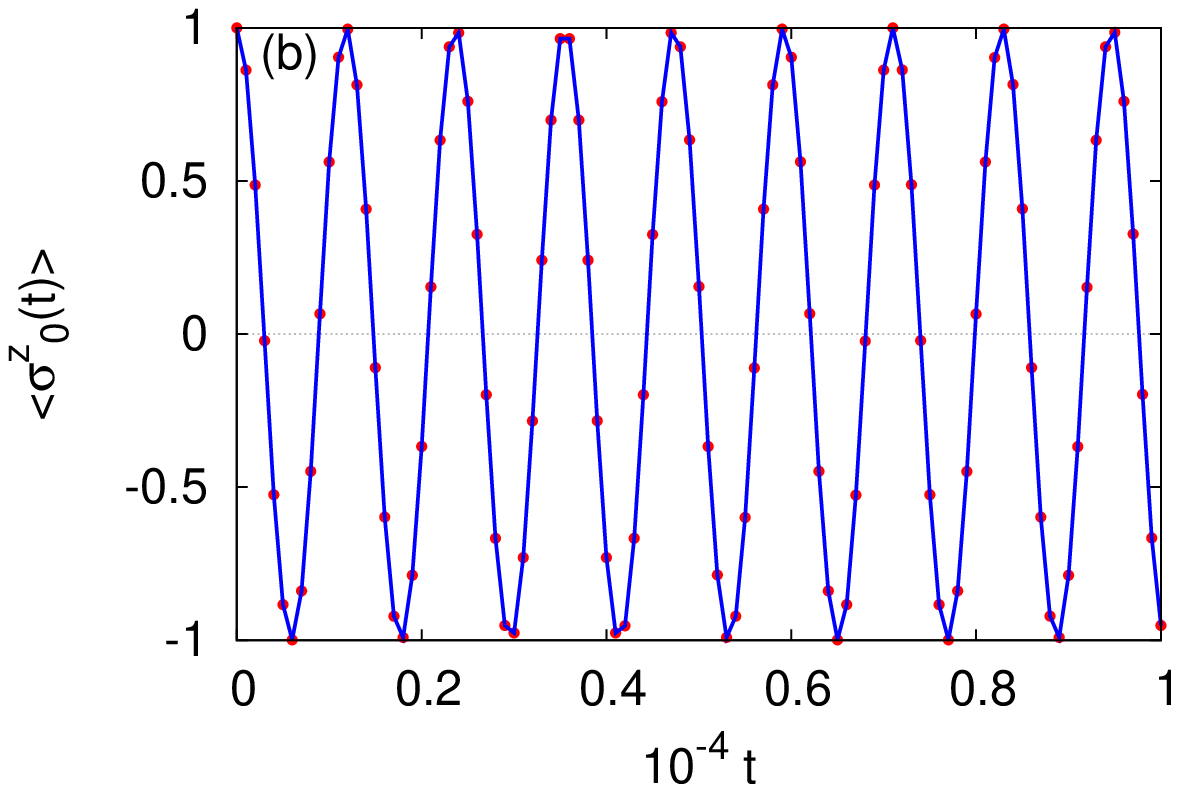}
\caption{(color online) %
Comparison between the spin averages as obtained by solving the TDSE (solid lines)
and the QMEQ (solid circles) with
$e^{\tau\mathbf{A}}$ and $\mathbf{B}$ extracted from the TDSE data.
(a) initial state $|y\rangle |\phi\rangle$;
(b) initial state $|\uparrow\rangle |\phi\rangle$.
The model parameters are:
$\lambda=0$, $N_\mathrm{B}=13$, $\beta=0$, $K=-1/4$, $\Delta=1$ and $h^x_\mathrm{B}=h^z_\mathrm{B}=1/8$.
For clarity, the system-spin averages are shown with a time interval of 100.
The markers represent the data obtained by least-square fitting to 15000 numbers generated by the TDSE solver.
}
\label{fig1}
\end{center}
\end{figure}

\subsection*{Test of the procedure to fit Eq.~(\ref{GBE}) to TDSE data}\label{section7a}

If the system does not interact with the bath ($\lambda=0$),
the system spin simply performs Larmor rotations in the magnetic field $\mathbf{H}=(h^x,0,0)$.
Therefore, the $\lambda=0$ case provides a simple, but as mentioned in Appendix~\ref{section6}
from the numerical viewpoint the most difficult case for the fitting procedure.

In Fig.~\ref{fig1}, we present simulation results of the $y$- and $z$-components
of the system spin as obtained by solving the TDSE with initial states
$|y\rangle |\phi\rangle$ and $|\uparrow\rangle |\phi\rangle$, respectively.
Looking at the time interval shown in Fig.~\ref{fig1} and recalling
that the spin components perform oscillations with a period
$\pi/ h^x$, it is clear that Fig.~\ref{fig1} does not show these rapid oscillations.
Instead, not to clutter the plots too much,
we only plotted the values at regular intervals, as indicated by the markers.
For the initial state $|x\rangle |\phi\rangle$, the $x$-component
is exactly constant (both for the TDSE and time evolution using the estimated $e^{\tau\mathbf{A}}$ and $\mathbf{B}$) and therefore not shown.
The difference between the spin averages obtained from the TDSE and from time evolution according to Eq.~(\ref{QMEQ7}) (using
the estimated $e^{\tau\mathbf{A}}$ and $\mathbf{B}$)
is rather small ($e_{\mathrm{max}}(t)< 10^{-5}$ for $0\le t \le 10000$) and is therefore not shown either.

The small values of $e_{\mathrm{max}}(t)$ are reflected in the excellent agreement between the TDSE and
QMEQ (Eq.~(\ref{GBE})) data shown in Fig.~\ref{fig1}.
From these simulation data we conclude that for $\lambda=0$, the matrix $e^{\tau\mathbf{A}}$ and vector $\mathbf{B}$
obtained by least-square fitting to the TDSE data
define a QMEQ that reproduces the correct values of the spin averages.

\begin{figure}[t]
\begin{center}
\includegraphics[width=0.33\hsize]{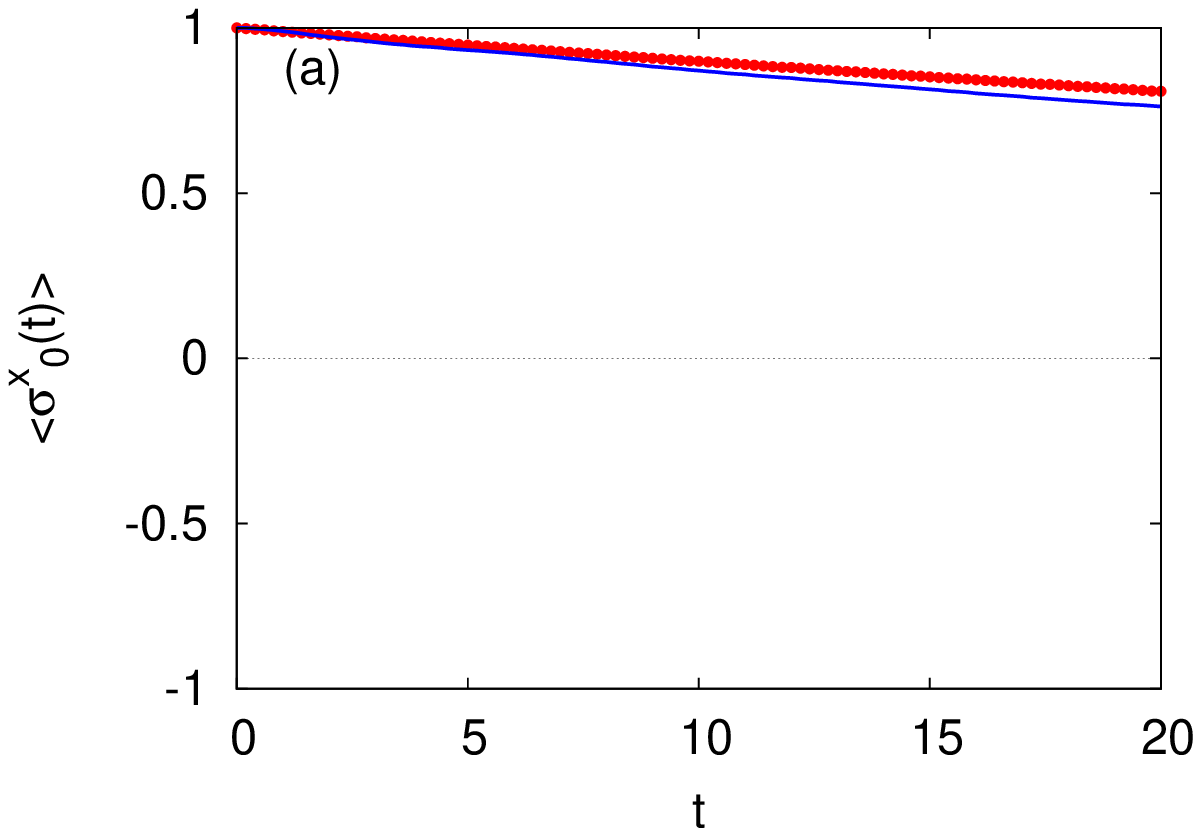}
\includegraphics[width=0.33\hsize]{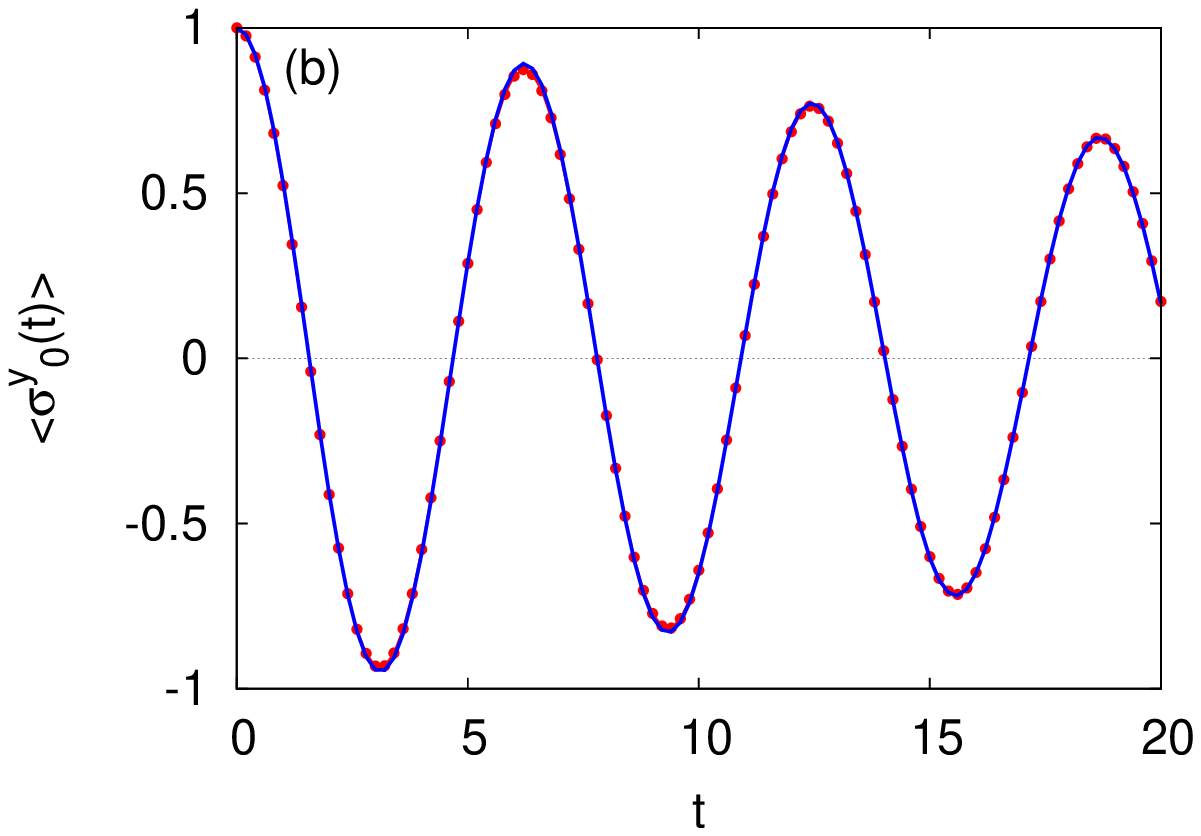}
\includegraphics[width=0.33\hsize]{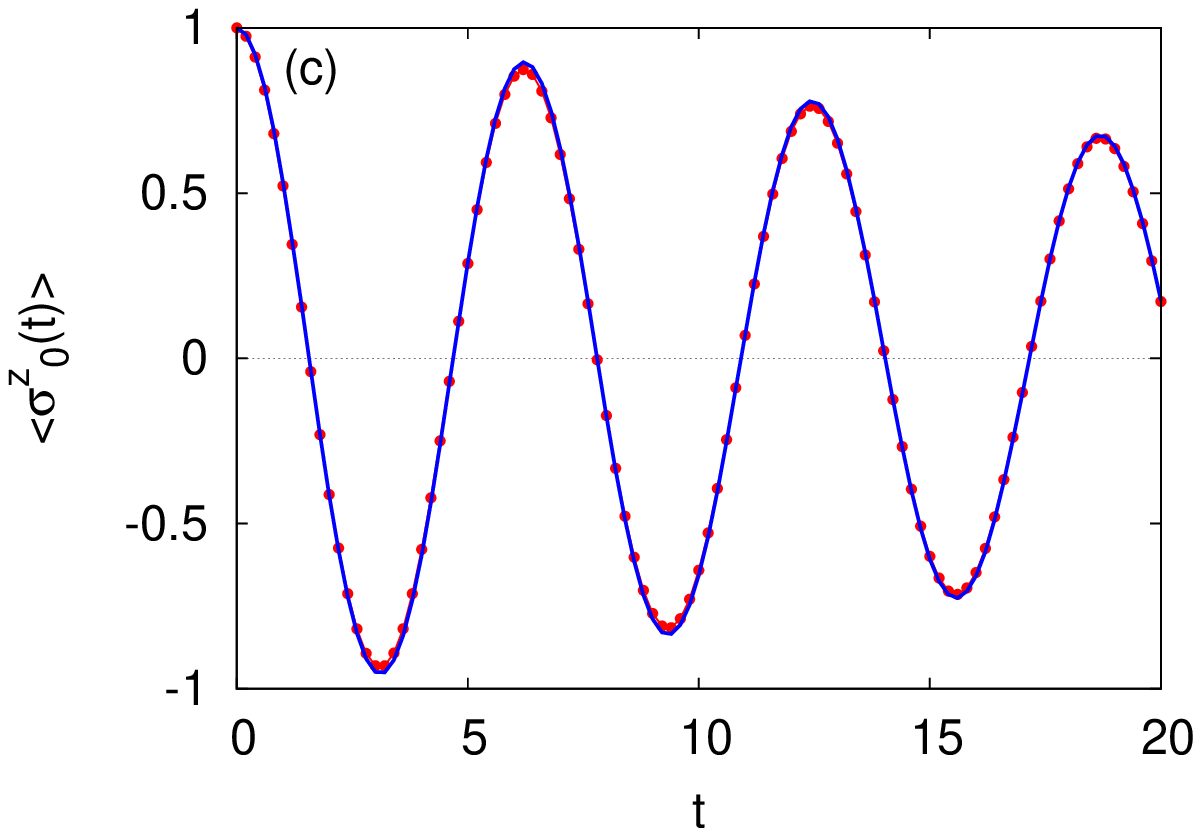}
\includegraphics[width=0.33\hsize]{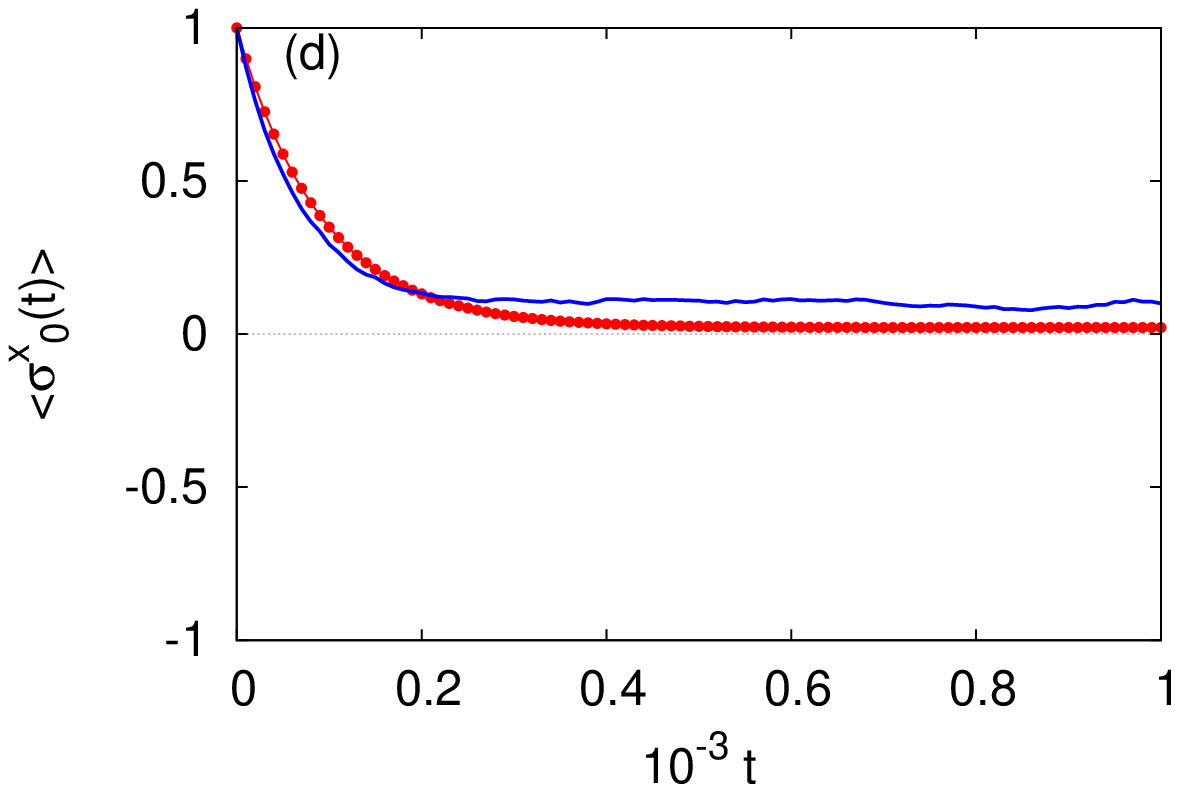}
\includegraphics[width=0.33\hsize]{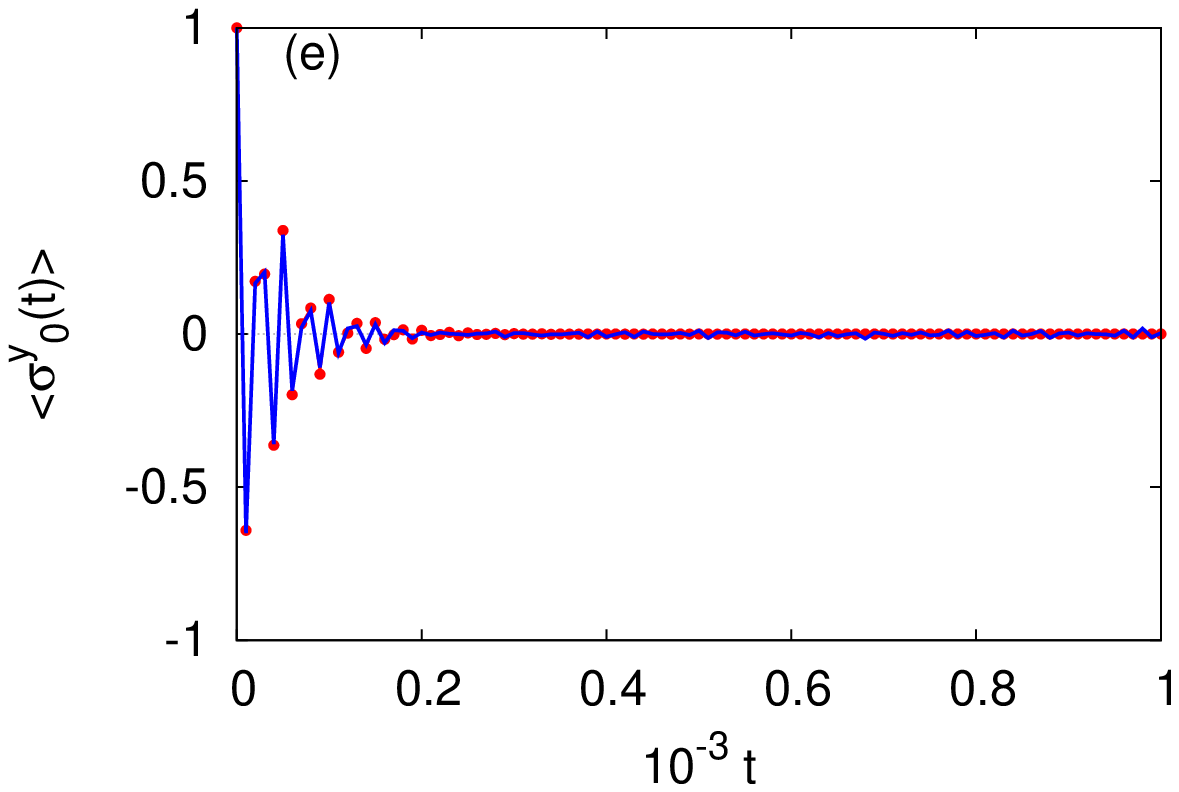}
\includegraphics[width=0.33\hsize]{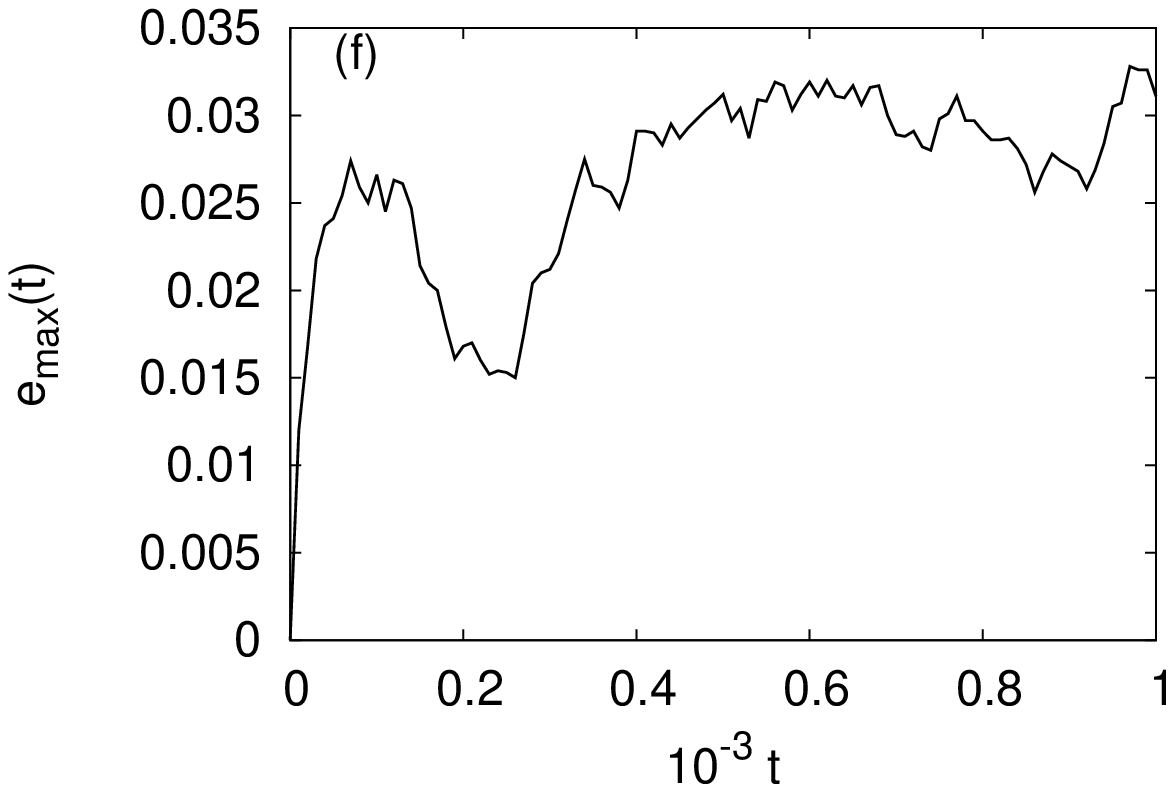}
\caption{(color online) %
Comparison between the spin averages as obtained by solving the TDSE (solid lines)
and the QMEQ (solid circles) with
$e^{\tau\mathbf{A}}$ and $\mathbf{B}$ extracted from the TDSE data.
(a)--(c) show how the TDSE data (solid line) are being sampled,
namely at times indicated by the $t$-values of the markers, which in the case corresponds
to a time steps of 0.2.
(d)-- (f): the sampled data of the whole interval $[0,1000]$, in this case
15000 numbers, are used to determine by the least-square procedure described in Sec.~\ref{section5},
the parameters that enter the time-evolution of the Markovian master equation Eq.~(\ref{GBE}).
The latter is then used to compute the time-evolution of the spin components,
the data being represented by the markers.
For clarity, in the bottom figures, the data are shown with a time interval of 10.
The model parameters are: $h^x_\mathrm{B}=h^z_\mathrm{B}=1/8$
and $\lambda=0.1$, $N_\mathrm{B}=13$, $\beta=0$, $K=-1/4$, and $\Delta=1$.
(a),(d) initial state $|x\rangle |\phi\rangle$;
(b),(e) initial state $|y\rangle |\phi\rangle$;
(c) initial state $|z\rangle |\phi\rangle$;
(f) the error $e_{\mathrm{max}}(t)$.
}
\label{fig3}
\end{center}
\end{figure}

The next step is to repeat the analysis for the case of weak system-bath interaction $\lambda=0.05$
(recall that we already found that $\lambda=0.1$ corresponds to a weak interaction).
To head off misunderstandings, recall that our least-square procedure
estimates the best $e^{\tau\mathbf{A}}$ and $\mathbf{B}$ using the data of three different solutions of the TDSE.
It does not fit data for individual spin components separately
nor does it fit data obtained from a TDSE solution of one particular choice of the initial state.
Our procedure yields the best {\sl global} estimates for $e^{\tau\mathbf{A}}$ and $\mathbf{B}$ in the least-square sense.

In Fig.~\ref{fig3} we illustrate the procedure for sampling and processing the TDSE data
and for plotting these data along with the data obtained from Eq.~(\ref{QMEQ7}) using
the estimated $e^{\tau\mathbf{A}}$ and $\mathbf{B}$.
We present data for short times (top figures) and for the whole time interval (bottom figures).
The TDSE data (solid line) is being sampled, namely at times indicated by the $t$-values of the markers,
which in the case corresponds to a time steps of 0.2 [see Figs.~\ref{fig3}(a)--\ref{fig3}(c)].
The sampled data of the whole interval $[0,1000]$ are used to determine
$e^{\tau\mathbf{A}}$ and $\mathbf{B}$
by the least-square procedure described in Sec.~\ref{section5}.
In this particular case, the TDSE solver supplies 15000 numbers to the least-square procedure.
The estimated $e^{\tau\mathbf{A}}$ and $\mathbf{B}$ thus obtained
are then used to compute the time-evolution of the spin components,
the data being represented by the markers.

From Fig.~\ref{fig3}(d), it is clear that although the QMEQ produces the correct qualitative behavior
of the $x$-component of the system spin,
the difference with the TDSE data is significant (as is also clear from $e_{\mathrm{max}}(t)$).
In particular, the TDSE data of the $x$-component of the system spin
do not show relaxation to the thermal equilibrium value, which is zero for $\beta=0$.
At first sight, this could be a signature that the fitting procedure breaks down
because it is certainly possible to produce a much better fit to the TDSE data
of the $x$-component {\sl if we would fit a curve to this data only}.
But, as explained above, we estimate $e^{\tau\mathbf{A}}$ and $\mathbf{B}$
by fitting to the nine (three spin components $\times$ three different initial states)
of such curves simultaneously.
Apparently, the mismatch in the $x$-component is compensated for by the close match
of the $y$--component [see Fig.~\ref{fig3}(e) and $z$-component (not shown)].

Remarkably, the matrix $e^{\tau\mathbf{A}}$ and vector $\mathbf{B}$
extracted from the TDSE data yield a QMEQ that
does indicate that the system spin relaxes to a state that is close to thermal equilibrium:
The QMEQ yields a value of 0.04 for the expectation value of the
$x$-component of the system spin and values less than $10^{-4}$ for the other two components.
From the general theory of the QMEQ in the Markovian approximation~\cite{BREU02}, we know that
if the correlations of the bath-operators Eq.~(\ref{bc0}) satisfy the Kubo-Martin-Schwinger
condition, the stationary state solution of the QMEQ is exactly the same
as the thermal equilibrium state of the system (ignoring corrections of ${\cal O}(\lambda)$ [see
Ref.~\onlinecite{MORI08} for a detailed discussion)].

The mismatch between the QMEQ and TDSE data of the $x$-component
can be attributed to the fact that a bath of $N_\mathrm{B}=13$ spins is too small
to act as a bath in thermal equilibrium.
However, the argument that leads to this conclusion is somewhat subtle.
As shown in Sec.~\ref{section3b}, the random state approach applied to the system + bath yields the correct
thermal equilibrium properties.
In particular, in the case at hand ($\beta=0$, $N_\mathrm{B}=13$), within the usual statistical fluctuations it yields
$\langle\Phi(\beta=0)| \sigma^\alpha_0(t)|\Phi(\beta=0)\rangle\approx 0$ for $\alpha=x,y,z$.
Note that in this kind of calculation, the initial state $|\Phi(\beta=0)\rangle$ is a random state of the system + bath.
In contrast, the data shown in Fig.~\ref{fig3}(d) are obtained by solving the TDSE
with the initial state $|\Psi(0)\rangle_x = |x\rangle |\phi\rangle$ (see Sec.~\ref{section7}).
Therefore, the results of Fig.~\ref{fig3}(d) demonstrate that for $N_\mathrm{B}=13$, the statement that
\begin{equation}
|x\rangle |\phi\rangle \longrightarrow \mathrm{\hbox{TDSE evolution}} \longrightarrow
|\widetilde\Phi\rangle
,
\nonumber \\
\label{v7a}
\end{equation}
where $\widetilde\Phi$ denotes an (approximate) random state of the whole system, {\sl is not necessarily true}.
Otherwise, we would have $\langle\widetilde\Phi| \sigma^x_0(t)|\widetilde\Phi\rangle\approx 0$
for $t$ large enough, in contradiction with the data shown in Fig.~\ref{fig3}(d).
Roughly speaking, one could say that a bath of $N_\mathrm{B}=13$ is not sufficiently ``complex''
to let the TDSE evolve certain initial states towards a random state of the whole system.
For a discussion of the fact that in general, Eq.~(\ref{v7a}) does not necessarily hold, see Ref.~\onlinecite{JIN10x}.

\begin{figure}[t]
\begin{center}
\includegraphics[width=0.33\hsize]{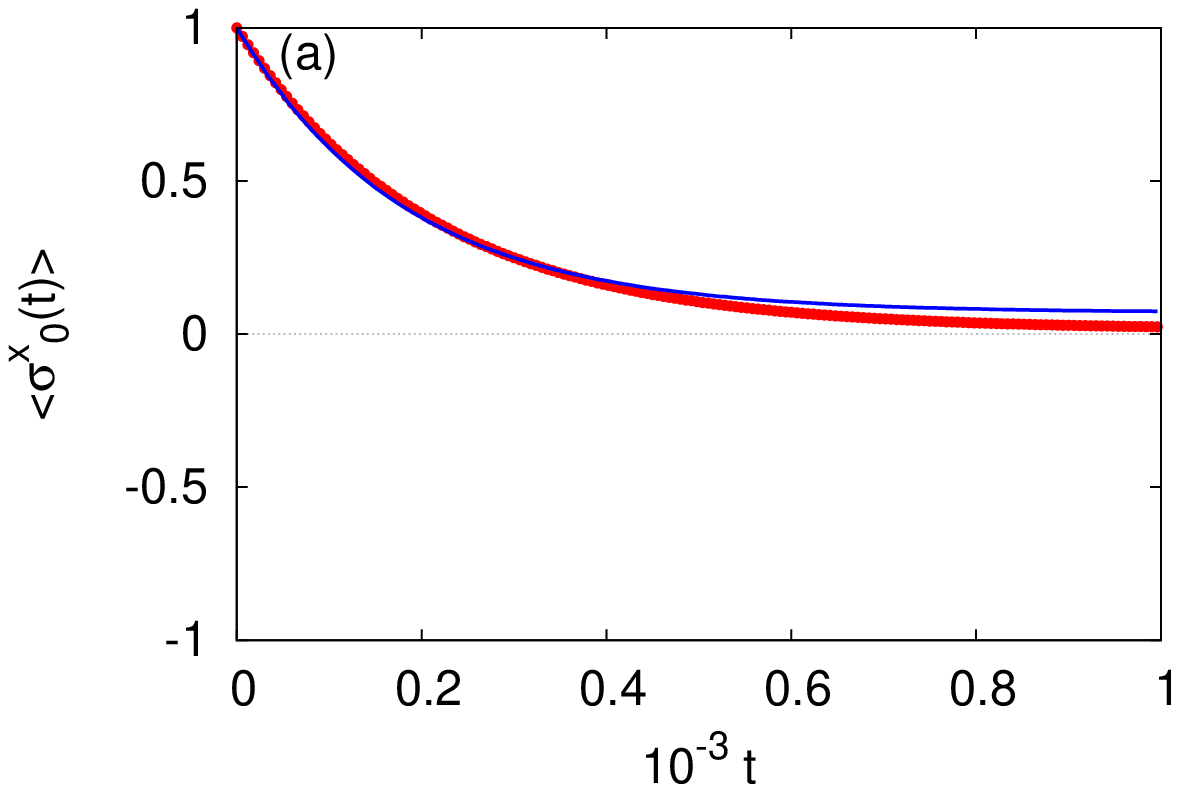}
\includegraphics[width=0.33\hsize]{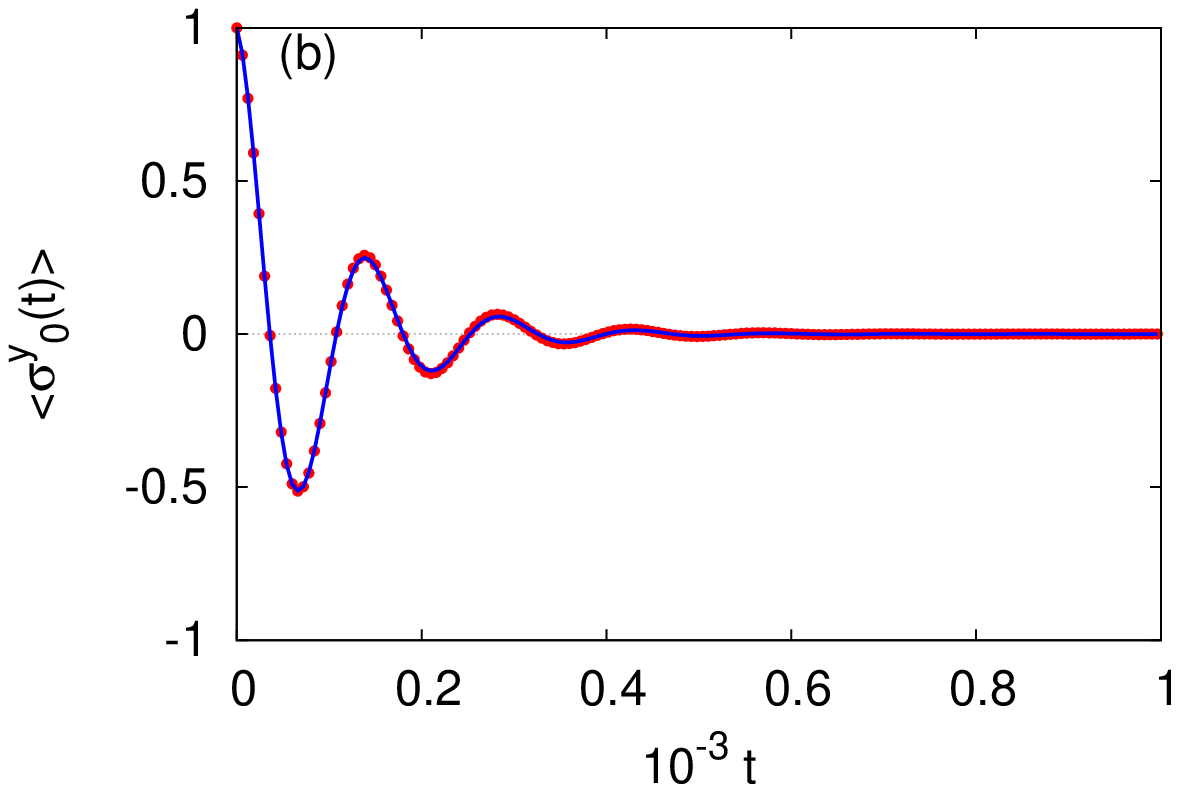}
\includegraphics[width=0.33\hsize]{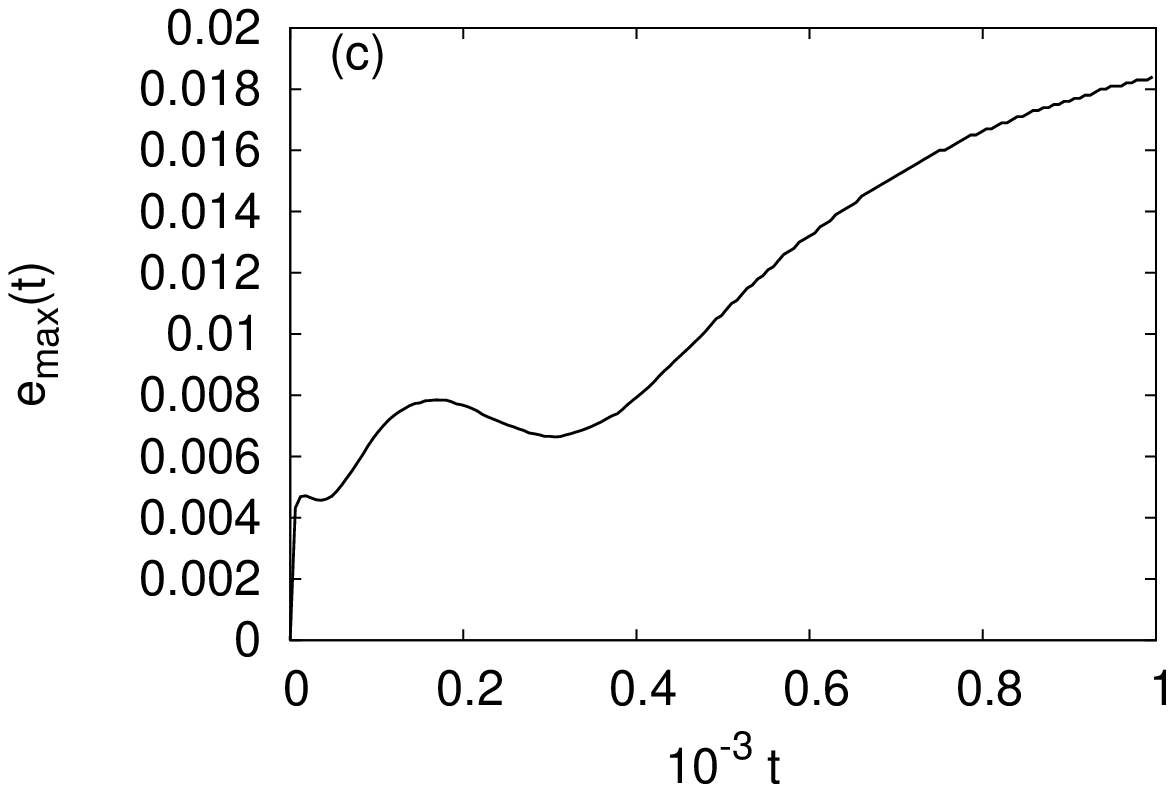}
\caption{(color online) %
Same as Fig.~\ref{fig3}, except that the bath contains $N_\mathrm{B}=24$ spins and $\lambda=0.05$.
The markers represent the data obtained by least-square fitting to 15000 numbers generated by the TDSE solver.
For clarity the data is shown with a time interval of 6.
}
\label{fig2j}
\end{center}
\end{figure}

\begin{figure}[t]
\begin{center}
\includegraphics[width=0.33\hsize]{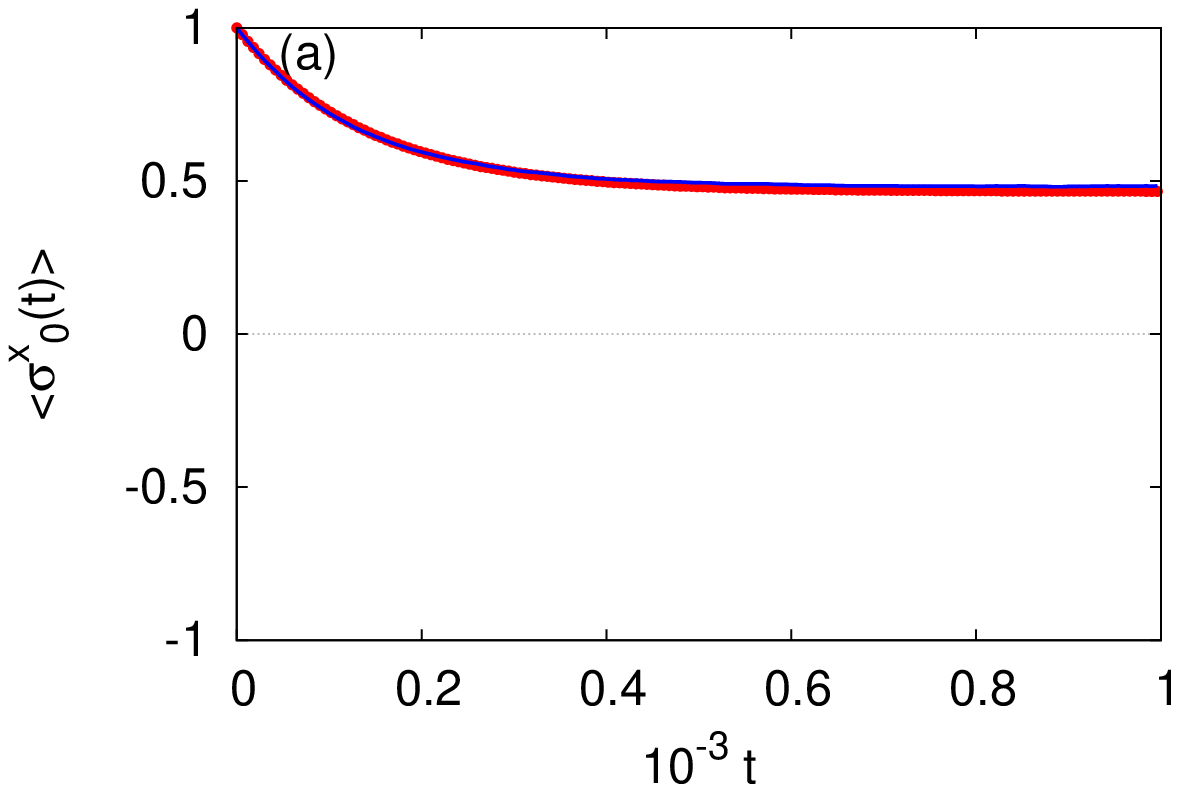}
\includegraphics[width=0.33\hsize]{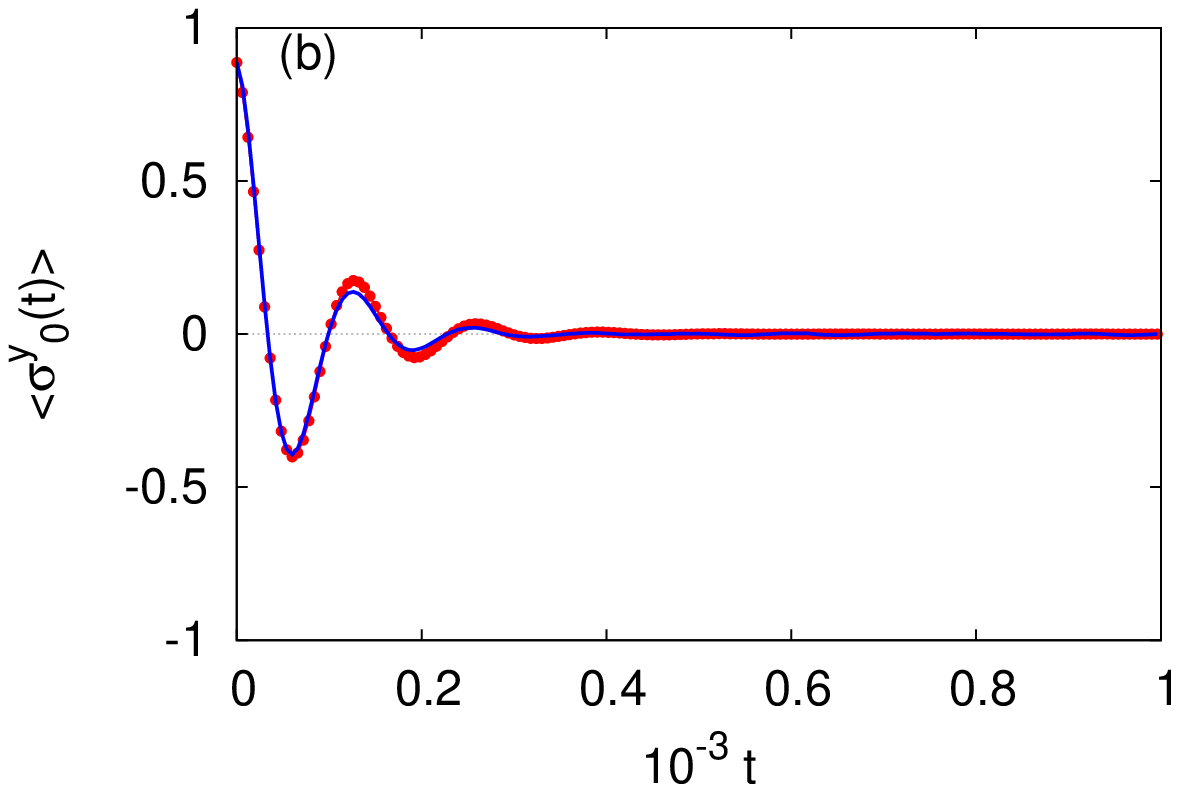}
\includegraphics[width=0.33\hsize]{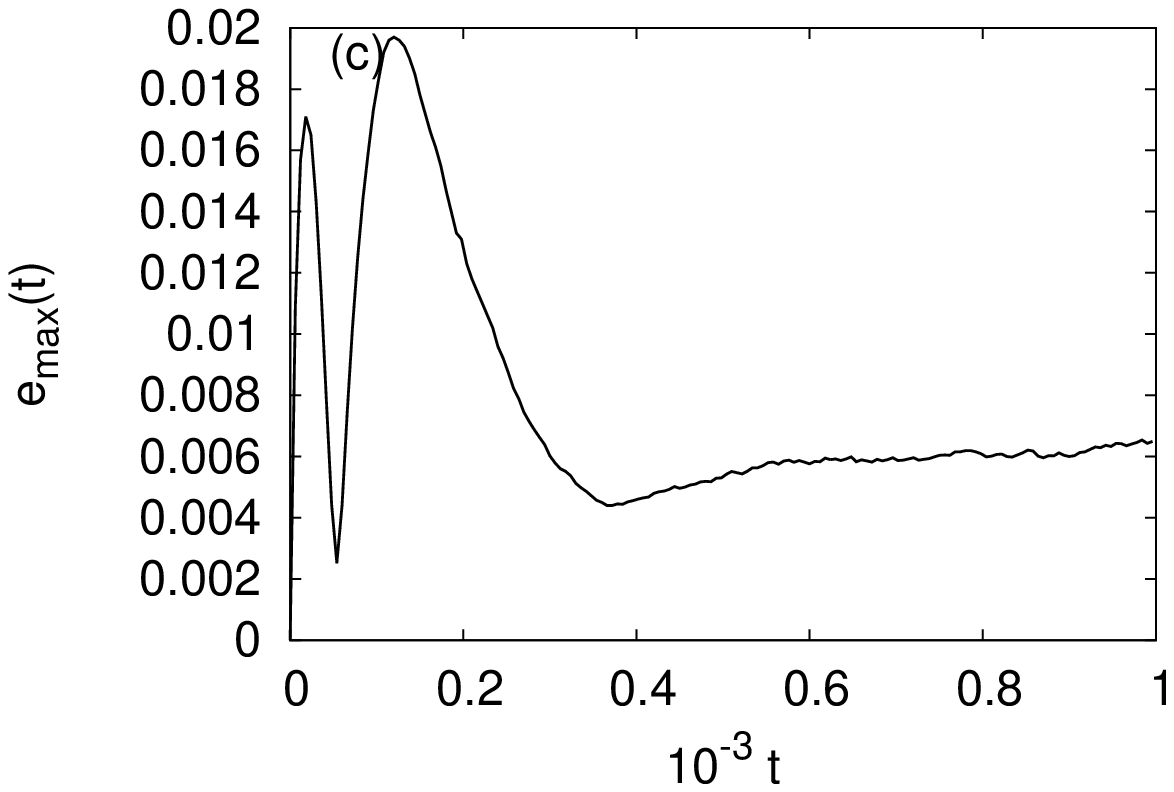}
\caption{(color online) %
Same as Fig.~\ref{fig2j}, except that the bath is initially at $\beta=1$.
}
\label{fig2k}
\end{center}
\end{figure}

As a check on this argument, we repeat the simulation with a bath $N_\mathrm{B}=24$ spins.
The results are shown in Fig.~\ref{fig2j}.
Comparing Figs.~\ref{fig3} and \ref{fig2j}, it is clear that for long times
the value of the $x$-component decreases as the number of spins in the bath increases
and that the agreement between the TDSE data and the fitted QMEQ data has improved considerably.
This suggests that as the size of the bath increases and with the
bath initially in a random state, the TDSE evolution can drive the state to an (approximate) random state of the whole system,
meaning that the whole system relaxes to the thermal equilibrium state.
However, as discussed in Sec.~\ref{section10} there are exceptions~\cite{JIN10x}.

In general, we may expect that for short times, a Markovian QMEQ cannot
represent the TDSE evolution very well~\cite{SUAR92,GASP99}.
But if we follow the evolution for times much longer than the typical correlation times of the
bath-operators, the difference between the QMEQ and TDSE data for short times does
not affect the results of fitting the data over the whole, large time-interval in a significant manner.
Hence there is no need to discard the short-time data in the fitting procedure.
As a matter of fact, the data shown in Fig.~\ref{fig3} indicate that the least-square procedure
applied to the whole data set yields a Markovian master equation that reproduces the short-time behavior
quite well.

Finally, we check that the conclusions reached so far for a bath at $\beta=0$ also hold when $\beta>0$.
In Fig.~\ref{fig2k}, we show the simulation results for $\beta=1$, for the same system and bath as
the one used to obtain the data shown in Fig.~\ref{fig2j}.
From Fig.~\ref{fig2k} we conclude that the agreement between the TDSE and QMEQ data is quite good.

\begin{figure}[t]
\begin{center}
\includegraphics[width=0.33\hsize]{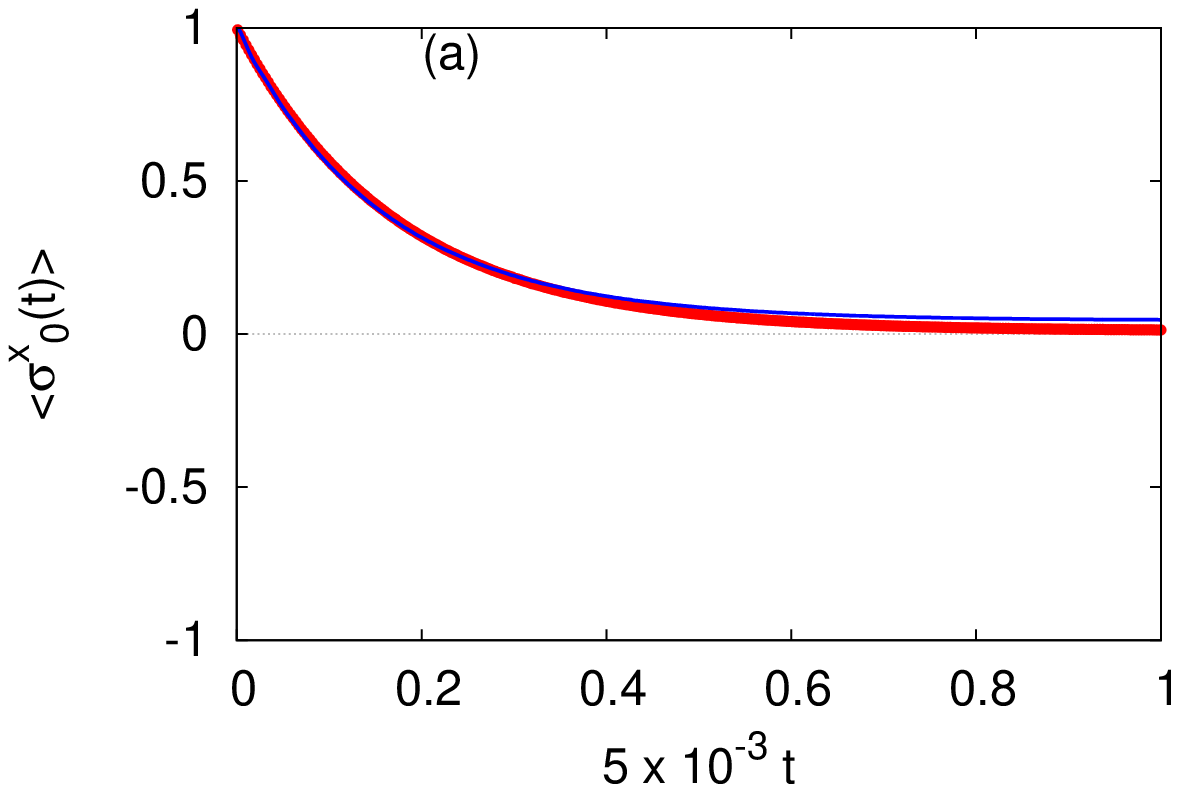}
\includegraphics[width=0.33\hsize]{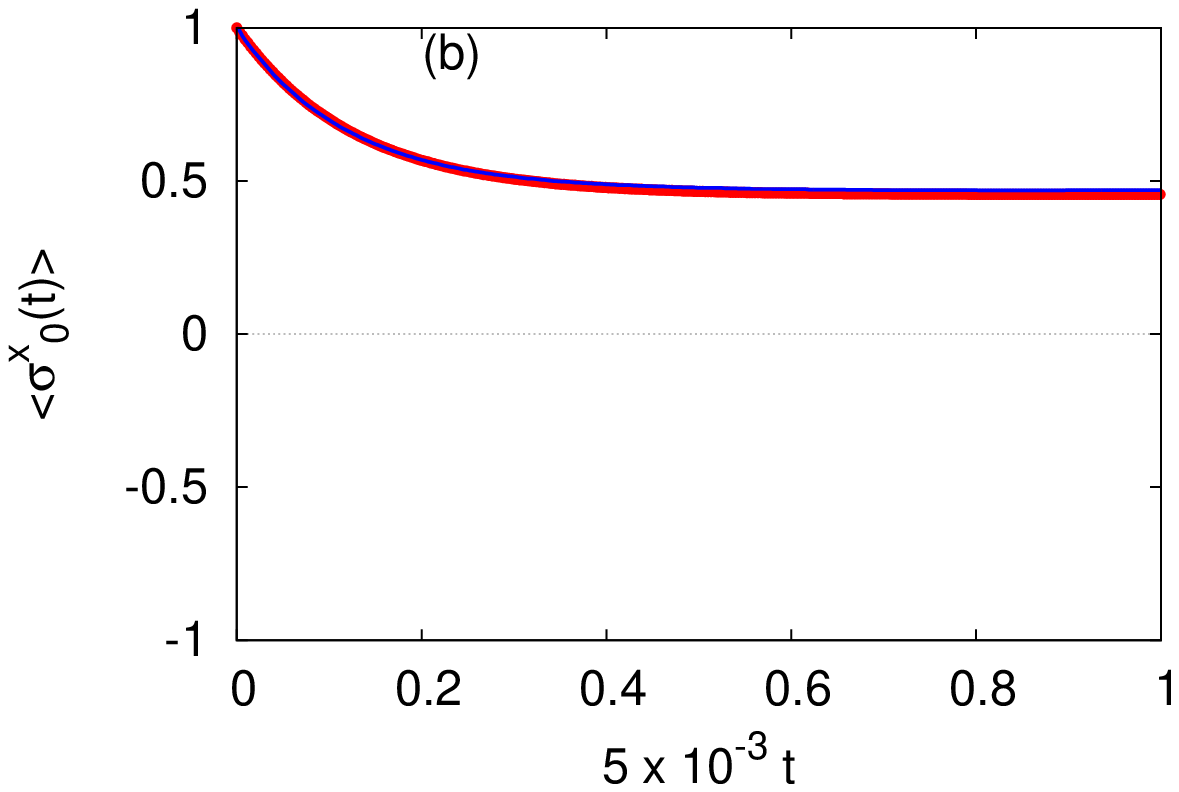}
\includegraphics[width=0.33\hsize]{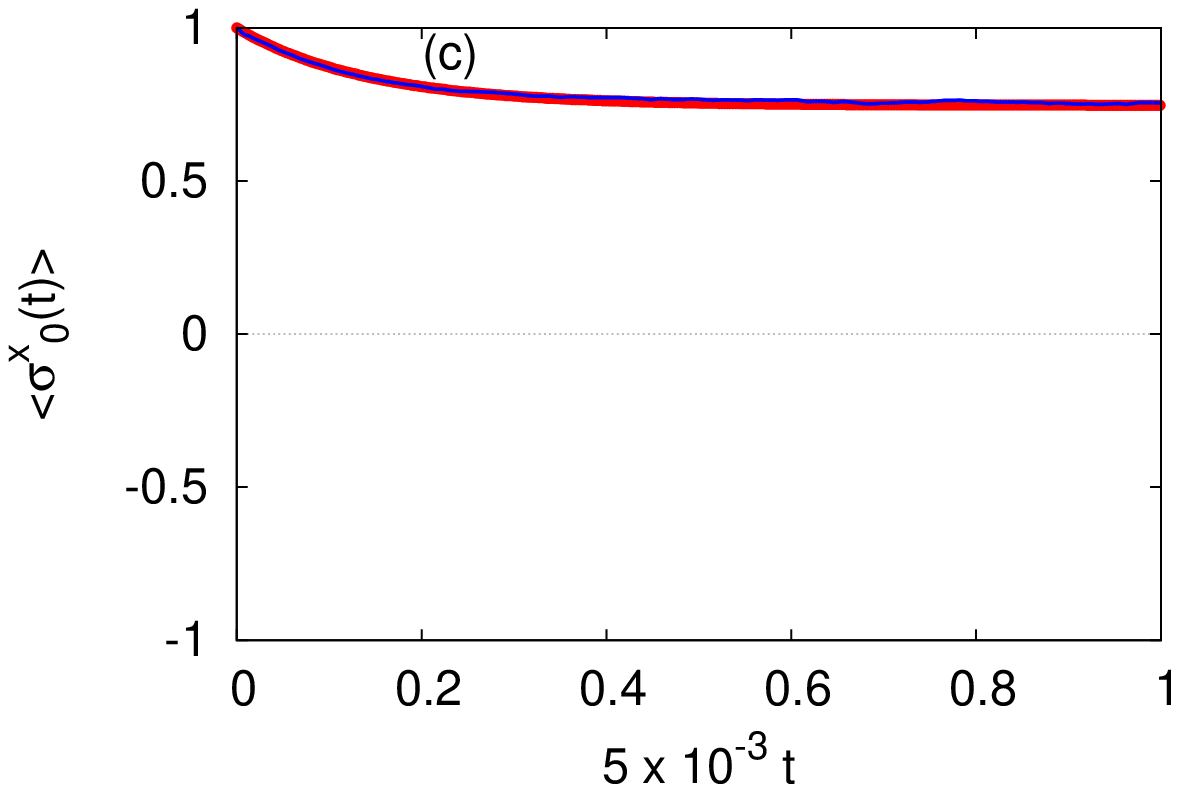}
\includegraphics[width=0.33\hsize]{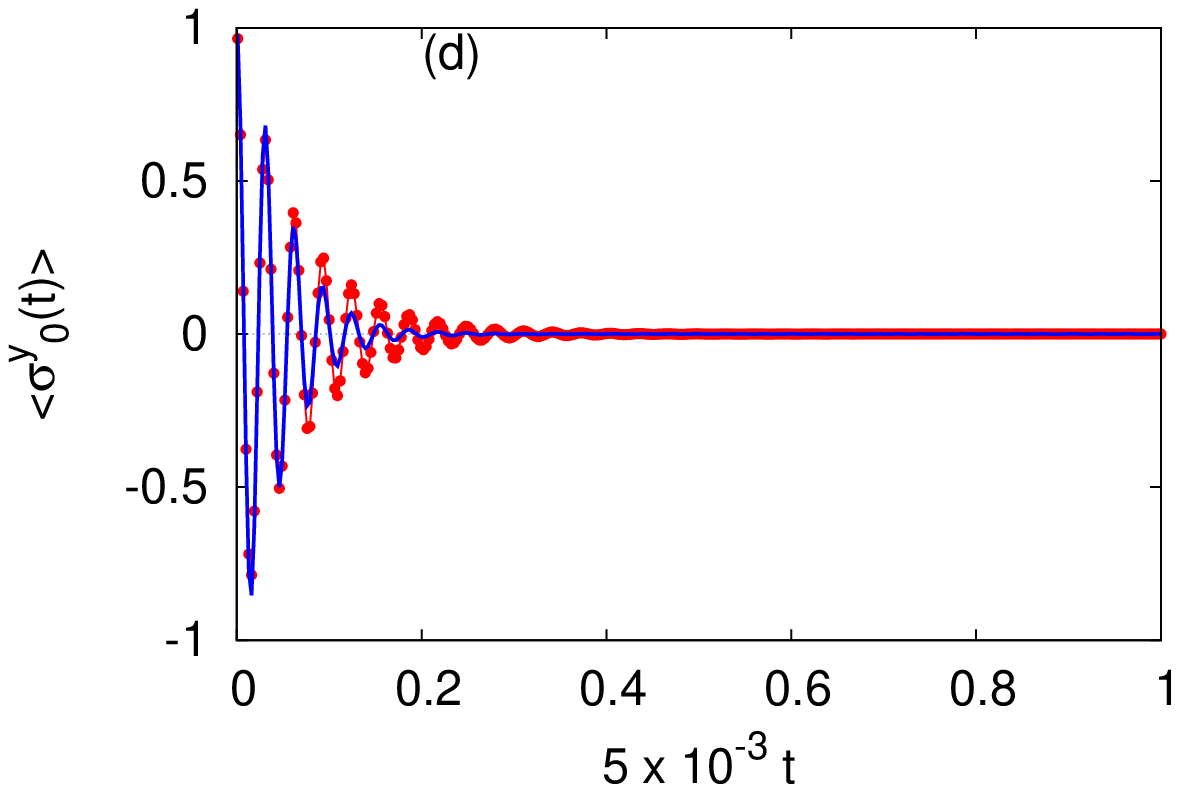}
\includegraphics[width=0.33\hsize]{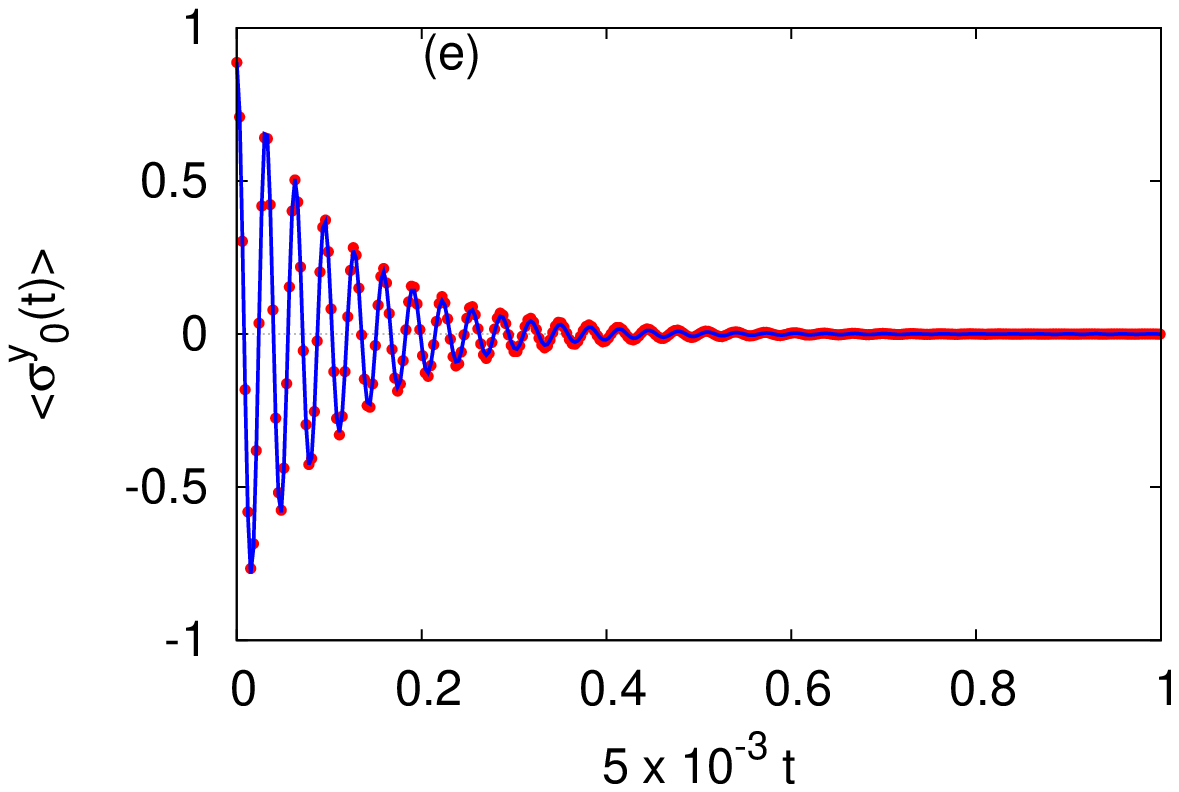}
\includegraphics[width=0.33\hsize]{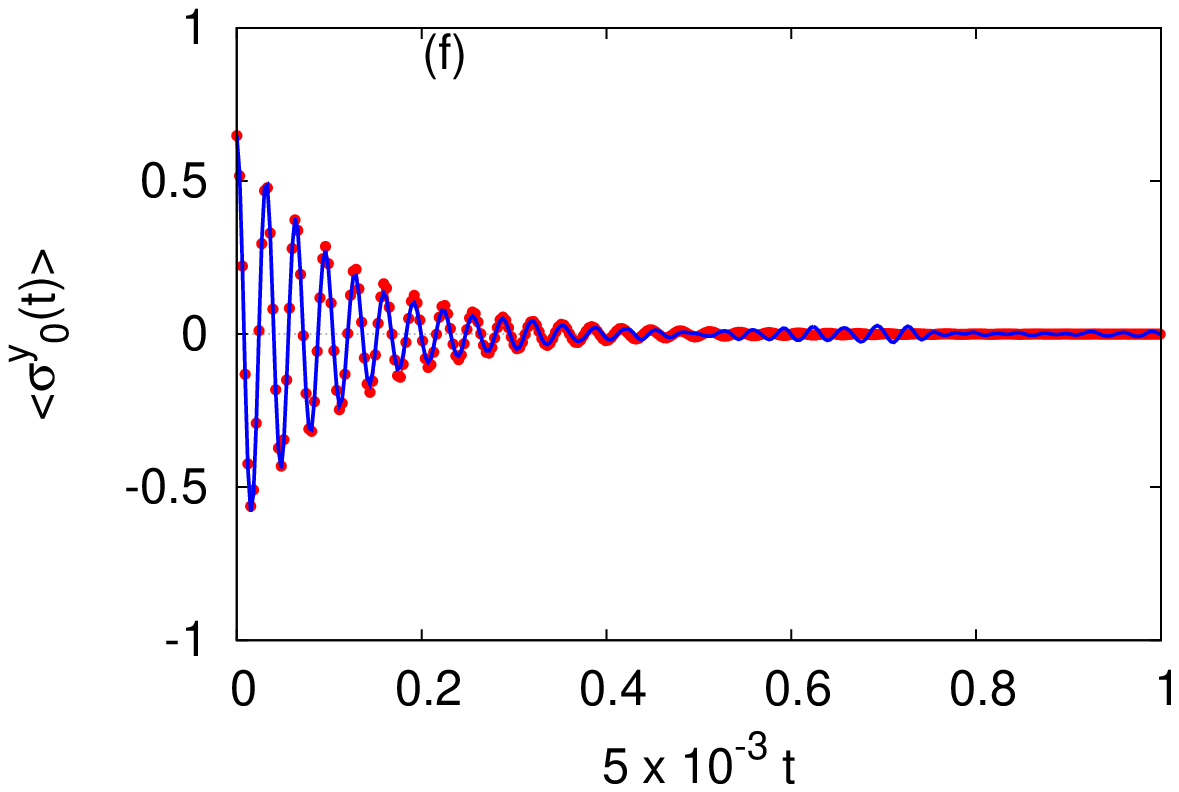}
\caption{(color online) %
Simulation data for a bath with $N_\mathrm{B}=28$ spins and system-bath interaction $\lambda=0.1$.
The model parameters are: $h^x_\mathrm{B}=h^z_\mathrm{B}=1/8$, $K=-1/4$, and $\Delta=1$.eps.
Solid lines: TDSE data; solid circles: QMEQ data.
Top row: $\langle \sigma^x(t)\rangle$ as obtained by starting from the initial state $|x\rangle |\phi\rangle$,
(a)--(c) corresponding to $\beta=0,1,2$, respectively.
Bottom row: $\langle \sigma^y(t)\rangle$ as obtained by starting from the initial state $|y\rangle |\phi\rangle$,
(d)--(f) corresponding to $\beta=0,1,2$, respectively.
The TDSE simulations yield
$\langle| \sigma^x_0(t=200)|\rangle = 0.044$,
$\langle| \sigma^x_0(t=200)|\rangle = 0.475$,
and
$\langle| \sigma^x_0(t=200)|\rangle = 0.756$
for $\beta=0,1,2$, respectively,
whereas for the system in equilibrium we have
$\langle\sigma^x_0\rangle = 0,0.462,0.762$ for $\beta=0,1,2$, respectively.
For clarity, the data are shown with a time interval of 0.6.
The TDSE solver provided 3000 numbers as input to the least-square procedure.
}
\label{fig3j}
\end{center}
\end{figure}

\begin{table}[ht]
\caption{
The parameters that appear in Eq.~(\ref{GBEx}) as obtained by
fitting the QMEQ to the TDSE data shown in Fig.~\ref{fig3j}.
}
%\begin{ruledtabular}
%\begin{tabular*}{}
\begin{tabular*}{\textwidth}{@{\extracolsep{\fill}} ccllll}
\noalign{\medskip}
\hline\hline%\noalign{\smallskip}
$\beta$ & $i$ & \hfil $A_{i,1}$ \hfil  & \hfil $A_{i,2}$ \hfil & \hfil $A_{i,3}$ \hfil & \hfil $b_i$ \hfil  \\
\hline%\noalign{\smallskip}
$0 $&$ 1 $&$ -0.29\times10^{-1} $&$ +0.57 \times10^{-3} $&$ -0.11\times10^{-2} $&$ -0.31\times10^{-3}$  \\
$0 $&$ 2 $&$ -0.55\times10^{-2} $&$ -0.73 \times10^{-1} $&$ +1.01 $&$ -0.95\times10^{-4}$  \\
$0 $&$ 3 $&$ -0.73\times10^{-3} $&$ -1.01  $&$ -0.74 \times10^{-1}$&$ -0.56\times10^{-4}$  \\
\hline%\noalign{\smallskip}
$1 $&$ 1 $&$ -0.40\times10^{-1} $&$ +0.11 \times10^{-1} $&$ -0.11\times10^{-3} $&$ -0.18\times10^{-1}$  \\
$1 $&$ 2 $&$ -0.11\times10^{-1} $&$ -0.36 \times10^{-1} $&$ +0.99 $&$ -0.29\times10^{-3}$  \\
$1 $&$ 3 $&$ -0.54\times10^{-3} $&$ -0.99  $&$ -0.53 \times10^{-1}$&$ -0.32\times10^{-3}$  \\
\hline%\noalign{\smallskip}
$2 $&$ 1 $&$ -0.35\times10^{-1} $&$ +0.29 \times10^{-1} $&$ +0.75\times10^{-3} $&$ -0.27\times10^{-1}$  \\
$2 $&$ 2 $&$ -0.22\times10^{-1} $&$ -0.45 \times10^{-1} $&$ +0.98 $&$ -0.47\times10^{-2}$  \\
$2 $&$ 3 $&$ -0.84\times10^{-3} $&$ -0.98  $&$ -0.40 \times10^{-1}$&$ -0.16\times10^{-3}$  \\
\hline%\noalign{\smallskip}
\end{tabular*}
%\end{ruledtabular}
\label{tab2}
\end{table}

\section{Simulation results: $N_\mathrm{B}=28,32$}\label{section8}

As already mentioned in Sec.~\ref{section3}, in practice, there is a limitation
on the sizes and time intervals that can be explored.
By increasing the system-bath interaction $\lambda$, we can shorten
the time needed for the system to relax to equilibrium.
On the other hand, $\lambda$ should not be taken too large
because when we leave the perturbative regime, the
QMEQ of the form Eq.~(\ref{GBE}) cannot be expected to capture the true quantum dynamics.
From our exploratory simulations, we know that $\lambda=0.1$
is still within the perturbative regime, hence we will
adopt this value when solving the TDSE for baths with up to
$N_\mathrm{B}=32$ spins.

\begin{figure}[t]
\begin{center}
\includegraphics[width=0.33\hsize]{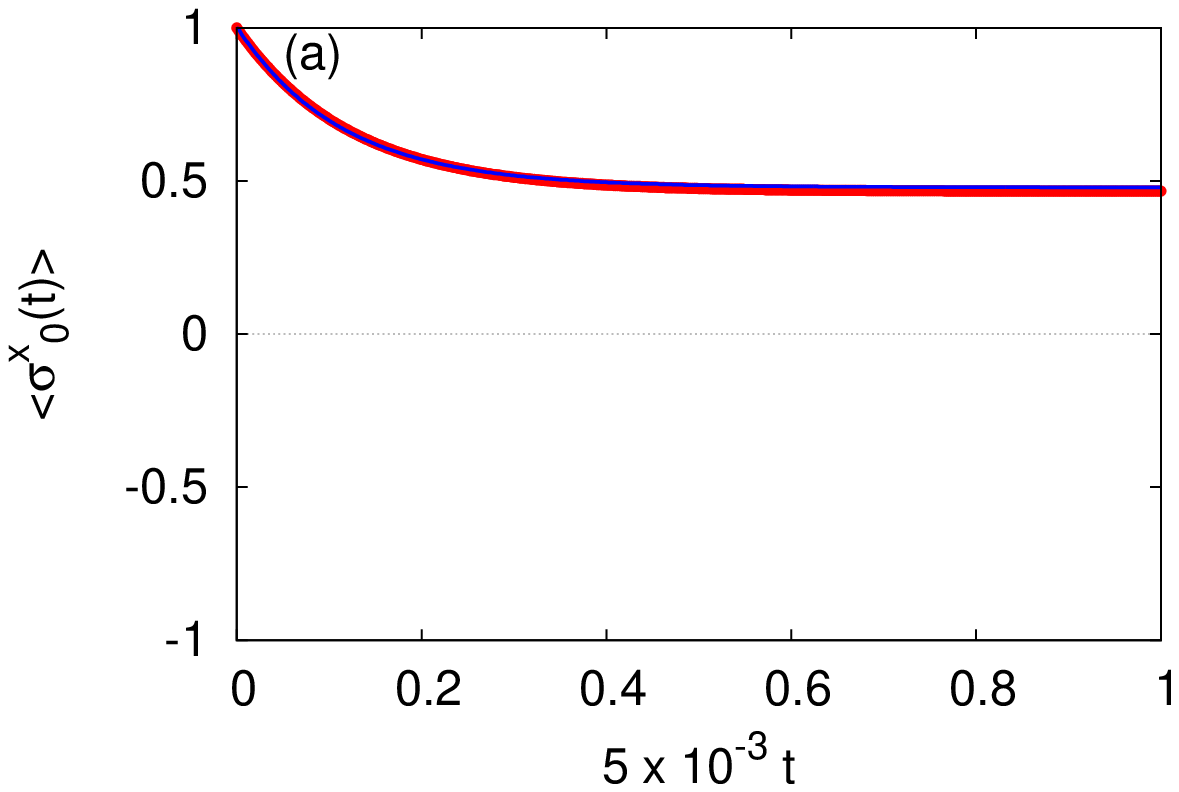}
\includegraphics[width=0.33\hsize]{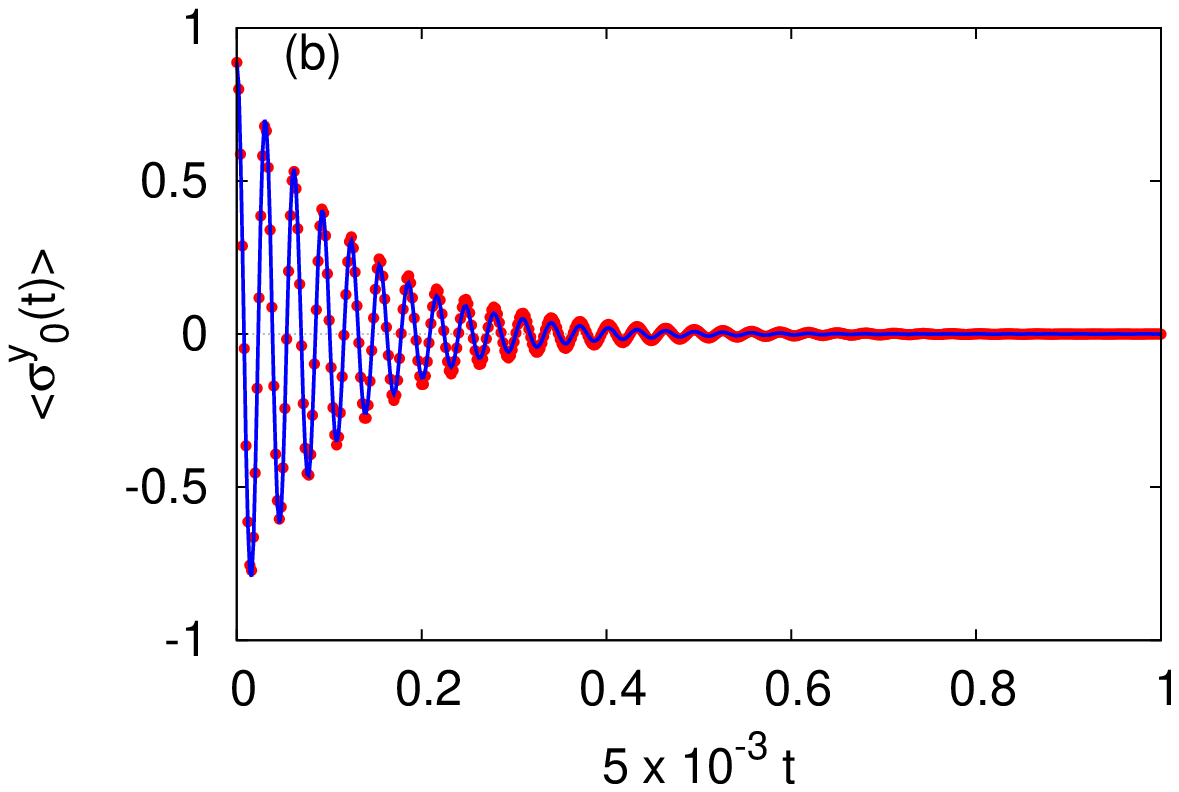}
\includegraphics[width=0.33\hsize]{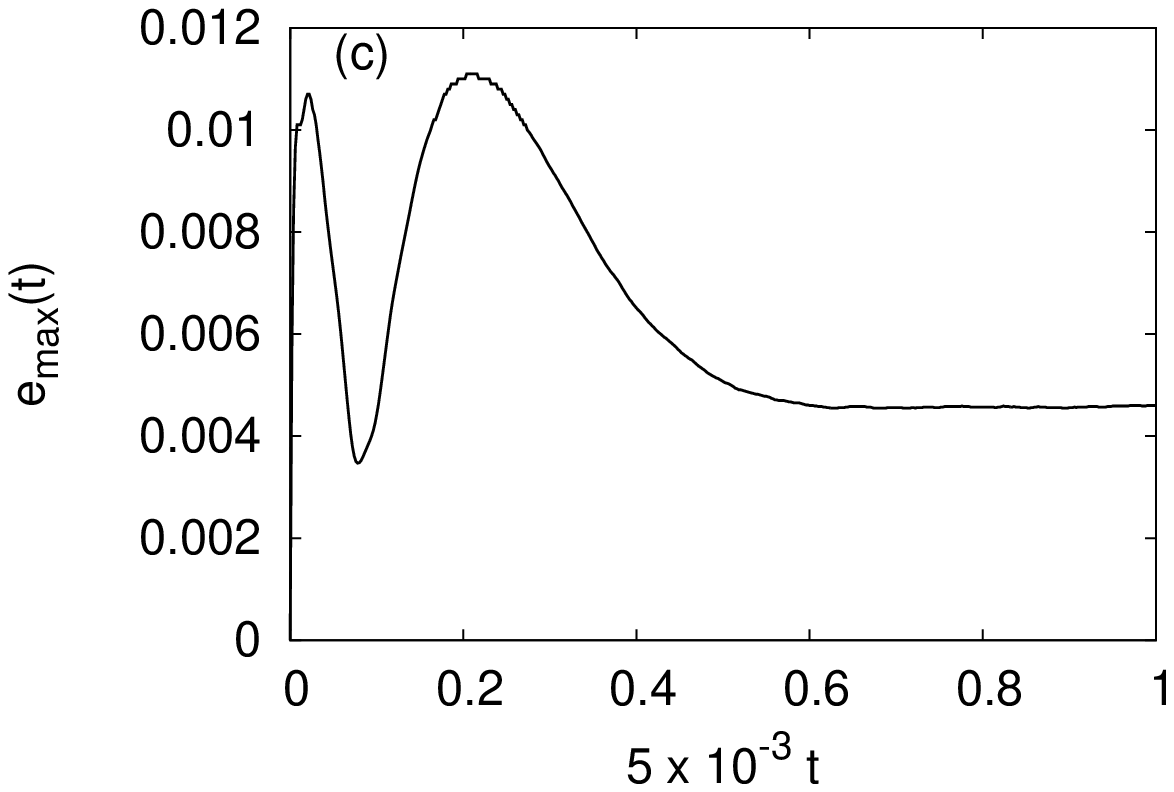}
\caption{(color online) %
Simulation data for a bath with $N_\mathrm{B}=32$ spins prepared at $\beta=1$ and system-bath interaction $\lambda=0.1$.
The model parameters are: $h^x_\mathrm{B}=h^z_\mathrm{B}=1/8$, $K=-1/4$, and $\Delta=1$.
Figures (a,b) show TDSE data (solid lines) and QMEQ data (solid circles).
(a) initial state $|x\rangle |\phi\rangle$;
(b) initial state $|y\rangle |\phi\rangle$;
(c) the error $e_{\mathrm{max}}(t)$.
The data obtained with the initial state $|\uparrow\rangle |\phi\rangle$ is
very similar as the data obtained with the initial state $|y\rangle |\phi\rangle$
and are therefore not shown.
For clarity, the data are shown with a time interval of 0.4.
The TDSE solver provided 3000 numbers as input to the least-square procedure.
}
\label{fig4j}
\end{center}
\end{figure}

In Fig.~\ref{fig3j} we present the results as obtained with a bath containing
$N_\mathrm{B}=28$ spins, prepared at $\beta=0,1,2$.
Although Fig.~\ref{fig3j} may suggest otherwise,
the maximum error $\max_k e_{\mathrm{max}}(t)\approx 0.05,0.1,0.2$ for $\beta=0,1,2$, respectively,
indicating that the difference between the TDSE data and the QMEQ approximation
increases with $\beta$.
The results presented in Fig.~\ref{fig4j} for a bath of $N_\mathrm{B}=32$ spins
and $\beta=1$ provide additional evidence for the observation that a bath of
$N_\mathrm{B}=28,32$ spins are sufficiently large to mimic an infinite thermal bath.
At any rate, in all cases, there is very good qualitative agreement between the TDSE and QMEQ data.

From the TDSE data, we can, of course, also extract the values
of the entries in the matrix $\mathbf{A}$ and vector $\mathbf{b}$, see Eq.~(\ref{GBE}).
Writing Eq.~(\ref{GBE}) more explicitly as
\begin{eqnarray}
\frac{\partial \langle \bm{\sigma}_0^x(t) \rangle}{\partial t}
&=& A_{1,1} \langle \bm{\sigma}_0^x(t) \rangle + A_{1,2} \langle \bm{\sigma}_0^y(t) \rangle
+ A_{1,3} \langle \bm{\sigma}_0^z(t) \rangle +b_1
\nonumber \\
\frac{\partial \langle \bm{\sigma}_0^y(t) \rangle}{\partial t}
&=& A_{2,1} \langle \bm{\sigma}_0^x(t) \rangle + A_{2,2} \langle \bm{\sigma}_0^y(t) \rangle
+ A_{2,3} \langle \bm{\sigma}_0^z(t) \rangle +b_2
\nonumber \\
\frac{\partial \langle \bm{\sigma}_0^z(t) \rangle}{\partial t}
&=& A_{3,1} \langle \bm{\sigma}_0^x(t) \rangle + A_{3,2} \langle \bm{\sigma}_0^y(t) \rangle
+ A_{3,3} \langle \bm{\sigma}_0^z(t) \rangle +b_3
,
\label{GBEx}
\end{eqnarray}
and using, as an example, the data shown in Fig.~\ref{fig3j}, we obtain
the values of the coefficients as given in Table~\ref{tab2}.
From Table~\ref{tab2}, we readily recognize that (i) $A_{2,3}\approx -A_{3,2} \approx 1$
represents the precession of the system spin in the magnetic field $h^x=1/2$,
(ii) there is a weak coupling between the $x$- and $(y,z)$- components
of the system spin and (iii) the three spin components have different relaxations times.

\begin{figure}[t]
\begin{center}
\includegraphics[width=0.33\hsize]{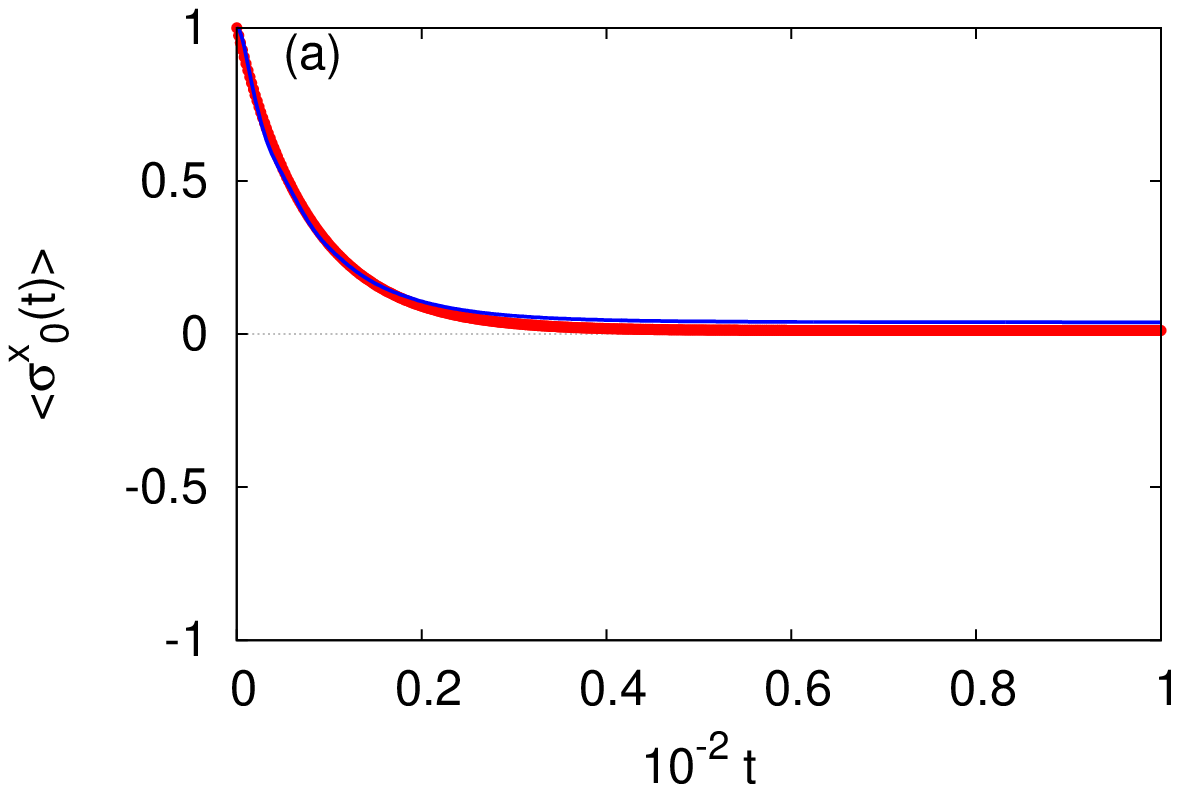}
\includegraphics[width=0.33\hsize]{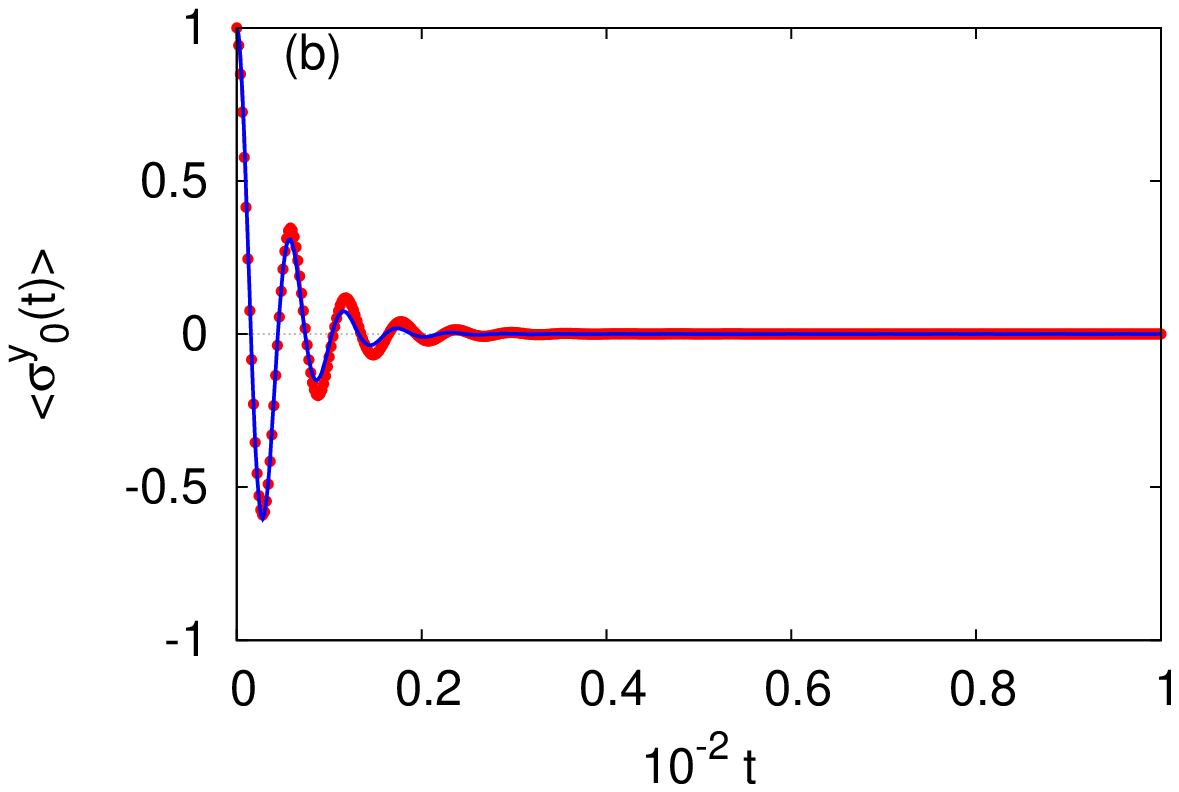}
\includegraphics[width=0.33\hsize]{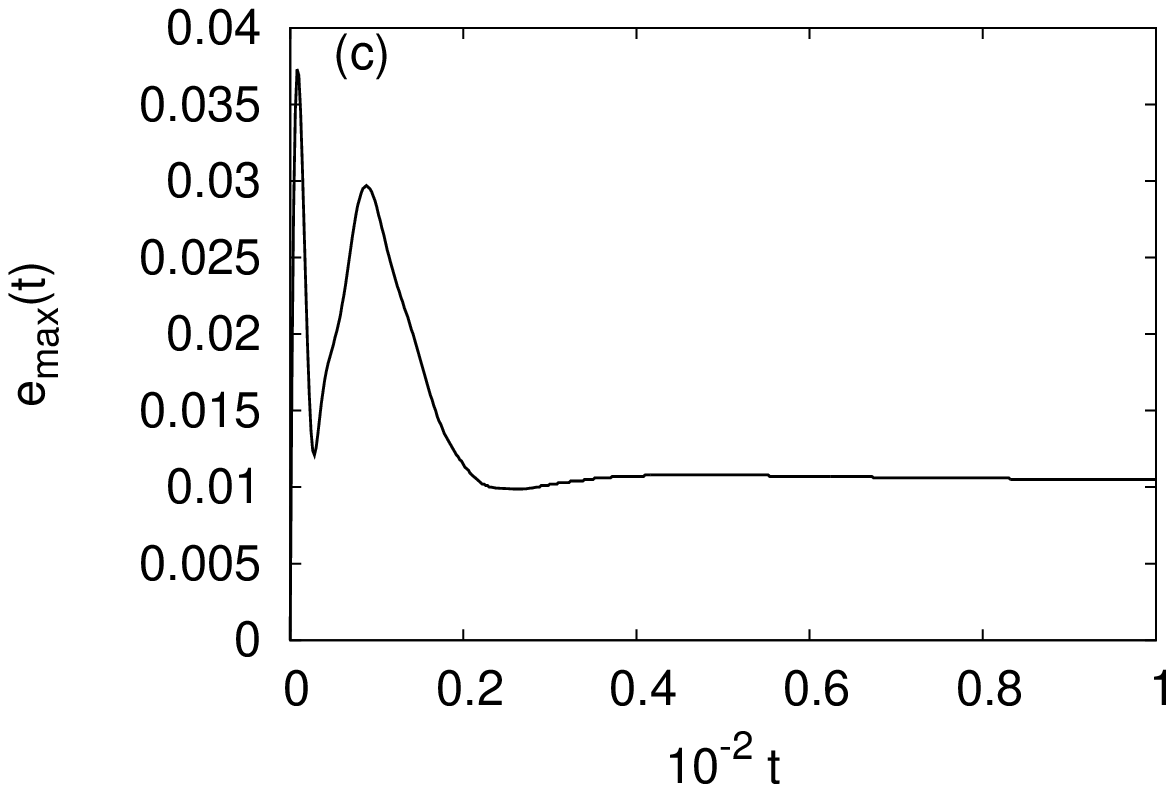}
\caption{(color online) %
Same as Fig.~\ref{fig4j} except that $\beta=0$ and $\lambda=0.2$.
}
\label{fig4k}
\end{center}
\end{figure}

As a final check whether $\lambda=0.1$ is well within the perturbative regime,
we repeat the simulations for a bath containing $N_\mathrm{B}=32$ spins
and system-bath interaction $\lambda=0.2$ and $\beta=0$.
The simulation data are presented in Fig.~\ref{fig4k}.
Clearly, there still is good qualitative agreement between the TDSE and QMEQ data but,
as expected, $\max_k e_{\mathrm{max}}(t)$ has become larger (by a factor of about 3).

In Table~\ref{tab3} we present results (first three rows) for the least-square estimates of the
parameters that enter the QMEQ, as obtained from the TDSE data shown in Fig.~\ref{fig4j}.
Taking into account that with each run, the random values of the model parameters
change, the order-of-magnitude agreement between the data for $N_\mathrm{B}=28$
(Table~\ref{tab2}, rows 4--6) and the $N_\mathrm{B}=32$ data is rather good.
We also present results (middle and last three rows) for the parameters that enter the Redfield equation Eq.~(\ref{QMEQ4}),
as obtained from the TDSE data of the bath-operator correlations $C(i,j,t)$
for $0\le t \le 40$ [see Figs.~\ref{figbc0}(a) and \ref{figbc0}(c) for a picture of some of these data].
From Table~\ref{tab3}, it is clear that there seems to be little quantitative agreement between a description based on
the Redfield quantum master equation (\ref{QMEQ4}) obtained by using the bath-operator correlations $C(i,j,t)$ data
and the parameters obtained from the least-square fit of Eq.~(\ref{GBEx}) to the TDSE data.
Simulations using the 3D bath Hamiltonian (\ref{s42b}) support this conclusion (see Appendix~\ref{3D}).

Although our results clearly demonstrate that QMEQ Eq.~(\ref{GBE}) quantitatively
describes the true quantum dynamics of a spin interacting with a spin bath rather well,
the Redfield quantum master equation Eq.~(\ref{QMEQ4}) in the Markovian approximation,
which is also of the form Eq.~(\ref{GBE}), seems to perform rather poorly in comparison.
The estimates of the diagonal matrix elements of the matrix $\mathbf{A}$
as obtained from the expressions in terms of the bath-operator correlations $C(i,j,t)$
are too small by factors 3--7.
This suggests that the approximations involved in the derivation of Eq.~(\ref{QMEQ4})
are not merely of a perturbative nature but affect the dynamics in
a more intricate manner [see Ref.~\onlinecite{BREU06} for an in-depth discussion of these aspects].

\begin{table}[tb]
\caption{
First three data rows: coefficients that appear in Eq.~(\ref{GBEx}) as obtained by
fitting the QMEQ to the TDSE data shown in Fig.~\ref{fig4j}.
Middle three rows: the corresponding coefficients as obtained by numerically
calculating the parameters $r_{jk}$ that appear in the Redfield quantum master equation Eq.~(\ref{QMEQ4})
according to Eq.~(\ref{QMEQ2b}), using the TDSE data of the bath-operator correlations
shown in Fig.~\ref{figbc0}(a).
Last three rows: same as the middle three rows except that
the used TDSE data of the bath-operator correlations are shown in Fig.~\ref{figbc0}(c).
Note that the baths used in these simulations are very different (see Fig.~\ref{figbc0}), yet the
relevant numbers (those with absolute value larger than $10^{-4}$) are in the same ballpark.
}
%\begin{ruledtabular}
%\begin{tabular*}{}
\begin{tabular*}{\textwidth}{@{\extracolsep{\fill}} cllll}
\noalign{\medskip}
\hline\hline%\noalign{\smallskip}
$i$ & \hfil $A_{i,1}$ \hfil  & \hfil $A_{i,2}$ \hfil & \hfil $A_{i,3}$ \hfil & \hfil $b_i$ \hfil  \\
\hline%\noalign{\smallskip}
$ 1 $&$ -0.49\times10^{-1} $&$ +0.82 \times10^{-2} $&$ -0.56\times10^{-3} $&$ -0.19\times10^{-1}$  \\
$ 2 $&$ -0.80\times10^{-2} $&$ -0.42 \times10^{-1} $&$ +1.02 $&$ -0.14\times10^{-4}$  \\
$ 3 $&$ -0.38\times10^{-3} $&$ -1.01  $&$ -0.41 \times10^{-1}$&$ -0.40\times10^{-3}$  \\
\hline%\noalign{\smallskip}
% Eq.(5) xxx model ; qmeq13.x-ns1ne32b1jzz1hxs1hzs0xy1hx0.25hz0.25l0.1-t200-i2.b.bath
% # & $-0.71\times 10^{-2} $ & $-0.15\times 10^{-3} $ & $+0.18\times 10^{-3} $ & $-0.29\times 10^{-2} $\\
% # & $-0.13\times 10^{-3} $ & $-0.15\times 10^{-1} $ & $+1.00\times 10^{+0} $ & $-0.63\times 10^{-4} $\\
% # & $+0.16\times 10^{-3} $ & $-1.00\times 10^{+0} $ & $-0.15\times 10^{-1} $ & $+0.75\times 10^{-4} $\\
$ 1 $  & $-0.71\times 10^{-2} $ & $-0.15\times 10^{-3} $ & $+0.18\times 10^{-3} $ & $-0.29\times 10^{-2} $\\
$ 2 $  & $-0.13\times 10^{-3} $ & $-0.15\times 10^{-1} $ & $+1.00 $ & $-0.63\times 10^{-4} $\\
$ 3 $  & $+0.16\times 10^{-3} $ & $-1.00 $ & $-0.15\times 10^{-1} $ & $+0.75\times 10^{-4} $\\
\hline%\noalign{\smallskip}
% Eq.(6), random bath ; qmeq13.x-ns1ne32b1jzz1hxs1hzs0xy1zz1hx0.25hz0.25l0.1-t200-i2.b.bath
% # & $-0.64\times 10^{-2} $ & $+0.16\times 10^{-3} $ & $+0.14\times 10^{-3} $ & $-0.26\times 10^{-2} $\\
% # & $+0.75\times 10^{-3} $ & $-0.14\times 10^{-1} $ & $+1.00\times 10^{+0} $ & $+0.64\times 10^{-4} $\\
% # & $-0.16\times 10^{-3} $ & $-1.00\times 10^{+0} $ & $-0.15\times 10^{-1} $ & $+0.64\times 10^{-4} $\\
$ 1 $  & $-0.64\times 10^{-2} $ & $+0.16\times 10^{-3} $ & $+0.14\times 10^{-3} $ & $-0.26\times 10^{-2} $\\
$ 2 $  & $+0.75\times 10^{-3} $ & $-0.14\times 10^{-1} $ & $+1.00 $ & $+0.64\times 10^{-4} $\\
$ 3 $  & $-0.16\times 10^{-3} $ & $-1.00 $ & $-0.15\times 10^{-1} $ & $+0.64\times 10^{-4} $\\
\hline%\noalign{\smallskip}
%\end{ruledtabular}
\end{tabular*}
\label{tab3}
\end{table}

%%%%%%%%%%%%%%%%%%%%%%%%%%%%%%%%%%%%%%%%%%%%%%%%%%%%%%%%%%%%%%%%%%%%%%%%%%%%%%%%%%%%%%%%%%%%%%%%%%%

\section{Exceptions}\label{section9}

The simulation results presented in Secs.~\ref{section7} and \ref{section8} strongly
suggest that, disregarding some minor quantitative differences,
the complicated Schr\"odinger dynamics of the system interacting with the bath
can be modeled by the much simpler QMEQ of the form (\ref{GBE}).
But, as mentioned in Sec.~\ref{section4}, there are several
approximations involved to justify the reduction of the Schr\"odinger dynamics to a QMEQ.
In this section, we consider a few examples for which this reduction may fail.

\begin{figure}[t]
\begin{center}
\includegraphics[width=0.33\hsize]{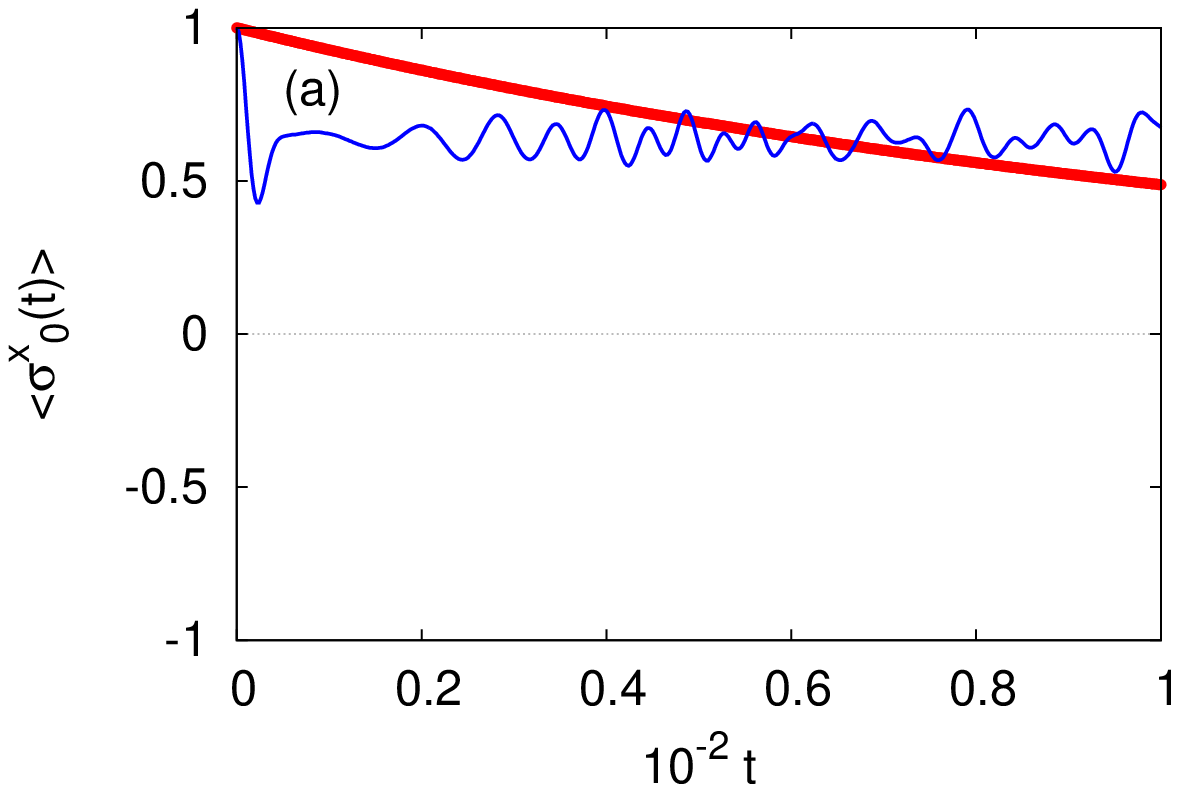}
\includegraphics[width=0.33\hsize]{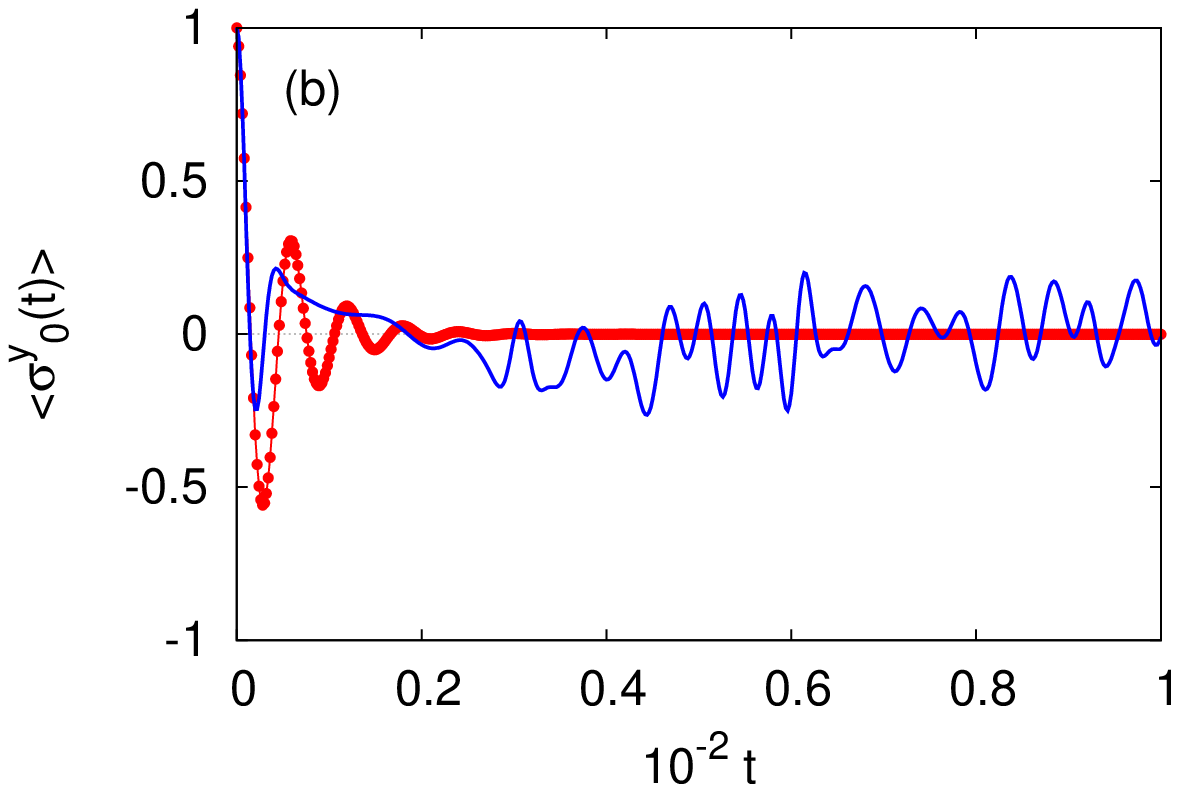}
\includegraphics[width=0.33\hsize]{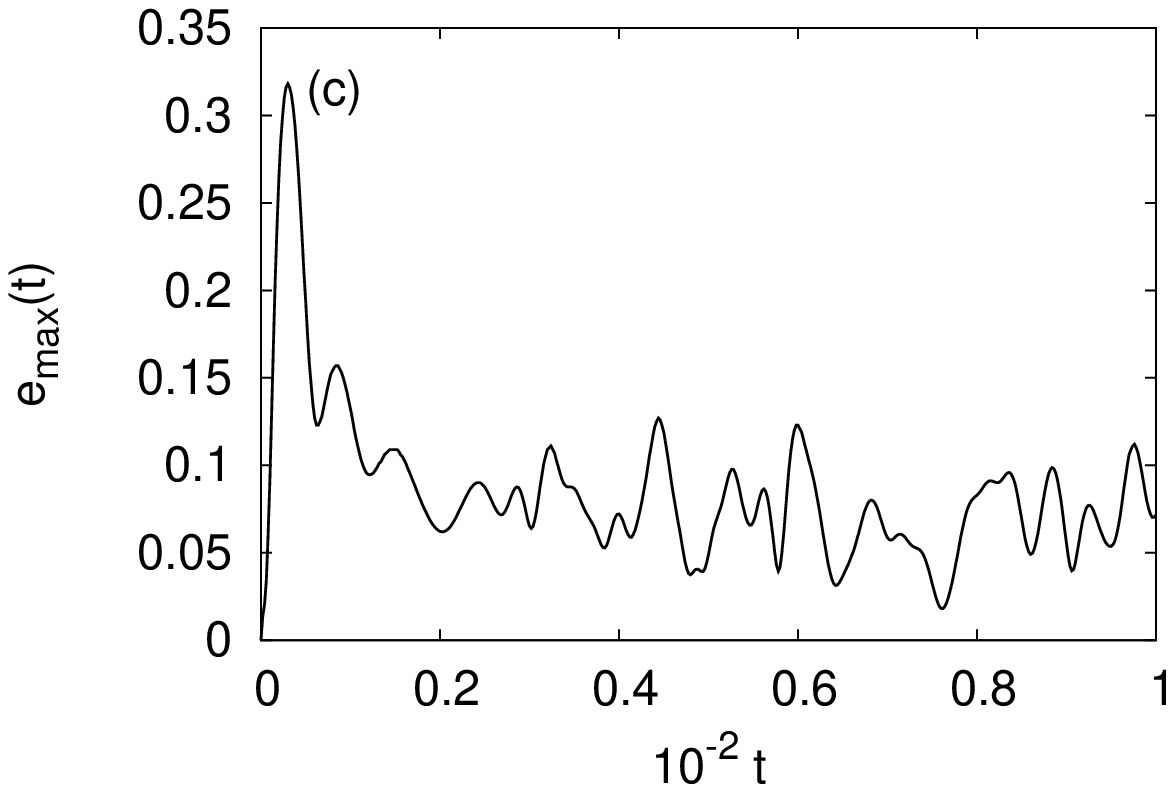}
\caption{(color online) %
Simulation data for a bath with $N_\mathrm{B}=32$ spins prepared at $\beta=0$ and system-bath interaction $\lambda=0.2$.
The system Hamiltonian is given by Eq.~(\ref{s41}).
The system-bath interaction is given by Eq.~(\ref{s43}) with $J^x_n=J^y_n=J^z_n=1/4$.
The bath Hamiltonian is given by Eq.~(\ref{s42}) with $K=-1/4$, $\Delta=1$ and $h^x_n=h^z_n=0$.
The full Hamiltonian is given by Eq.~(\ref{eq9a}).
Figures (a,b) show TDSE data (solid lines) and QMEQ data (solid circles).
(a) initial state $|x\rangle |\phi\rangle$;
(b) initial state $|y\rangle |\phi\rangle$;
(c) the error $e_{\mathrm{max}}(t)$.
The data obtained with the initial state $|\uparrow\rangle |\phi\rangle$ is
very similar as the data obtained with the initial state $|y\rangle |\phi\rangle$
and are therefore not shown.
With this choice of parameters of bath and system-bath Hamiltonians, the system does not relax
to its thermal equilibrium state
$\lim_{t\rightarrow\infty}\langle \sigma^x_0(t)\rangle=
\lim_{t\rightarrow\infty}\langle \sigma^y_0(t)\rangle=
\lim_{t\rightarrow\infty}\langle \sigma^z_0(t)\rangle=0$.
For clarity, the data is shown with a time interval of 0.4.
The TDSE solver provided 3000 numbers as input to the least-square procedure.
}
\label{fig9a}
\end{center}
\end{figure}

The first case that we consider is defined by the Hamiltonian
\begin{eqnarray}
H&=& -h^x \sigma^x_{0}+\frac{\lambda}{4}\sum_{n=1}^{N_{\mathrm{B}}}\left(\sigma^x_{n}\sigma^x_{0}+\sigma^y_{n}\sigma^y_{0}+\sigma^z_{n}\sigma^z_{0}\right)
+\frac{1}{4}\sum_{n=1}^{N_{\mathrm{B}}}\left( \sigma^x_{n}\sigma^x_{n+1}+\sigma^y_{n}\sigma^y_{n+1}+\sigma^z_{n}\sigma^z_{n+1}\right)
.
\label{eq9a}
\end{eqnarray}
In other words, both the system-bath and intra-bath interactions are of the isotropic antiferromagnetic Heisenberg type
and all interaction strengths are constant.
The simulation results for this case are presented in Fig.~\ref{fig9a}.
From Fig.~\ref{fig9a}(a), it is immediately clear that the system does not relax to its
thermal equilibrium state at $\beta=0$ (for which $\lim_{t\rightarrow\infty}\langle \sigma^x_0(t)\rangle=0$).
Apparently, the bath Hamiltonian is too ``regular'' to drive the system to its thermal equilibrium state, hence
it is also not surprising that the attempt to let the QMEQ describe the Schr\"odinger dynamics fails.

\begin{figure}[t]
\begin{center}
\includegraphics[width=0.33\hsize]{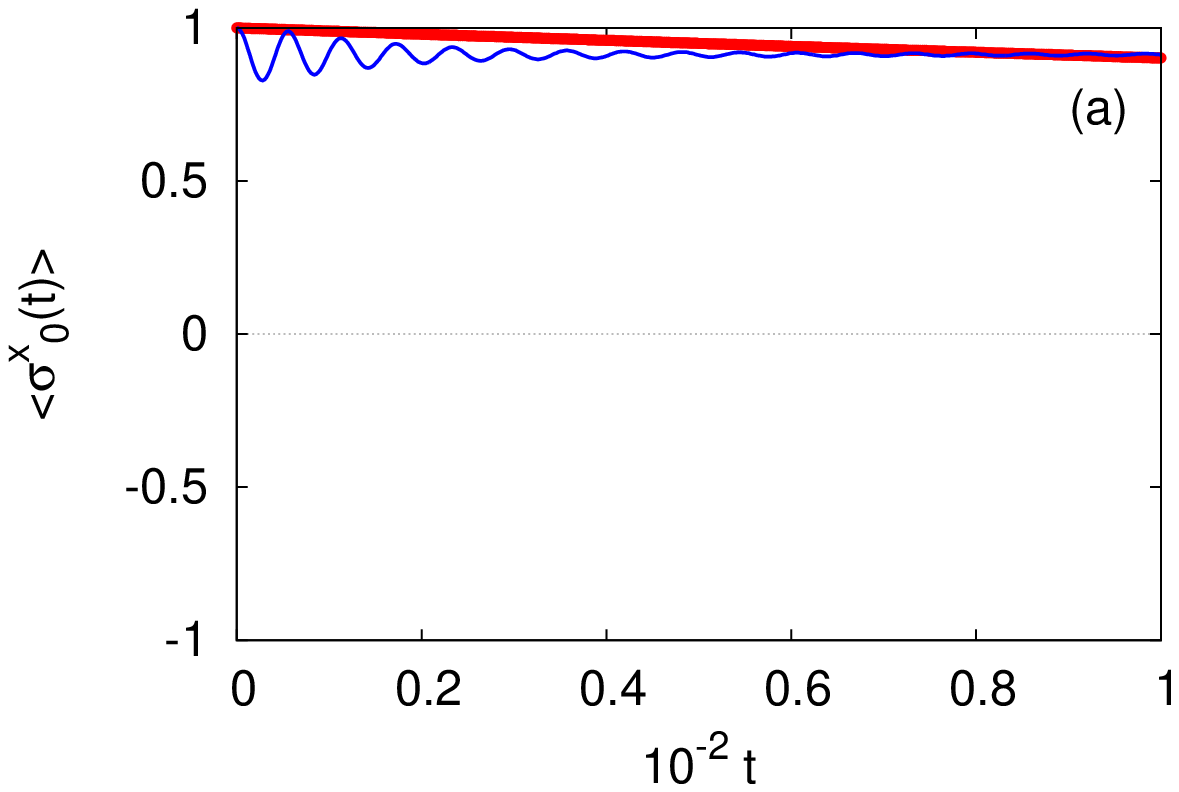}
\includegraphics[width=0.33\hsize]{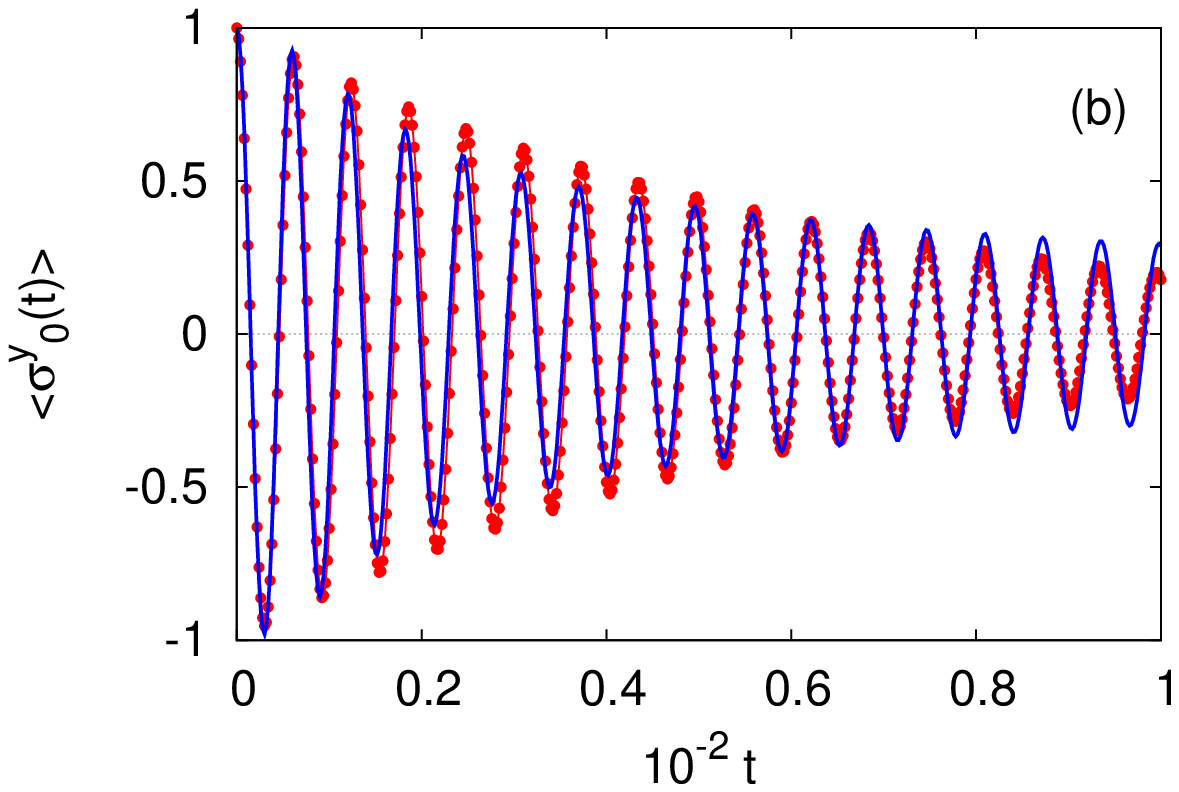}
\includegraphics[width=0.33\hsize]{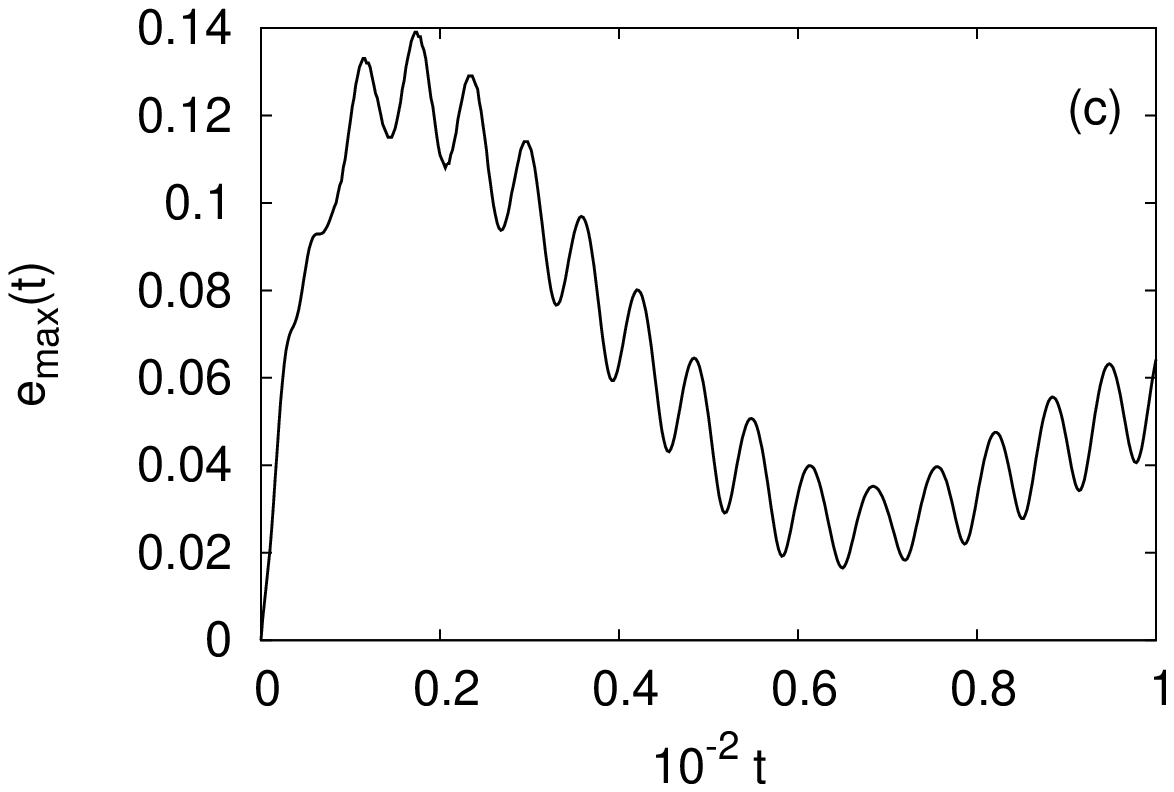}
\caption{(color online) %
Simulation data for a bath with $N_\mathrm{B}=32$ spins prepared at $\beta=0$
and system-bath interaction $\lambda=0.2$.
The system Hamiltonian is given by Eq.~(\ref{s41}).
The system-bath interaction is given by Eq.~(\ref{s43}) with $J^x_n=J^y_n=0$ and $J^z_n$
uniformly random between $-1/4$ and $1/4$, in which case the interaction
of the system and bath spins is through the coupling of the $z$-components of the spins only.
The bath Hamiltonian is given by Eq.~(\ref{s42a}) with $K^x_n=K^y_n=h^x_n=h^z_n=0$
and $K^z_n$ uniformly random between $-1$ and $1$.
The full Hamiltonian is given by Eq.~(\ref{eq9b}).
(a),(b) Show TDSE data (solid lines) and QMEQ data (solid circles).
(a) initial state $|x\rangle |\phi\rangle$;
(b) initial state $|y\rangle |\phi\rangle$;
(c) the error $e_{\mathrm{max}}(t)$.
The data obtained with the initial state $|\uparrow\rangle |\phi\rangle$ are
very similar as the data obtained with the initial state $|y\rangle |\phi\rangle$
and are therefore not shown.
With this choice of bath and system-bath Hamiltonians, the system does not relax
to its thermal equilibrium state
$\lim_{t\rightarrow\infty}\langle \sigma^x_0(t)\rangle=
\lim_{t\rightarrow\infty}\langle \sigma^y_0(t)\rangle=
\lim_{t\rightarrow\infty}\langle \sigma^z_0(t)\rangle=0$.
For clarity, the data are shown with a time interval of 0.4.
The TDSE solver provided 3000 numbers as input to the least-square procedure.
}
\label{fig9b}
\end{center}
\end{figure}

The second case that we consider is defined by the Hamiltonian
\begin{eqnarray}
H&=& -h^x \sigma^x_{0}+\frac{\lambda}{4}\sum_{n=1}^{N_{\mathrm{B}}} J^z_n\sigma^z_{n}\sigma^z_{0}
+\frac{1}{4}\sum_{n=1}^{N_{\mathrm{B}}} \sigma^z_{n}\sigma^z_{n+1}
,
\label{eq9b}
\end{eqnarray}
with system-bath interactions $J^z_n$ chosen at random and distributed uniformly over the interval $[-1,1]$
and the bath is modeled by an Ising Hamiltonian.
The model Eq.~(\ref{eq9b}) is known to exhibit quantum oscillations in the absence of quantum coherence~\cite{DOBR03z}.
As the bath Hamiltonian commutes with all other terms of the Hamiltonian,
the only non-zero bath correlation $C(3,3,t)$ is constant in time,
hence one of the basic assumptions in deriving the QMEQ Eq.~(\ref{GBE}) does not hold.

Because of the special structure of the Hamiltonian Eq.~(\ref{eq9b}) it
is straightforward to compute closed form expressions for the
expectation values of the system spin.
For $\beta=0$ we find
\begin{eqnarray}
z\langle \sigma^x_0(t)\rangle&=&1-2\lambda^2
\left\langle \left\langle
\frac{{\cal B}^2\sin^2 t\sqrt{(h^x)^2+{\cal B}^2} }{ (h^x)^2+{\cal B}^2 }
\right\rangle\right\rangle
\quad,\quad |\Psi(t=0)\rangle=|x\rangle |\phi\rangle
,
\label{eq9c1}
\\
\langle \sigma^y_0(t)\rangle&=&
\left\langle \left\langle
\cos 2 t\sqrt{(h^x)^2+{\cal B}^2}
\right\rangle\right\rangle
\quad,\quad |\Psi(t=0)\rangle=|y\rangle |\phi\rangle
,
\label{eq9c2}
\\
\langle \sigma^z_0(t)\rangle&=&1-2\lambda^2(h^x)^2
\left\langle \left\langle
\frac{\sin^2 t\sqrt{(h^x)^2+{\cal B}^2} }{ (h^x)^2+{\cal B}^2 }
\right\rangle\right\rangle
\quad,\quad |\Psi(t=0)\rangle=|\uparrow\rangle |\phi\rangle
,
\label{eq9c3}
\end{eqnarray}
where ${\cal B}={\cal B}(\{s_n\})=\sum_{n=1}^{N_{\mathrm{B}}} J^z_n s_n$ and
\begin{eqnarray}
\left\langle \left\langle {\cal X} \right\rangle\right\rangle &\equiv&
%2^{-N_{\mathrm{B}}}
\sum_{\{s_1=\pm1\}}\ldots \sum_{\{s_{N_{\mathrm{B}}}=\pm1\}} \vert \langle s_1\ldots s_{N_{\mathrm{B}}}|\phi\rangle\vert^2  {\cal X}(\{s_n\})
,
\label{eq9d}
\end{eqnarray}
denotes the average over all the bath-spin configurations.

From Eq.~(\ref{eq9c1}) it follows immediately that if
the system+bath is initially in the state $|\Psi(t=0)\rangle=|x\rangle |\phi\rangle$,
we must have $\langle \sigma^x_0(t)\rangle\ge 1-2\lambda^2$.
Hence the system will never relax to its thermal equilibrium state [for which
$\lim_{t\rightarrow\infty}\langle \sigma^x_0(t)\rangle=0$].
Nevertheless, from Fig.~\ref{fig9b} it may still seem that the QMEQ captures the essential features
of the Schr\"odinger dynamics but the qualitative agreement is a little misleading.
More insight into this aspect can be obtained by considering the limit of a very larger number of bath spins $N_{\mathrm{B}}$,
by assuming $|\phi\rangle$ to be a uniform superposition of the $2^{N_{\mathrm{B}}}$ different bath states and by
approximating ${\cal B}$, being a sum of independent uniform random variables, by a Gaussian random variable.
Then we have (after substituting ${\cal B}=h^x u$)
\begin{eqnarray}
\langle \sigma^y(t)\rangle&=&
\frac{1}{\sigma\sqrt{2\pi}}\int_{-\infty}^{+\infty}\;du\; e^{-u^2\theta/2\sigma^2} \cos\left( 2 t h^x\sqrt{1+u^2}\right)
.
\label{eq9e}
\end{eqnarray}
For large $t$, we can evaluate Eq.~(\ref{eq9e}) by the stationary phase method and we find that
$\langle \sigma^y(t)\rangle$ decays as $1/\sqrt{t}$.
Such a slow algebraic decay cannot result from a time evolution described by
a single matrix exponential $e^{t\mathbf{A}}$.
In other words, the apparent agreement shown in Fig.~(\ref{fig9b}) is due
to the relatively short time interval covered.
On the other hand, as already mentioned, the model defined by Eq.~(\ref{eq9b})
is rather exceptional in the sense that the bath correlations do not
exhibit any dynamics.
Hence it is not a surprise that the QMEQ cannot capture the $1/\sqrt{t}$ dependence.

\begin{figure}[t]
\begin{center}
\includegraphics[width=0.33\hsize]{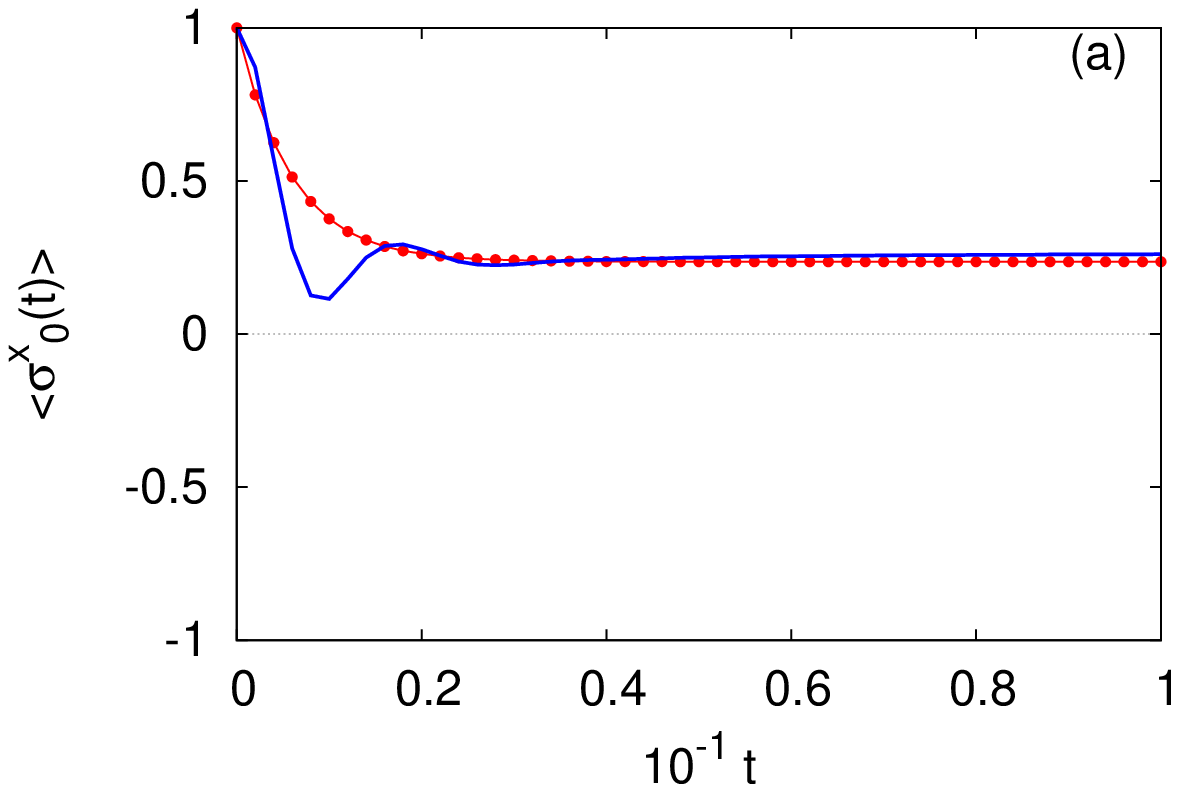}
\includegraphics[width=0.33\hsize]{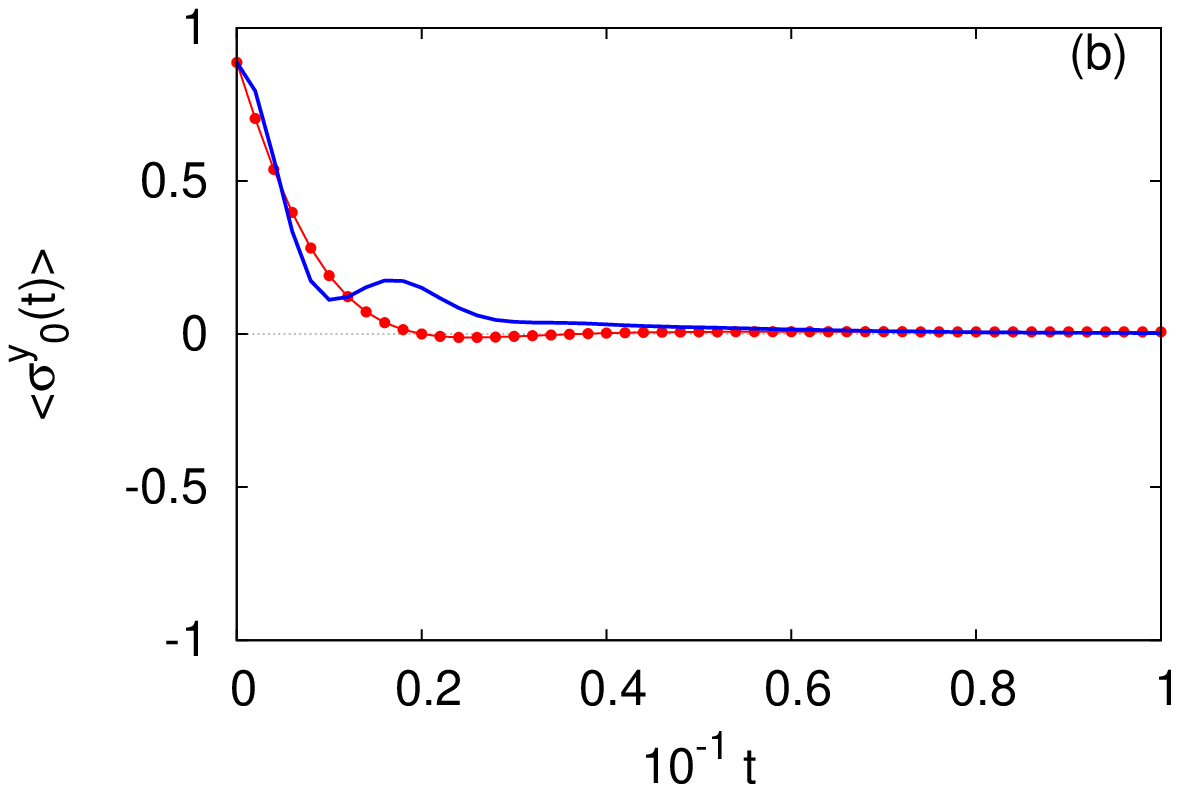}
\includegraphics[width=0.33\hsize]{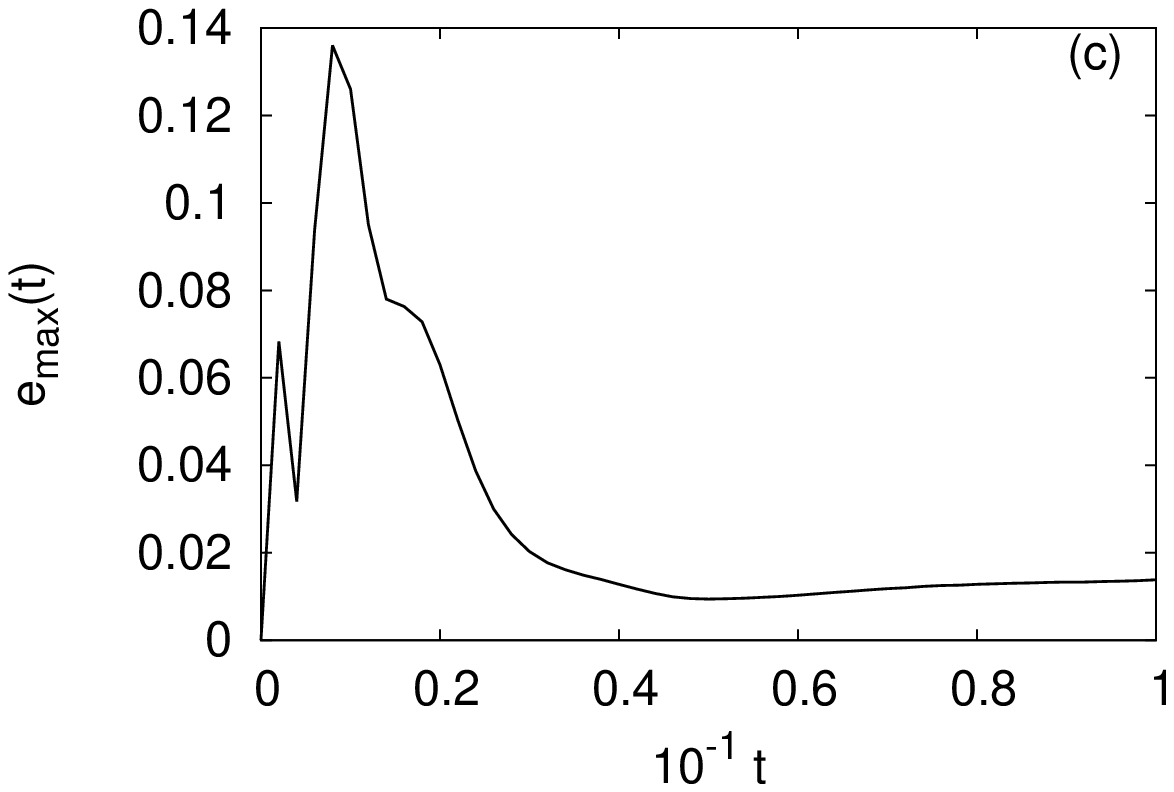}
\caption{(color online) %
Same as Fig.~\ref{fig4j} except that $\beta=1$ and $\lambda=1$.
}
\label{fig9c}
\end{center}
\end{figure}

Finally, in Fig.~\ref{fig9c} we illustrate what happens if the $\lambda=1$,
that is if the system-bath interaction becomes comparable to the other energy scales $h^x$ and $K$.
Then, the perturbation expansion that is used to derive the QMEQ of the form Eq.~(\ref{GBE})
is no longer expected to hold~\cite{BREU02}.
The data presented in Fig.~\ref{fig9c} clearly show that
even though the time it takes for the system to reach the stationary state is rather short
(because $\lambda=1$), the QMEQ fails to capture, even qualitatively, the dynamic behavior
of the system.
Note that the Schr\"odinger dynamics drives the system to a stationary state
which is far from the thermal equilibrium state of the isolated system.
The TDSE solution yields
$\langle \sigma^x_0(t=100)\rangle=0.264$ [$|\langle \sigma^z_0(t=100)\rangle|\le 10^{-2}$
$|\langle \sigma^z_0(t=100)\rangle|\le 10^{-2}$],
whereas from statistical mechanics for the isolated system at $\beta=1$  we expect
$\langle \sigma^x_0\rangle=\tanh(1/2)=0.462$ [$\langle \sigma^y_0(t=100)\rangle=\langle \sigma^z_0(t=100)\rangle=0$],
a significant difference which, in view of the strong system-bath interaction, is not entirely unexpected.

\section{Summary}\label{section10}

We have addressed the question to what extent a quantum master equation of the form \ref{GBE})
captures the salient features of the exact Schr\"odinger equation dynamics of a single spin coupled to a bath of spins.
The approach taken was to solve the time-dependent Schr\"odinger equation
of the whole system and fit the data of the expectation values of the spin
components to those of a quantum master equation of the form (\ref{GBE}).

In all cases in which the approximations used to derive a quantum master equation of the form (\ref{GBE})
seem justified, it was found that the quantum master equation (\ref{GBE})
extracted from the solutions of the time-dependent Schr\"odinger equation
describes these solutions rather well.
The least-square procedure that is used to fit the quantum master equation (\ref{GBE}) data to the
time-dependent Schr\"odinger data accounts for non-Markovian effects and nonperturbative contributions.
Quantitatively, we found that differences between the data produced by the quantum master equation, obtained by least-square fitting to the
time-dependent Schr\"odinger data, and the latter data increases with decreasing temperature.

The main finding of this work is that the exact Schr\"odinger dynamics
of a single spin-1/2 object interacting with a spin-1/2 bath can be accurately and effectively
described by Eq.~(\ref{GBE}) which, for convenience of the reader, is repeated here and reads as
\begin{eqnarray}
\frac{\partial \bm{\rho}(t)}{\partial t}= \mathbf{A}\bm{\rho}(t) + \mathbf{b}
,
\label{GBEz}
\end{eqnarray}
where the $3\times3$ matrix $\mathbf{A}$ and the three elements of the vector $\mathbf{b}$ are time independent.
As the mathematical structure of the (Markovian) quantum master equation (\ref{GBEz})
is the same as that of the Bloch equation (\ref{BE2}),
as a phenomenological description, the quantum master equation (\ref{GBEz}) offers no advantages over the latter.
Of course, when the system contains more than one spin, the Bloch equation
can no longer be used whereas the quantum master equation (\ref{GBEz}) still has the potential
to describe the dynamics.
We relegate the assessment of the quantum master equation approach
to systems of two or more spins to a future research project.

\section*{Acknowledgements}
We thank Dr. Takashi Mori for valuable discussions.
Work of P.L. Zhao is supported by the China Scholarship Council (No.201306890009).
Work of S.M. was supported by Grants-in-Aid for Scientific Research C (25400391) from MEXT of Japan, and the Elements
Strategy Initiative Center for Magnetic Materials under the outsourcing project of MEXT.
The authors gratefully acknowledge the computing time granted by the JARA-HPC
Vergabegremium and provided on the JARA-HPC Partition part of the supercomputer JUQUEEN~\cite{JUQUEEN} at Forschungszentrum J{\"u}lich.

\appendix
\section{Bloch equations}\label{section6}

Whatever method we use to extract $e^{\tau\mathbf{A}}$ and $\mathbf{B}$, it is necessary to validate the method by
applying it to a non-trivial problem for which we know the answer for sure.
The Bloch equations, originally introduced by Felix Bloch~\cite{BLOC46} as
phenomenological equations to describe the equations of motion of nuclear magnetization,
provide an excellent test bed for the extraction algorithm presented in Sec.~\ref{section5}.

In matrix notation the Bloch equations read as
\begin{eqnarray}
\frac{d\mathbf{M}(t)}{d t}
&=&
\mathbf{\widehat{A}}\mathbf{M}(t) +\mathbf{\widehat{b}}
,
\label{BE2}
\end{eqnarray}
where $\mathbf{M}$ is the magnetization,
\begin{eqnarray}
\mathbf{\widehat{A}}&=&
\left(\begin{array}{rrr}
-1/T_2 & h_z & -h_y \\
-h_z & -1/T_2 & h_x\\
h_y &-h_x & -1/T_1
\end{array}\right)
,
\label{BE3}
\end{eqnarray}
and $\mathbf{\widehat{b}}=\mathbf{M}_{0}/T_1$ where $\mathbf{M}_{0}$ is the steady state magnetization.
The transverse and longitudinal relaxation times $T_2$ and $T_1$ are strictly larger than zero.
The special but interesting case in which there is no relaxation corresponds to $1/T_1=1/T_2=0$,

Obviously Eq.~(\ref{BE2}) has the same form as Eq.~(\ref{GBE}).
Hence we can use Eq.~(\ref{BE2}) to generate the data
$\bm{\rho(t)}=\mathbf{M}(t)$ that is needed to test the algorithm described in Sec.~\ref{section5}.
In order that the identification $\bm{\rho(t)}=\mathbf{M}(t)$ makes sense
in the context of the quantum master equation
we have to impose the trivial condition that $\Vert\mathbf{M}(t=0)\Vert\le1$
and $\Vert\mathbf{M}_0\Vert\le1$.

We generate the test data by integrating Eq.~(\ref{BE2}).
In practice, we compute $e^{\tau \mathbf{\widehat{A}}}$ using the second-order product-formula~\cite{SUZU85}
\begin{eqnarray}
e^{\tau \mathbf{\widehat{A}}} &\approx& e^{\tau \widetilde{\mathbf{A}}}
= \left(e^{\tau \mathbf{A}_1/2m}e^{\tau \mathbf{A}_2/m}e^{\tau \mathbf{A}_1/2m}\right)^m
,
\label{BE3a}
\end{eqnarray}
where $\mathbf{\widehat{A}}=\mathbf{A}_1+\mathbf{A}_2$ and
\begin{eqnarray}
\mathbf{A}_1
&=&
\left(\begin{array}{rrr}
-1/T_2&0&0\\
0&-1/T_2&0\\
0&0&-1/T_1
\end{array}\right)
,
%\nonumber
\\
\mathbf{A}_2
&=&
\left(\begin{array}{rrr}
0&h_z&-h_y\\
-h_z&0&h_x\\
h_y&-h_x&0
\end{array}\right)
.
\label{BE7}
\end{eqnarray}
The second-order product-formula approximation satisfies the bound
%\begin{eqnarray}
$\Vert e^{\tau \mathbf{\widehat{A}}}- e^{\tau \widetilde{\mathbf{A}}}\Vert \le c_2\tau^3/m^2$
%,
%\label{BE3b}
%\end{eqnarray}
where the constant $c_2={\cal O}(\Vert[A_1,A_2]\Vert)$.
Hence the error incurred by the approximation is known and can be reduced systematically by increasing $m$.

It is straightforward to compute the closed form expressions
of the matrix exponentials that appear
in the second-order product-formula.
We have
\begin{eqnarray}
e^{\tau \mathbf{A}_1}&=&
\left(\begin{array}{rrr}
e^{-\tau/T_2}&0&0\\
0&e^{-\tau/T_2}&0\\
0&0&e^{-\tau/T_1}
\end{array}\right)
\nonumber \\
e^{\tau \mathbf{A}_2}&=&
\frac{1}{\Omega^2}
\left(\begin{array}{ccc}
h_x^2 + (h_y^2+h_z^2)\cos\tau \Omega&
h_x h_y (1-\cos\tau \Omega) + h_z\Omega \sin\tau \Omega &
h_x h_z (1-\cos\tau \Omega) - h_y\Omega \sin\tau \Omega \\
h_x h_y (1-\cos\tau \Omega) - h_z\Omega \sin\tau \Omega &
h_y^2 + (h_x^2+h_z^2)\cos\tau \Omega&
h_y h_z (1-\cos\tau \Omega) + h_x\Omega \sin\tau \Omega \\
h_x h_z (1-\cos\tau \Omega) + h_y\Omega \sin\tau \Omega &
h_y h_z (1-\cos\tau \Omega) - h_x\Omega \sin\tau \Omega &
h_z^2 + (h_x^2+h_y^2)\cos\tau \Omega
\end{array}\right)
,
\label{BE8}
\end{eqnarray}
where $\Omega^2=h_x^2 + h_y^2+h_z^2$.

Summarizing, the numerical solution of the Bloch equations Eq.~(\ref{BE2}) is given by
\begin{eqnarray}
\bm{\rho}(t+\tau) &=&e^{\tau\widetilde{\mathbf{A}}}\bm{\rho}(t) + \widetilde{\mathbf{B}}
,
\label{QMEQ7c}
\end{eqnarray}
where $\bm{\rho}(t)=\mathbf{M}(t)$
and
the trapezium rule was used to write
\begin{eqnarray}
\mathbf{\widehat{B}}&=&\int_{0}^{\tau} e^{(\tau-u){\mathbf{\widehat{A}}}}\mathbf{\widehat{b}}\;du
\approx\frac{\tau}{2}\left(\openone+e^{\tau\widetilde{\mathbf{A}}}\right)\mathbf{\widehat{b}}=\widetilde{\mathbf{B}}
.
\label{QMEQ7b}
\end{eqnarray}
The approximate solution obtained from Eqs.~(\ref{QMEQ7c}) and~(\ref{QMEQ7b})
will converge to the solution of Eq.~(\ref{BE2}) as $\tau\rightarrow0$.
Clearly, Eq.~(\ref{QMEQ7c}) has the same structure as Eq.~(\ref{QMEQ7}) and
hence we can use the solution of the Bloch equations as input data for
testing the extraction algorithm.
Note that the extraction algorithm
is expected to yield $e^{\tau \widetilde{\mathbf{A}}}$ and $\widetilde{\mathbf{B}}$,
not  $e^{\tau {\mathbf{\widehat{A}}}}$ and $\mathbf{\widehat{B}}$.

\subsection{Validation procedure}\label{section6a}

We use the Bloch equation model to generate the data set
${\cal D}=\{\bm\rho(k\tau)|\; 0 \le k \le N-1\}$.
The validation procedure consists of the following steps:
\begin{enumerate}
\item
Choose the model parameters $h_x$, $h_y$, $h_z$, $1/T_1$, $1/T_2$ and the steady-state magnetization $\mathbf{M}_0$.
\item
Choose $\tau$ and $m$.
\item
For each of the three initial states
$\rho^{(1)}(0)=(1,0,0)^T$, $\rho^{(2)}(0)=(0,1,0)^T$, and $\rho^{(3)}(0)=(0,0,1)^T$
repeat the operation
\begin{eqnarray}
\rho^{(j)}((k+1)\tau)\leftarrow e^{\tau \widetilde{\mathbf{A}}}\rho^{(j)}(k\tau) + \widetilde{\mathbf{B}}
\quad,\quad k=0,\dots,N-1\quad,\quad j=1,2,3,
\nonumber
\end{eqnarray}
and store these data.
\item
Use the data $\{\rho^{(j)}(k\tau)\}$ to construct the matrices
$3\times 3N$ matrix $\mathbf{Z}=(\mathbf{Z}^{(1)}\,\mathbf{Z}^{(2)}\,\mathbf{Z}^{(3)})$
and the $4\times 3N$ matrix $\mathbf{X}=(\mathbf{X}^{(1)}\,\mathbf{X}^{(2)}\,\mathbf{X}^{(3)})$.
Then use the singular value decomposition of $\mathbf{X}$
to compute the matrix $\mathbf{Y}$ according to Eq.~(\ref{QMEQ9})
and extract the matrix $e^{\tau\mathbf{A}}$ and vector $\mathbf{B}$ from it, see Eq.~(\ref{QMEQ11}).
If one or more of the singular values are zero, the extraction failed.
\item
Compute the relative errors
\begin{eqnarray}
e_\mathrm{A}&=&\Vert e^{\tau \widetilde{\mathbf{A}}} -  e^{\tau \mathbf{\widehat{A}}} \Vert/\Vert e^{\tau \mathbf{\widehat{A}}}\Vert,
\\
e_\mathrm{B}&=&\Vert \widetilde{\mathbf{B}}- \mathbf{\widehat{B}}\Vert/\Vert  \mathbf{\widehat{B}} \Vert,
\\
e_{\bm\rho}&=&\max_k\Vert \bm{\rho}((k+1)\tau) - e^{\tau\widetilde{\mathbf{A}}}\bm{\rho}(k\tau) -\widetilde{\mathbf{B}}\Vert/
\Vert \bm{\rho}(k\tau)\Vert
.
\end{eqnarray}
\end{enumerate}
A necessary condition for the algorithm to yield reliable results is that
the errors $e_\mathrm{A}$ and $e_\mathrm{B}$ are small, of the order of $10^{-10}$.
Indeed, if one or more of the singular values are zero
and the extraction has failed, $e_{\bm\rho}$ may be (very) small but $e_\mathrm{A}$ or $e_\mathrm{B}$ is not.

In the case that is of interest to us, the case in which the whole system evolves according
to the TDSE, we do not know $e^{\tau \mathbf{A}}$ nor $\mathbf{B}$
and a small value of  $e_{\bm\rho}$ is, by itself, no guarantee that the extraction process worked properly.
Hence, it also is important to check that all singular values are nonzero.

\subsection{Numerical results}\label{section6b}

In Table~\ref{tab1} we present some representative results for the errors incurred by
the extraction process.
In all cases, the relative errors on the estimate of the time evolution operator and
the constant term are for the present purpose, rather small.
Therefore, the algorithm to extract the time evolution operator $e^{\tau\widetilde{\mathbf{A}}}$ and
constant term $\widetilde{\mathbf{B}}$ appearing in the time evolution equation Eq.~(\ref{QMEQ7})  from
the data obtained by solving the TDSE yields accurate results
when the data are taken from the solution of the Bloch equations.
No exceptions have been found yet.

\begin{table}[ht]
\caption{
The errors $e_\mathbf{A}$, $e_\mathbf{B}$, and $e_{\bm\rho}$  as obtained
fitting the matrix $e^{\tau\mathbf{A}}$ and the constant term $\mathbf{B}$,
to the data of the numerical solution of the Bloch equation with three different
initial conditions (see text).
The Bloch equations are solved for $N=500$ steps with the time step $\tau$.
The value of the vector $\mathbf{M}_0=(0,0,0.4)^T$.
The data of the whole time interval $[0,N-1]$ were used for the least-square fitting procedure.
The column labeled $\Sigma_i\not=0$ indicates whether all singular values are nonzero or not.
For the meaning of all other symbols, see text.
}
%\begin{ruledtabular}
%\begin{tabular*}{}
\begin{tabular*}{\textwidth}{ c @{\extracolsep{\fill}} ccccccc|cccccc}
\noalign{\medskip}
\hline\hline%\noalign{\smallskip}
 &$h_x$ & $h_y$ & $h_z$ & $1/T_1$ & $1/T_2$ & $\tau$ &  &$\Sigma_i\not=0$ &
$e_\mathbf{A}<10^{-10}$ & $e_\mathbf{B}<10^{-10}$ & $e_{\bm\rho}<10^{-10}$ \\
\hline%\noalign{\smallskip}
 &$0.5$ & $1.5$ & $0.7$ & $0.05$ & $0.3$ & 0.1 & &\checkmark& $\checkmark$ & $\checkmark$ &$\checkmark$\\
 &$0.5$ & $1.5$ & $0.7$ & $0$ & $0$      & 0.1 & &\checkmark& $\checkmark$ & $\checkmark$ &$\checkmark$\\
 &$0.5$ & $1.5$ & $0.7$ & $0.01$ & $0$   & 1.0 & &\checkmark& $\checkmark$ & $\checkmark$ &$\checkmark$\\
 &$0$ & $0$ & $1$ & $0$ & $0$            & 1.0 & &\checkmark& $\checkmark$ & $\checkmark$ &$\checkmark$\\
\hline%\noalign{\smallskip}
\end{tabular*}
%\end{ruledtabular}
\label{tab1}
\end{table}

\section{Simulation results using the 3D bath Hamiltonian Eq.~(\ref{s42b})}\label{3D}

In this appendix, we present some additional results in support of the conclusions
drawn from the simulations of using the 1D bath Hamiltonians (\ref{s42}) and (\ref{s42a}).

Table~\ref{tab5} summarizes the results of the analysis of TDSE data, as obtained
with the 3D bath Hamiltonian Eq.~(\ref{s42b}) with random intra-bath couplings
and random $h$-fields for the bath spins.
The model parameters that were used to compute the TDSE data
are the same as those that yield the results for the 1D bath presented in Table~\ref{tab3}.
Comparing the first three rows (the parameters that appear in the Markovian master equation (\ref{GBEx})
with the corresponding last three rows
(the parameters $r_{jk}$ that appear in the Redfield quantum master equation Eq.~(\ref{QMEQ4})),
we conclude that changing the connectivity of the bath does not significantly improve
(compared to the data shown in Table~\ref{tab2})
the quantitative agreement between the data in the two sets of three rows.

In Table~\ref{tab6}, we show the effect of increasing the energy scale of the
bath spins by a factor of 10, reducing the relaxation times of the bath-correlations
by a factor of 10, i.e., closer to the regime of the Markovian limit in which Eq.~(\ref{QMEQ4}) has been derived.
The differences between the
QMEQ estimates (first three rows) and the Redfield equation estimates (second three rows)
values of $A_{2,2}$ and $A_{3,3}$ are significantly smaller than in those for the case
shows in e.g. Table~\ref{tab5} but the $A_{1,1}$ elements differ by a factor of four
and the $A(2,1)$ elements differ even much more.
Although the results presented in Tables~\ref{tab5} and \ref{tab6}
indicate that the data extracted from the TDSE
through Eq.~(\ref{GBEx}) and those obtained by calculating the parameters $r_{jk}$
that appear in the Redfield quantum master equation Eq.~(\ref{QMEQ4}) in the Markovian limit
will converge to each other, it becomes computationally very expensive to approach that limit closer.
The reason is simple: by increasing the energy-scale of the bath, it is necessary to reduce
the time step (or equivalently increase the number of terms in the Chebyshev polynomial expansion)
in order to treat the fast oscillations properly.
Keeping the same relaxation times roughly the same but taking a smaller time step requires more computation.
For instance, it takes about 4 (20) h CPU time of 16384 BlueGene/Q processors
to produce the TDSE data from which the numbers in Table~\ref{tab5} (Table~\ref{tab6}) have been obtained.

\begin{table}[tb]
\caption{
First three data rows: coefficients that appear in Eq.~(\ref{GBEx}) as obtained by
fitting the QMEQ Eq.~(\ref{GBE}) to the TDSE data for
$h^x=1/2$, $\lambda=0.1$, $N_{\mathrm{B}}=27$, the 3D bath Hamiltonian Eq.~(\ref{s42b})
with random couplings ($K=1/4$) and random $h$-fields ($h^x_{\mathrm{B}}=h^z_{\mathrm{B}}=1/4)$.
Last three rows: the corresponding coefficients as obtained by numerically
calculating the parameters $r_{jk}$ that appear in the Redfield quantum master equation Eq.~(\ref{QMEQ4})
according to Eq.~(\ref{QMEQ2b}) from the TDSE data of the bath-operator correlations.
}
%\begin{ruledtabular}
%\begin{tabular*}{}
\begin{tabular*}{\textwidth}{@{\extracolsep{\fill}} cllll}
\noalign{\medskip}
\hline\hline%\noalign{\smallskip}
$i$ & \hfil $A_{i,1}$ \hfil  & \hfil $A_{i,2}$ \hfil & \hfil $A_{i,3}$ \hfil & \hfil $b_i$ \hfil  \\
\hline%\noalign{\smallskip}
% A = log(E)/tau :
% (-0.249E-01, 0.000E+00), (-0.820E-02,-0.237E-18), ( 0.668E-03, 0.108E-17)
% ( 0.107E-01,-0.678E-19), (-0.466E-01,-0.694E-16), ( 0.992E+00,-0.867E-17)
% (-0.161E-03, 0.108E-17), (-0.998E+00, 0.000E+00), (-0.467E-01, 0.139E-15)
%
% b :
% (-0.112E-01, 0.419E-21), ( 0.911E-04,-0.833E-20), ( 0.514E-03, 0.416E-19)
$1$ &$ -0.25\times10^{-1} $&$ -0.82 \times10^{-2} $&$ +0.67\times10^{-3} $&$ -0.11\times10^{-1}$  \\
$2$ &$ +0.11\times10^{-1} $&$ -0.47 \times10^{-1} $&$ +0.99              $&$ +0.91\times10^{-4}$  \\
$3$ &$ +0.16\times10^{-3} $&$ -1.00               $&$ -0.47\times10^{-1} $&$ +0.51\times10^{-3}$  \\
\hline%\noalign{\smallskip}
% Eq.(6), random bath ; spi16mpi x-ns1ne27b1jzz1hxs1hzs0xy1zz1hx0.25hz0.25l0.1-t1000-i2d3.b
$1$ & $-0.49\times 10^{-2} $ & $+0.49\times 10^{-5} $ & $-0.80\times 10^{-4} $ & $-0.20\times 10^{-2} $\\
$2$ & $+0.15\times 10^{-3} $ & $-0.19\times 10^{-1} $ & $+1.00\times 10^{+0} $ & $+0.59\times 10^{-5} $\\
$3$ & $+0.53\times 10^{-2} $ & $-0.99\times 10^{+0} $ & $-0.19\times 10^{-1} $ & $-0.50\times 10^{-4} $\\
\hline%\noalign{\smallskip}
%\end{ruledtabular}
\end{tabular*}
\label{tab5}
\end{table}

\begin{table}[tb]
\caption{
The same as Table~\ref{tab5} except that the random couplings ($K=10/4$)
and random h-fields ($h^x_{\mathrm{B}}=h^z_{\mathrm{B}}=10/4)$.
}
%\begin{ruledtabular}
%\begin{tabular*}{}
\begin{tabular*}{\textwidth}{@{\extracolsep{\fill}} cllll}
\noalign{\medskip}
\hline\hline%\noalign{\smallskip}
$i$ & \hfil $A_{i,1}$ \hfil  & \hfil $A_{i,2}$ \hfil & \hfil $A_{i,3}$ \hfil & \hfil $b_i$ \hfil  \\
\hline%\noalign{\smallskip}
% A = log(E)/tau :
% (-0.773E-02,-0.339E-19), ( 0.202E-01, 0.000E+00), (-0.777E-04, 0.217E-17)
% (-0.194E-01, 0.000E+00), (-0.986E-02, 0.000E+00), ( 0.985E+00, 0.000E+00)
% ( 0.430E-02, 0.108E-17), (-0.985E+00, 0.542E-18), (-0.879E-02,-0.694E-16)
%
% b :
% (-0.372E-03, 0.355E-23), (-0.519E-04,-0.496E-21), (-0.211E-04, 0.512E-23)
$1$ &$ -0.77\times10^{-2} $&$ +0.20 \times10^{-1} $&$ -0.77\times10^{-4} $&$ -0.37\times10^{-3}$  \\
$2$ &$ -0.19\times10^{-1} $&$ -0.99 \times10^{-2} $&$ +0.99              $&$ -0.52\times10^{-4}$  \\
$3$ &$ +0.43\times10^{-2} $&$ -0.99               $&$ -0.88\times10^{-2} $&$ -0.21\times10^{-4}$  \\
\hline%\noalign{\smallskip}
% Eq.(7), random bath ; spi16mpi qmeq14 x-ns1ne27b0.1jzz1hxs1hzs0xy10zz10hx2.5hz2.5l0.1-t1000-i10d3.b
$1$ & $-0.16\times 10^{-2} $ & $+0.19\times 10^{-4} $ & $+0.38\times 10^{-4} $ & $-0.66\times 10^{-4} $\\
$2$ & $+0.87\times 10^{-5} $ & $-0.64\times 10^{-2} $ & $+1.00               $ & $+0.13\times 10^{-4} $\\
$3$ & $-0.67\times 10^{-2} $ & $-1.01               $ & $-0.66\times 10^{-2} $ & $+0.17\times 10^{-4} $\\
\hline%\noalign{\smallskip}
%\end{ruledtabular}
\end{tabular*}
\label{tab6}
\end{table}

%%%%%%%%%%%%%%%%%%%%%%%%%
\bibliography{../../../../all16,../qmeq}
\end{document}